\documentclass[fleqn,10pt]{wlscirep_arXiv}
\usepackage[utf8]{inputenc}
\usepackage[T1]{fontenc}
\usepackage{subcaption}

\usepackage{bibunits}

\newcommand{\tmean}[1]{\left\langle #1\right\rangle}

\def\taum{\tau_\mathrm{min}}
\def\Nina{N_\mathrm{inactive}}
\def\Nmiss{N_\mathrm{missing}}

\title{Lineage topology, replication kinetics and cell cycle synchronization reveal regulated growth dynamics in human bone marrow stromal cell colonies}

\author[1]{Alessandro Allegrezza}
\author[1]{Riccardo Beschi}
\author[1,2]{Domenico Caudo}
\author[1,3,4]{Andrea Cavagna}
\author[5]{Alessandro Corsi}
\author[3]{Antonio Culla}
\author[5,a]{Samantha Donsante}
\author[6]{Giuseppe Giannicola}
\author[1,3,4]{Irene Giardina}
\author[2,b]{Giorgio Gosti}
\author[3,7,8,9]{Tom\'as S. Grigera}
\author[1,3]{Stefania Melillo}
\author[3,5,*]{Biagio Palmisano}
\author[1,3]{Leonardo Parisi}
\author[3,c]{Lorena Postiglione}
\author[5]{Mara Riminucci}
\author[1]{Francesco Saverio Rotondi}

\affil[1] {Physics Department, Sapienza University, Rome, Italy} 
\affil[2] {Center for Life Nano \& Neuro Science, Italian Institute of Technology, Rome, Italy}
\affil[3] {Institute for Complex Systems, National Research Council, Rome, Italy} 
\affil[4] {Istituto Nazionale di Fisica Nucleare, Sezione Roma 1, Rome, Italy} 
\affil[5] {Department of Molecular Medicine, Sapienza University, Rome, Italy} 
\affil[6] {Department of Anatomical, Histological, Medico Legal and Orthopaedic Sciences, Sapienza University, Rome, Italy} 
\affil[7] {Instituto de F\'\i{}sica de L\'\i{}quidos y Sistemas Biol\'ogicos, CONICET and Universidad Nacional de La Plata,  La Plata, Argentina} 
\affil[8] {CCT CONICET La Plata, Consejo Nacional de Investigaciones Cient\'\i{}ficas y T\'ecnicas, Argentina} 
\affil[9] {Departamento de F\'\i{}sica, Facultad de Ciencias Exactas, Universidad Nacional de La Plata, Argentina \newline} 

\affil[*]{Correspondence: biagio.palmisano@uniroma1.it \newline}
\affil[a]{Current address: Tettamanti Center, Fondazione IRCCS San Gerardo dei Tintori, Monza, Italy}
\affil[b]{Current address: Institute of Heritage Science, National Research Council, Rome, Italy}
\affil[c]{Current address: Dipartimento di Ingegneria Chimica, dei Materiali e della Produzione Industriale, Universit\`a degli Studi di Napoli Federico II, Napoli, Italy}


\begin{abstract}
Bone marrow stromal cells (BMSC) -- which include skeletal stem cells -- are a promising tool in regenerative medicine. However, their heterogeneous and unpredictable {\it in vivo} behaviour remains a critical barrier preventing the development of standardized therapeutic approaches for skeletal tissue regeneration. Several studies have attempted to identify {\it in vitro} features that could correlate with the {\it in vivo} differentiation properties, yet the mechanisms ruling BMSC heterogeneity remain poorly understood. Here, using time-lapse imaging, we lineage-trace 32 single-cell-derived BMSC colonies through seven generations. We observe significant inter-colony and intra-colony heterogeneity in lineage topology (determined by the number of senescent or apoptotic cells) and in replicative kinetics (measured from proliferating cells only). Interestingly, topology and kinetics result strongly correlated, suggesting the existence of regulatory factors linking the non-dividing/apoptotic subpopulations with proliferating cells. Furthermore, BMSCs display highly synchronized cell cycles during early generations, indicating stage-specific regulatory mechanisms through which cells influence each other. By employing a non-interacting population growth model, we demonstrate that the observed synchronisation cannot be explained by an uncorrelated branching process; instead, cell-to-cell correlation of division times must exist. Our findings reveal fundamental mechanisms governing BMSC heterogeneity and growth dynamics that may inform strategies to control their regenerative potential.
\end{abstract}

\begin{document}
%\begin{bibunit}[plain]

\flushbottom
\maketitle

\thispagestyle{empty}

\vfill\eject

%TC:endignore
\section*{Introduction}

Bone marrow stromal cells (BMSCs) can be isolated from the post-natal bone marrow and cultured {\it in vitro} as adherent, fibroblast-like cells. When transplanted {\it in vivo} at heterotopic sites, they are able to generate skeletal tissue and establish a functional bone marrow microenvironment \cite{friedenstein1966osteogenesis}.  BMSCs contain a small subset of cells that are able to proliferate even in low density cultures -- i.e.\ in the absence of paracrine stimuli produced by neighbouring cells -- to form colonies \cite{friedenstein1966osteogenesis,friedenstein1974stromal}.  This clonogenic BMSCs include skeletal stem cells (SSCs) and progenitor cells normally residing in the post-natal bone marrow \cite{owen2007stromal,kuznetsov1997single, donsante2021stem}. Therefore, they are the repository of the regenerative properties observed in {\it in vivo} assays and are critical to any potential therapeutic application of BMSCs.  Colony-forming BMSCs share many biological features, including the expression of specific surface markers that can be used for prospective purification \cite{sacchetti2007self,tormin2011cd146,rennerfeldt2016concise}. However, the colonies that they generate are rather heterogeneous and show considerably different {\it in vivo} behaviour.  Indeed, colonies formed by BMSCs corresponding to genuine SSCs undergo multi-lineage differentiation and organise a bone marrow microenvironment, whereas colonies derived from BMSCs corresponding to more committed progenitor cells produce only bone \cite{kuznetsov1997single, satomura1998receptor}.  Furthermore, some colony-forming BMSCs generate a progeny that does not perform any specific function \cite{kuznetsov1997single}. 

The heterogeneous {\it in vivo} behaviour of BMSC colonies is one of the most important factors preventing the development of effective therapeutic approaches for skeletal tissue regeneration. Several studies have attempted to identify morphological and/or molecular features expressed {\it in vitro} that correlate with the differentiation properties displayed {\it in vivo} \cite{satomura1998receptor, sworder2015molecular, james2015multiparameter, makhija2024topological}.  Although specific surface markers for human SSCs have been identified \cite{chan2018identification}, {\it ex-vivo} expansion is a required step to achieve an appropriate cell number for clinical applications; hence, predictive parameters of BMSC colony behavior are needed to assess their {\it in vivo} potency \cite{samsonraj2017concise}. Therefore, new methods to sort BMSC colonies with similar potency need to be explored.  Lineage studies are a promising approach in this respect. By tracing the proliferation of colonies derived from different initial cells, we can try to identify the nature of the variability between initiator cells leading to the subsequent {\it inter-colony} heterogeneity we wish to control.  It is of particular interest whether the variability among initiator cells belonging to the same donor and expanded in the same conditions is any different from the variability across different experiments and different donors.  Moreover, different parts of the same colony may develop differently and it is important to understand whether these \emph{intra-colony} heterogeneities are strong enough to wash out any colony-level correlations, or whether these fluctuations are generated and structured in a way that keeps the collective identity of the clone.

Time-lapse microscopy has been used to reconstruct the lineage of both embryonal and adult cell colonies \cite{schroeder2011long,hilsenbeck2016software,shen2006timing,eilken2009continuous,ravin2008potency,plambeck2022heritable}. In the case of BMSCs  \cite{seiler2014time, whitfield2013onset, lee2014multivariate, rennerfeldt2019emergent}, the technical challenges are considerable, so that lineage studies have followed proliferation only for the first few generations (typically five). All studies discovered great heterogeneities and some proved that colonies may have significant variability in their differentiating potential \cite{seiler2014time, lee2014multivariate}.  An important aspect to keep in mind in this respect is that, even at low plating densities, BMSCs tend to cluster, so apparently isolated colonies may actually result from merged clones \cite{rennerfeldt2016concise}; undetected merging dangerously confounds inter-colony and intra-colony heterogeneity. Therefore, the study of clonal colonies should only be performed through lineage tracing of single-cell derived colonies, which is, however, very time-consuming; no studies to date presents more than 10 single-cell derived colonies, which is borderline for statistical analysis.

Here, we conduct a study of human BMSCs proliferation {\it in vitro} through time-lapse microscopy, reconstructing the full lineage of 32 single-cell derived clonal colonies, traced up to the seventh generation.  We characterise colonies according to two major traits: the topology of the lineage -- namely how many cells reach a specific generation and how many stop proliferating -- and the kinetics of the colony -- expressed by the statistics of the division times.  We show that, although both traits are strongly heterogeneous, fluctuations in topology and kinetics are not independent from each other, unveiling a strong correlation that connects the propensity of the colony to produce senescent and/or apoptotic cells to the proliferation properties of dividing cells. Moreover, we find that the variability observed among the colonies within each cell-culture experiment (i.e. same donor, age and passage) is as large as the overall heterogeneity across all donors, age and passages. Finally, we find a remarkable degree of synchronization among cellular divisions, which we characterise in the context of a branching process model for a population of non-interacting individuals. However, our data demonstrate that cell cycle in BMSC colonies is much more synchronized than what it would be expected from such model, therefore proving that cell-to-cell correlation must exist.
%TC:ignore

%%%%%%%%%%%%%%%%%%%%%%%%%%%%%%%%%%%%%%%%%%%%%%%%%%%%%%%%%%%%%%%%%%%%%
\section*{Methods}

\subsection*{Cell culture and Time-lapse imaging}
\label{sec:data}

\begin{figure}
\centering
\includegraphics[width=0.7\textwidth]{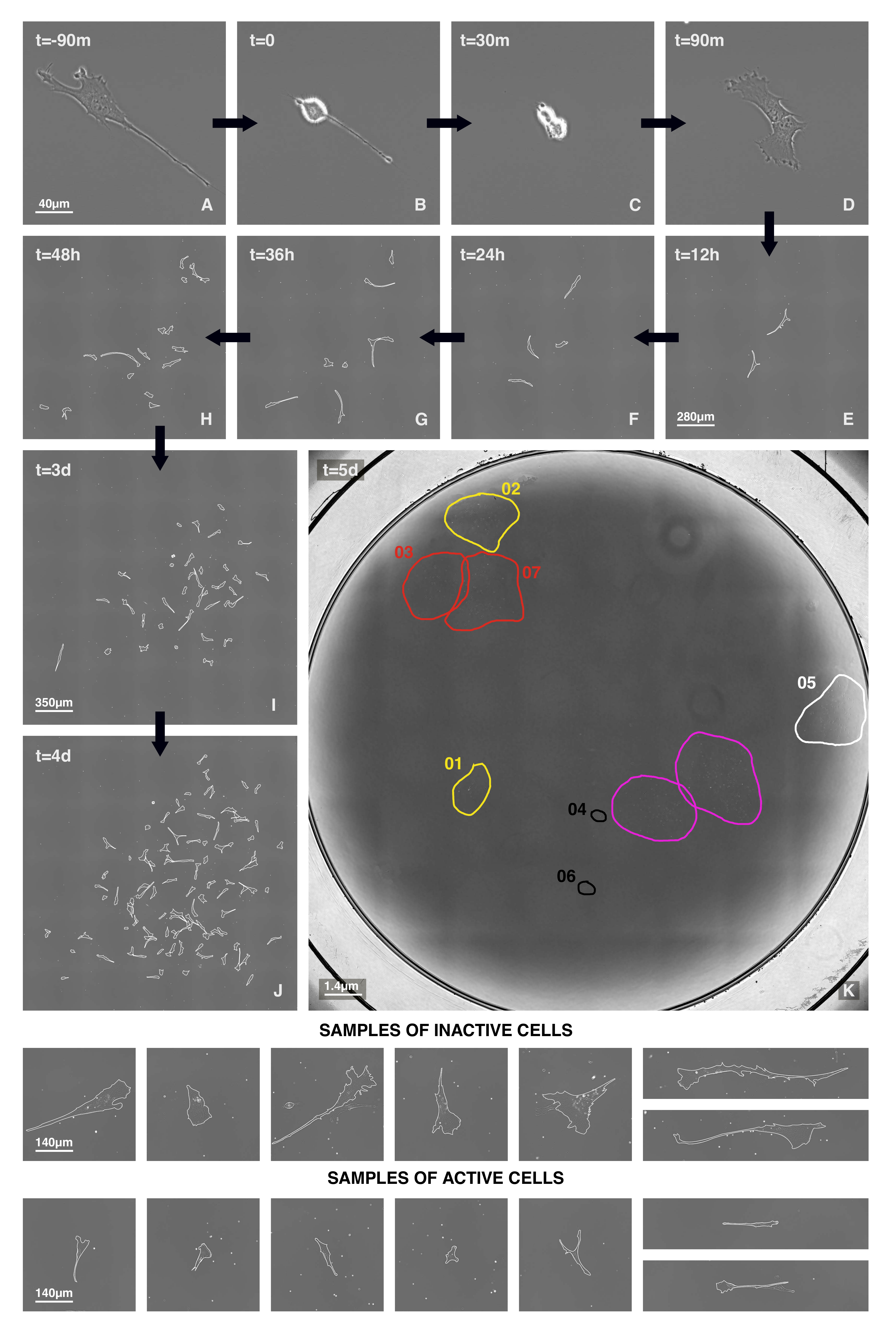}
\caption{ {\bf Time-lapse of a BMSC colony.} {\bf (A-K)} Images follow the evolution in time of colony \texttt{\texttt{20241112\_05}} from its single originating cell. 
Panels (A-J) are different scale crops of images acquired at magnification $20\times$.  {\bf (A)} The originating cell $90$ minutes before its mitosis. {\bf (B)} During the Mitotic Rounding Phase (MRP), cells assume a bright and almost spherical shape. We use the MRP as a marker for mitosis. The MRP of the originating cell marks the temporal beginning of the colony development defining the time origin ($t=0$) for the colony. {\bf (C)} Half an hour after the beginning of MRP, the two daughter cells still have a spherical shape. \textbf{(D)} At $t=90$ minutes, the two daughters are distinguishable by their well-separated nuclei. {\bf(E-H)} Within the first $48$ hours, the colony expands to the fourth generation. {\bf (I-J)} During the third and the fourth days the colony expands to the fifth and sixth generation. {\bf (K)} Overview at magnification $4\times$ of the entire dish at day five ($8\times 8$ photo tessellation). The white contour encloses the colony described in panels (A-J). Yellow contours indicate isolated colonies that we still follow on day five; red contours indicate infiltrated colonies (colonies that got in contact with each other), which we stop recording. Purple denotes colonies derived from cells that we did not select at the beginning of the experiment because of their close proximity. Finally, black curves correspond to cells that were selected but never formed colonies.\\ {\bf Bottom panels}. Samples of G$_0$ and active cells at the same magnification. G$_0$ cells have a different --flat-- morphology, and are larger than the spindle-shaped active cells.}
\label{fig:time-lapse}
\end{figure}

Detailed methods are given in the Supplementary Information (SI) for better manuscript readability; here we present a brief summary of the experimental protocol.  Data were collected during the course of 11 experiments, from January 2023 to June 2025.  Human bone marrow stromal cells (BMSCs) were isolated from bone samples harvested from 6 different healthy donors undergoing elbow orthopaedic surgery, and were plated at passage P0, P1, P2, or P3 depending on the experiment (see Supplementary Table~S1).  BMSCs were seeded in 35mm culture dishes  at very low density (10--20$\,$cells/dish), in order to maximise the probability to form colonies derived from single spatially isolated cells.  The growth of the colonies was studied with phase contrast time-lapse microscopy, recording high resolution images at 20$\times$ every 15--20$\,$minutes, depending on the experiment (see Supplementary Table~S1).  Colonies that got in contact with each other, or that had cells falling out of the field of view, were discarded.  We obtained in total 32 single-cell derived BMSC colonies.  In contrast to most time-lapse experiments, we do not fix the total time of the experiment, but the maximum generation $k$, keeping the experiment going until all proliferating cells in a colony have arrived to $k=7$.  In this way there is no need to correct for the bias in the statistics of the division times and in the reconstruction of the lineage tree (the main features we wish to investigate) that is introduced if the total observation time is fixed \cite{powell1955some}.  Fig.~\ref{fig:time-lapse} shows representative time-lapse images following the growth of one of the colonies, from the single cell scale up to the entire dish (see Supplementary Video S1 for the time-lapse of one colony).

\subsection*{Active and inactive cells}

We call {\it active} those cells which undergo mitosis and hence have a division time, $\tau$, defined as the time difference between their mother mitosis and their own mitosis. The maximum number of cells at the seventh generation is $2^7=128$, but not all cells continue to proliferate and reach this stage. Some cells stop dividing, yet remain alive, thus  entering the G$_0$ phase \cite{nachtwey1969cell, cheung2013molecular, mens2018cell}.  We classify a cell as G$_0$ if it does not divide within $84\,$hours ($3.5\,$days) from its birth; this threshold is much larger than the mean division time of BMSCs, which is $20 \pm 4\,$hours. In the Supplementary Notes section we demonstrate that this criterion is robust, so that our results do not depend on the specific threshold of $84$ hours. G$_0$ cells can be quiescent (able to reverse to a proliferating state) or irreversibly non-proliferating (senescent or differentiated) \cite{cheung2013molecular,stenderup2003aging}.  A previous study of BMSC colonies \cite{whitfield2013onset} established through the $\beta$-galactosidase assay that most G$_0$ cells were senescent ---although that study used passage P5, higher than our P0--P3.  Moreover, several studies \cite{wagner2008replicative,whitfield2013onset,he2024morphology} established a clear correlation between senescent state and morphology: senescent cells typically have a larger area and a flat morphology, very different from the spindle-like shape of proliferating cells; this is also the case for our G$_0$ cells (Fig.~\ref{fig:time-lapse}, bottom). Nevertheless, since we did not run the $\beta$-galactosidase assay, we prefer to use the less committed label G$_0$ for cells that stop dividing.

% Apo
In very few cases cells commit apoptosis and die (apoptosis is very clear in the imaging, as the cell bursts and detaches from the plate).   Despite the obvious biological difference between G$_0$ and dead cells, they have the same effect on lineage topology, namely to interrupt a branch.  It is therefore useful to use a term that encompasses both G$_0$ and dead cells: we call them \emph{inactive.}  In this way, the total population of a tree is given by the union of its active and inactive cells, while the inactive sub-population is the union of G$_0$ and dead cells.  In our case, most of the inactive cells are G$_0$: we recorded only 17 dead cells in our whole dataset, which is just 5\% of all 327 inactive cells (see Supplementary Table~S1).

\subsection*{Representation of the lineage trees}

\begin{figure}
\centering
\includegraphics[width=0.77\textwidth]{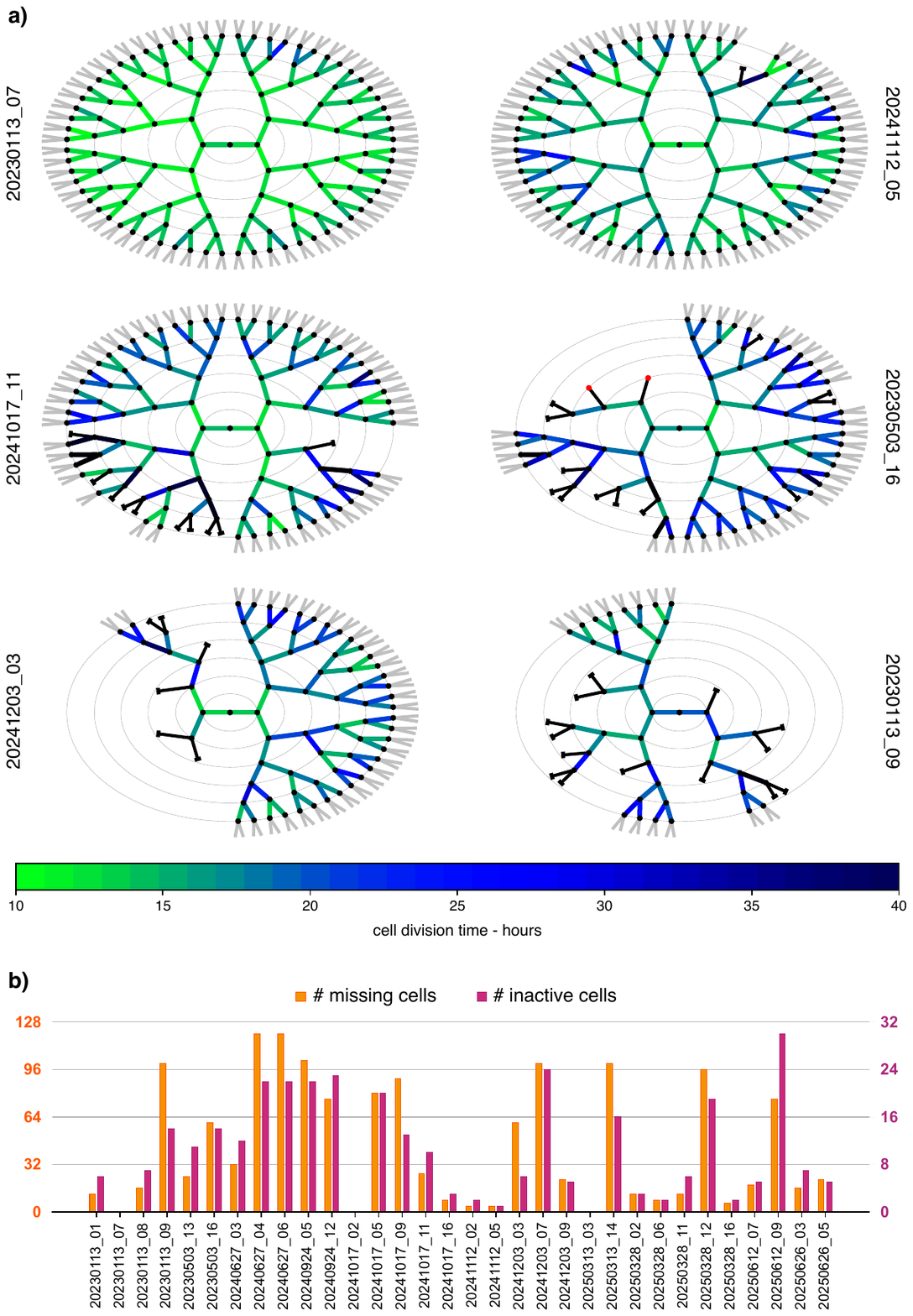}
\caption{ \textbf{Topology --- Lineage trees.}  {\bf a)} Abstract representation of the lineages of six BMSC colonies in the dataset.  Each node (black dots) represents a mitosis and each link between two nodes represents a cell; the central node is the mitosis of the originating cell (the originating cell is not represented in the tree). All links have the same length, while their color represents the division time of the corresponding cell: light-green for short division times, dark-blue for long times (the mean division time across the whole dataset is about 20 hours).  The thin elliptical lines separate different generations.  All colonies have been followed up to the seventh generation, which means that the last divisions we observe are those between generation $k = 6$ and $k = 7$; hence, we do not measure the division times of the $k = 7$ cells, which are therefore colored grey.  In some colonies all cells divide up to the seventh generation, giving rise to $2^7=128$ cells (\texttt{20230113\_07}, top left), but in most colonies some inactive cells (G$_0$ or dead) emerge.  We put a short black cap at the end of the link to represent a G$_0$ cell, or a red dot for a dead cell. Inactive cells do not have a division time and we color them in black. {\bf b)} To show the significant inter-colony heterogeneity of the topology, we report for each lineage in the dataset the number of inactive cells and the number of missing cells at the seventh generation. Both quantities have fluctuations that are almost as large as the mean.}
\label{fig:lineage-trees-topo}
\end{figure}

We adopt an abstract radial representation of the lineage trees (Fig.~\ref{fig:lineage-trees-topo}a), similar to that of Refs.~\cite{hormoz2015inferring, hicks2019maps}: each node represents a mitosis and each link represents a cell.  Unlike in more standard representations \cite{whitfield2013onset,rennerfeldt2019emergent}, all links have the same length (so that every lineage occupies the same area), and the division time (cycle) of the cell is represented by the link's color (light green for a fast cell cycle and dark blue for a slow cycle).  Since we do not know the time of birth for the first cell, there is no link to represent it; the mitosis of this first cell (the root of the tree) defines the origin of time for the colony ($t=0$ in Fig.~\ref{fig:time-lapse}).  Mitosis of cells belonging to the same generation lie on the same ellipse, while cells of the same generation are represented by segments lying in the region between two consecutive ellipses.  We trace cells up to generation $k=7$, so the last divisions we observe are those between $k=6$ and $k=7$; accordingly, $k=7$ cells are colored grey because they lack a recorded division time.

%{\bf The role of non-proliferating cells.}
In some cases all branches reach $k=7$ without inactive cells (3 out of 32 in our dataset), and the lineage is perfectly whole; but in most colonies inactive (mostly G$_0$) cells appear and some branches of the tree are broken.  An inactive cell has no mitosis or division time, hence we conventionally paint it black and put a black cap at the end of the link for G$_0$ cells, or a red circle for dead cells (Fig.~\ref{fig:lineage-trees-topo}a).  We stress that this abstract representation  contains no information about physical positions and mutual distances between cells.  Although there is some correlation between distance on the tree and physical distance, it is not very stringent (see \cite{allegrezza2025inheritance} for an in-depth analysis of this point).

%{\bf Structure is topology + kinetics.}
The bare backbone of a lineage ---how cells are connected to each other, independently of the color of the links (division times)---  is what we call the {\it topology} of the tree.  This is exclusively determined by the location of the inactive cells.  The overall structure of the colony is given by the distribution of links plus the division times (i.e.\ the links' color, or length if one uses such the standard representation \cite{rennerfeldt2019emergent}).  Thus different division times will produce different structures for the same topology.  In other words, the structure is the combination of topology (how inactive cells are distributed) and kinetics (how fast the active cells divide).  Since we aim here at disentangling the roles of these two elements, we have chosen to use the words `topology' and `kinetics', rather than `structure'.
%TC:endignore
%%%%%%%%%%%%%%%%%%%%%%%%%%%%%%%%%%%%%%%%%%%%%%%%%%%%%%%%%%%%%%%%%

\section*{Results}

\subsection*{Heterogeneity of topology}

%{\bf Inter-colony heterogeneity.}
In Fig.~\ref{fig:lineage-trees-topo}a we show a subset of our lineage trees: a qualitative inspection reveals great diversity in the trees' topologies, confirming, at the lineage level, what previous studies have already established for other parameters \cite{satomura1998receptor, sworder2015molecular, whitfield2013onset, lee2014multivariate, rennerfeldt2016concise, rennerfeldt2019emergent}, namely that BMSC populations display a great \emph{inter-colony} heterogeneity.  We find, on one hand, perfectly whole lineages, with all branches reaching the last recorded generation (e.g.\ \texttt{20230113\_07} in Fig.~\ref{fig:lineage-trees-topo}a), while on the other hand there are cases of very broken trees, with  many branches interrupted by inactive cells (e.g.\ \texttt{20230113\_09} in Fig.~\ref{fig:lineage-trees-topo}a).  The topology of a lineage depends entirely on the number and positions of the inactive cells.  Therefore, a first way to quantify the heterogeneity of the colonies is to measure the number of inactive cells, $\Nina$.  This number has fluctuations almost as large as the mean, $\Nina = 10.38 \pm 8.42$ (mean $\pm$ sd), i.e. relative fluctuation of $81.4\%$ (Fig.~\ref{fig:lineage-trees-topo}b).

% Inactive vs missing: the role of position of G$_0$
Given a certain number of inactive cells, their effect on the topology depends on their placement among generations: a G$_0$ emerging at the second generation cuts a far larger number of cells at generation $k=7$ than a G$_0$ at the sixth generation, which only cuts two. We can quantify this effect by counting the number of $k=7$ missing cells, $\Nmiss$, compared to a complete tree (which has $2^7=128$ leaves at $k=7$).  Naively one may expect that $\Nina$ and $\Nmiss$ are simply proportional to each other, but this is not necessarily true: although the two quantities are broadly related (in particular they are both zero in the complete tree), the value of $\Nmiss$ given $\Nina$ depends on the position of the inactive cells; hence, there is no function linking these two parameters.  We show in Fig.~\ref{fig:lineage-trees-topo}b that the number of missing cells varies even more strongly than the number of inactive cells, $\Nmiss= 44.4 \pm 40.77$, a relative fluctuation of $92\%$.  Hence, topology is very heterogeneous not only because of the number of inactive cells, but also because of the variability in the generation at which they emerge.

%{\bf Intra-colony heterogeneity.}
Fig.~\ref{fig:lineage-trees-topo}a also shows significant \emph{intra-colony} heterogeneity: if we look at a single lineage, we find branches with many interruptions and other branches that instead proliferate in a balanced manner. We examine potential mechanisms underlying heterogeneity among different branches of a single tree -- and their potential hereditary origin --  in a separate study \cite{allegrezza2025inheritance}. Here, we concern ourselves with another aspect of the lineages: whole trees have overall a lighter color than broken ones, i.e.\ the active sub-population is on average faster the more balanced the lineage (Fig.~\ref{fig:lineage-trees-topo}a).  This potential correlation is non-trivial, as the division time is a trait of the active cells, while the broken topology is caused by the inactive cells, these subsets being complementary. To make this observation quantitative, we turn our attention to the analysis of the division times of active cells.

\subsection*{Heterogeneity of kinetics}

\begin{figure}[h!]
\centering
\includegraphics[width=0.73\textwidth]{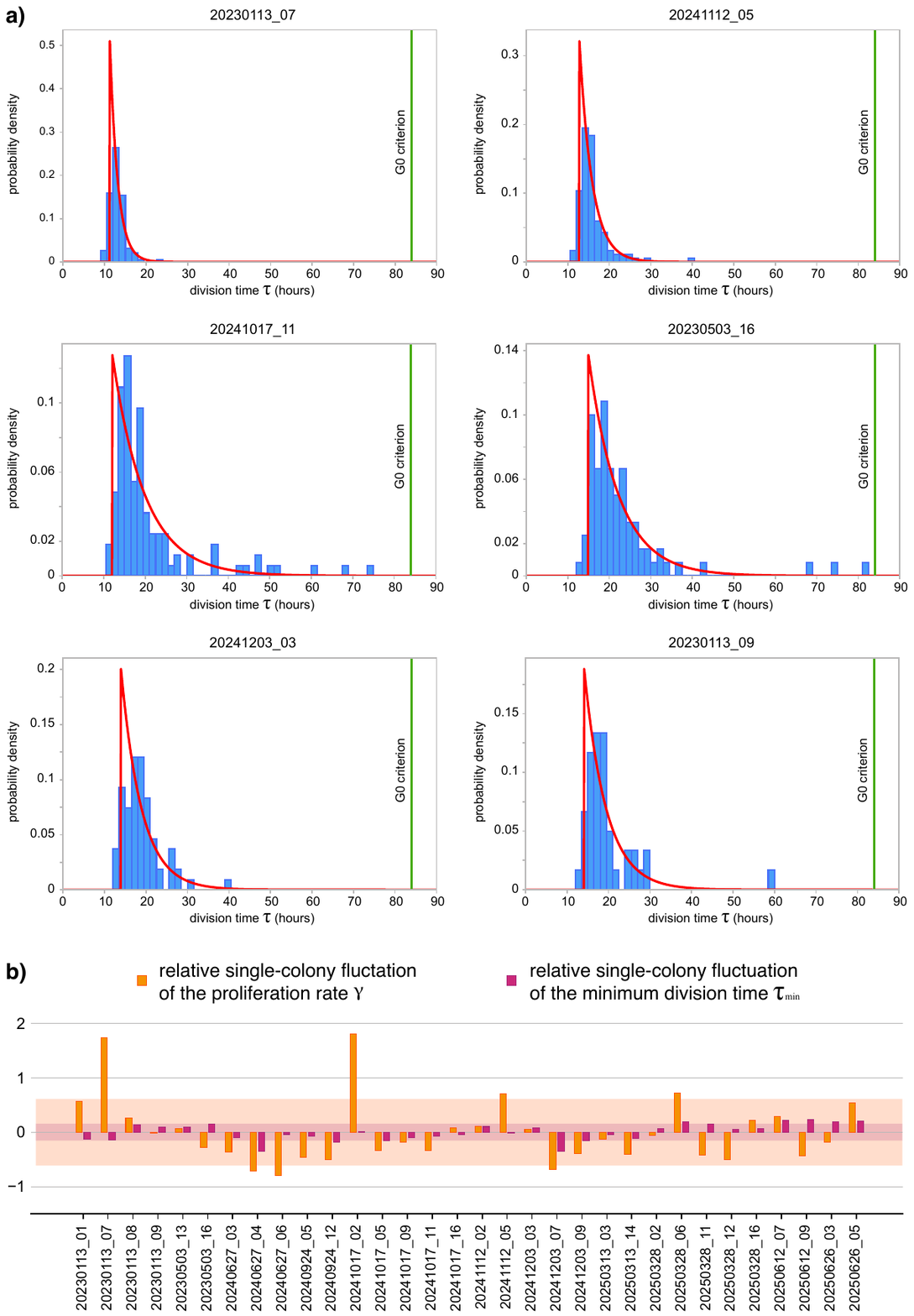}
\caption{ \textbf{Kinetics --- Statistics of the division times.} {\bf a)} We report the probability density of the cells' division times, $\tau$, for the same six colonies as in Fig.~\ref{fig:lineage-trees-topo}a.  The vertical green line in each histogram indicates the $84$ hours threshold that we use to define G$_0$ cells.  In all colonies there is a threshold below which we find no division times.  For this reason, a reasonably good fit of the data (full lines) is given by the gap-exponential form (see text), which is characterised by a minimum division time $\tau_\mathrm{min}$ and by a rate $\gamma$ of decay of the exponential part of the function (red line). {\bf b)} We report the relative single-colony fluctuations of the rate $\gamma$ and of the minimum time $\tau_\mathrm{min}$ defined as $(x - \bar{x})/\bar{x}$, where $x$ represents the observable (either $\tau_\mathrm{min}$ or $\gamma$) and the average $\bar{x}$ is computed over all the colonies. As in the topology, a significant heterogeneity is detected also in the kinetics, in particular in the fluctuations of the rate $\gamma$, while the minimum division time, $\tau_\mathrm{min}$, fluctuates much less. The orange and purple shading represent the standard deviation of the single-colony fluctuations of $\gamma$ and $\tau_{min}$ respectively over all the colonies.}
\label{fig:colony-kinetics}
\end{figure}

%{\bf The distributions of the division times rule out $t_c$ as a relevant factor.}
In Fig.~\ref{fig:colony-kinetics}a we show the normalized histogram (probability distribution) of the division times $\tau$ of active cells for the same colonies that appear in Fig.~\ref{fig:lineage-trees-topo}a (see Supplementary Fig. S1 for the complete dataset).  The green vertical line indicates the threshold used to classify cells as G$_0$ ($t_c=84\,$hours); clearly $t_c$ is quite far from the whole bulk of the distribution, hence the separation of the population into active and inactive sub-populations is a qualitative one (see Supplementary Notes for a careful assessment of this point).  Because we follow all cells up to a fixed generation we do not need to correct for the bias toward lower values of $\tau$ that results from the exclusion from the analysis of all the cells that divide after a fixed observation time \cite{powell1955some}.

%{\bf Gap-exponential fit.}
The large $\tau$ tails of the distribution decay exponentially, a feature consistent with simple models  of cell proliferation in which division events occur at a constant rate \cite{harris1963theory, haccou2005branching}.  However,  an exponential probability has its peak at $\tau=0$, while all the experimental histograms show a clear gap for small division times.  
This has been observed in several systems, inspiring a class of models \cite{smith1973cells, weber2014quantifying, lavalle2023fluctuations} based on the assumption that no divisions will happen below a minimum value, $\tau_\text{min}$, and that after this time cells can divide with a constant rate (probability per unit time) $\gamma$.  These assumptions lead to the following gap-exponential distribution of the division time $\tau$, 
\begin{equation}
  P(\tau) = \gamma\;  \theta(\tau-\tau_\mathrm{min}) \; e^{-\gamma(\tau-\tau_\mathrm{min})}  .
  \label{tonga}
\end{equation}
The meaning of equation~\eqref{tonga} is simple: the division time cannot be smaller than $\taum$ (the Heaviside function is $\theta(x)=1$ if $x\ge0$ and $\theta(x)=0$ if $x<0$), and beyond $\taum$ a pure exponential decay with rate $\gamma$ sets in.  $\gamma$ has units of the inverse of a time, and the factor $\gamma$ in front of the right-hand side of equation~\eqref{tonga} is required for normalization.

We fit equation~\eqref{tonga} to the division time data, finding the values of the proliferation rate $\gamma$ and minimum division time $\taum$ for each  colony (more precisely, we determine $\gamma$ and $\taum$ by fitting the cumulative distribution, which fluctuates less than the distribution itself).  The mean division time of distribution \eqref{tonga} is given by,
\begin{equation}
  \tmean{\tau}=\taum + \frac{1}{\gamma} \  ,
  \label{cip}
\end{equation}
where we use $\tmean{\tau}$ to distinguish it from the experimental mean division time, $\overline{\tau}$, which is obtained by performing the average over all experimental division times. Even though other forms could be proposed \cite{hawkins2007, weber2014quantifying, yates2017multi}, we see from Fig.~\ref{fig:colony-kinetics}a that distribution \eqref{tonga} fits reasonably well the data in most of the colonies.

\begin{figure}
\includegraphics[width=0.9\textwidth]{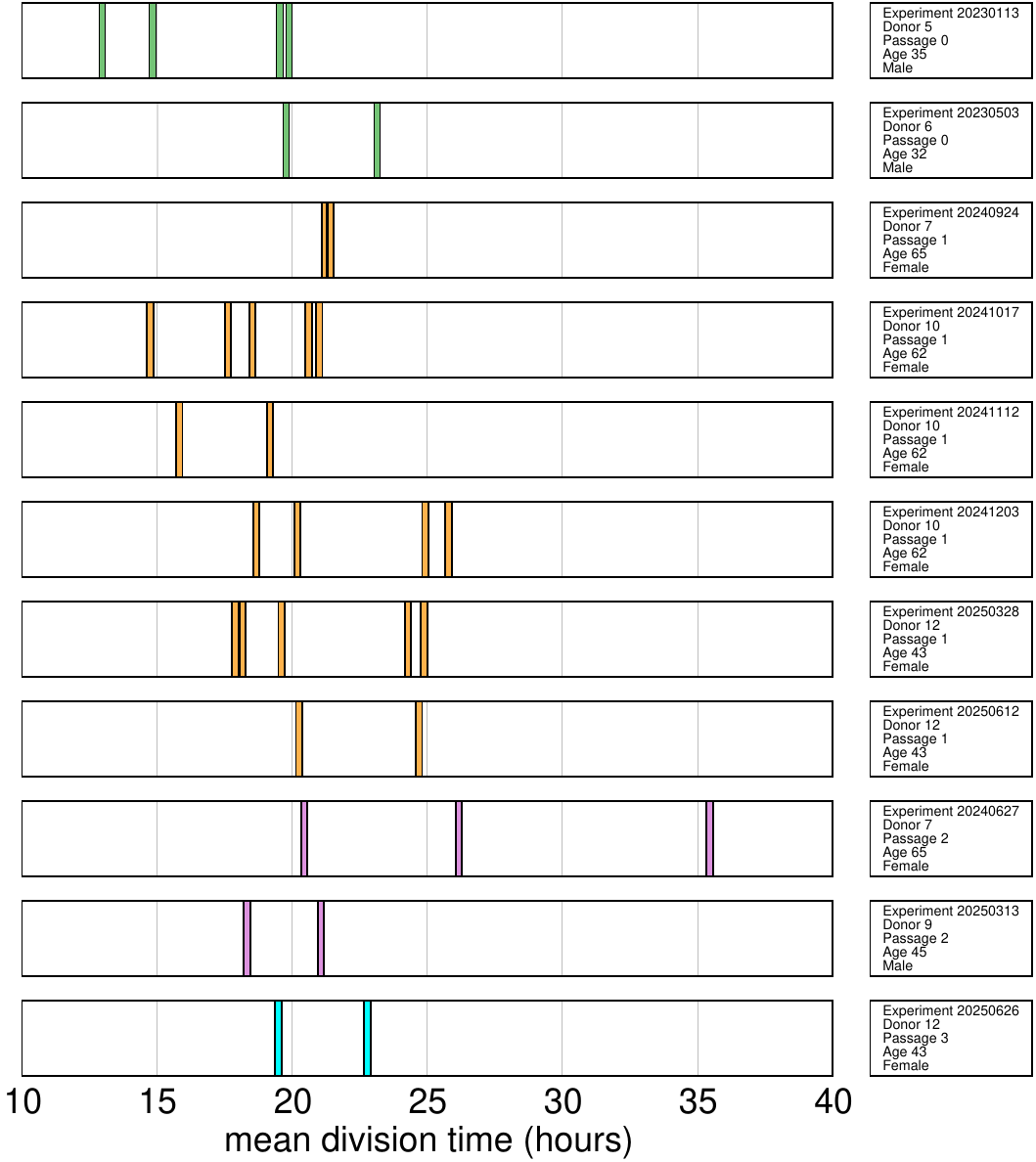}
\caption{ {\bf Variability with experiment, donor, age, sex, and passage.} 
We report here the mean division time for each colony in the dataset, ordered according to experiment, donor, age, sex, and the passage of the cell culture. Each vertical bar corresponds to a different colony; colour represents passage: P0-green, P1-orange, P2-pink, P3-cyan. Fluctuations within the same experiment are of the same order as the variability across the whole dataset. The only (very weak) correlation in the data is between division time and passage (Spearman correlation $\rho=0.30$, $P$-value$ = 0.046$). We notice that the only colony that seems an outlier in this plot is \texttt{20240627\_06} (largest division time of experiment \texttt{20240627}); interestingly, this is also the colony most separated from the bulk of points in the topology vs kinetics plot (see Fig.~\ref{fig:correlation-topo-kin}).}
\label{fig: passage vs tau}
\end{figure}

%{\bf Fluctuating vs non-fluctuating part of the division time.} 
The relative fluctuations of $\taum$ and $\gamma$ are given in Fig.~\ref{fig:colony-kinetics}b, whence we notice something interesting: while the proliferation rate $\gamma$  strongly fluctuates from colony to colony ($\gamma = 0.19 \pm 0.12\,$h$^{-1}$; relative fluctuation of $63\%$), the minimal division time, $\taum$, is significantly more stable ($\taum = 13 \pm 2\,$h;  relative fluctuation of $15\%$).  This suggests that the division time is made of two different contributions: a quasi-deterministic part, given by $\taum$, which is roughly the same for all cells, and a stochastic part,  which fluctuates from cell to cell and whose corresponding rate $\gamma$ fluctuates from colony to colony.  How  these two contributions are mutually located along the cell cycle, or whether there is any temporal order between them, cannot be inferred from these data.  What we can say is that there are certain processes within the cell cycle whose total duration $\taum$ is very stable and cannot be compressed, while the duration of the remaining processes fluctuates strongly.
% G1 phase
This result is consistent with the fact that cell cycle stages S, G$_2$ and M are quite stable in duration, while G$_1$ fluctuates much more \cite{nachtwey1969cell, smith1973cells},  both in embryonal and somatic stem cells \cite{liu2019cell, mens2018cell};  such variability is associated to proliferation regulation and cell fate determination  \cite{pauklin2013cell, dalton2015linking, mens2018cell}.  
We note that $\tau_\mathrm{min}$ takes approximately 63\% of BMSC cycle duration, while $1/\gamma$ takes 37\% of it; interestingly, in rapidly replicating human cells, the combined duration of the S, G$_2$ and M phases takes about 62\% of the cycle duration, while the G$_1$ phase takes about 38\% \cite{molecular2000lodish}.

%{Dependence on donor's age, sex, passage}
Considering that the colonies in our dataset are derived from cells coming from 6 different donors, of different age and sex, and that cells were plated at different passages (P0 to P3, see Supplementary Table S1), it is natural to ask how much of the observed variability is linked to these factors.  Interestingly, the data show that the degree of heterogeneity within one experiment (same donor, age, sex, passage, plate) is as large as the heterogeneity between different experiments; indeed, it may happen that two cells from the same donor, cultured within the same experiment on the same plate, give rise to two completely different colonies: for example, lineage \texttt{20230113\_07} has no inactive cells and is perfectly whole, while \texttt{20230113\_09} from the same experiment is highly broken (see Fig.~\ref{fig:lineage-trees-topo}a). In Fig.\ref{fig: passage vs tau} we report the mean division times $\overline{\tau}$ of all colonies in our dataset grouped by experiments, reporting donor, age, sex and passage. Although we cannot completely exclude some very weak correlation between $\overline{\tau}$ and passage (Spearman correlation $\rho=0.30$, with barely significant $P$-value$ = 0.046$), no clearcut correlations among the other parameters are observed. These results suggest that -- albeit passage seems to weakly affect the division time of BMSCs more than age and sex -- the intrinsic variability within each cell-culture experiment largely dominates the overall heterogeneity of the data (Fig.~\ref{fig: passage vs tau}).

\subsection*{Correlation between topology and kinetics}

\begin{figure}
\includegraphics[width=0.9\textwidth]{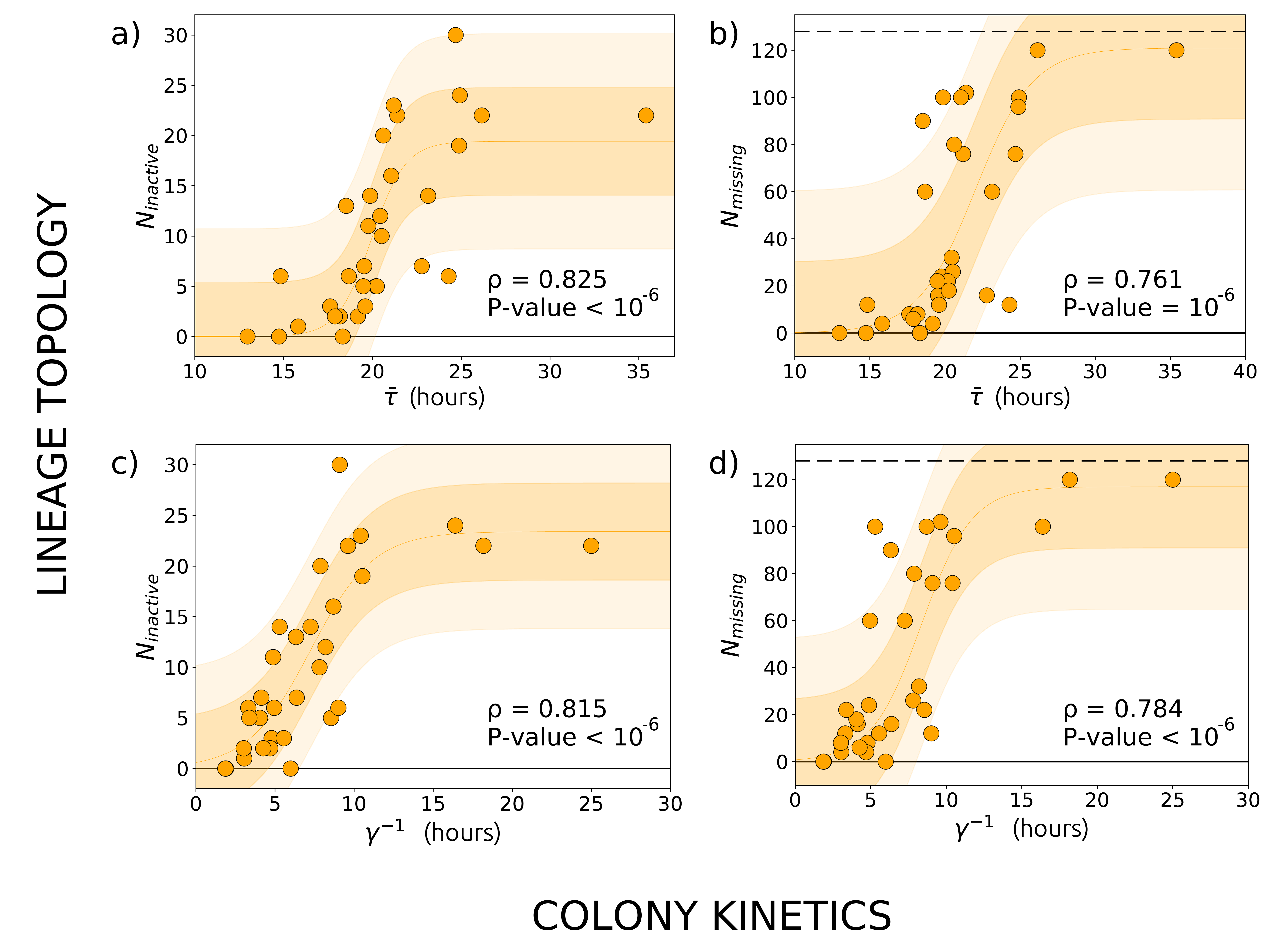}
\caption{ {\bf The correlation between topology and kinetics.} 
{\bf a)} We plot the number of inactive cells of each colony, $\Nina$ as a function of the mean division time of that colony, $\overline{\tau}$ (orange points). The Spearman correlation between $\Nina$ and $\bar\tau$ is both strong and significant: correlation coefficient $\rho=0.825$ and $P$-value $< 10^{-6}$. {\bf b)} A second characterisation of topology is given by the number of missing cells at generation $k=7$, $\Nmiss$, which also happens to be strongly correlated to the mean division time,  $\overline{\tau}$ (Spearman $\rho=0.761$, P-value $ =1.0 \times 10^{-6}$). The dashed line represents the maximum number of missing cells, namely $2^7=128$.  {\bf c-d)} The rate $\gamma$ extracted from the fit of the division times histograms is a second characterisation of kinetics.  We plot $\Nina$ and $\Nmiss$ vs.\ $1/\gamma$ and find in both cases a strong and significant correlation ($\rho=0.815$, P-value $< 10^{-6}$ and  $\rho=0.784$, P-value $< 10^{-6}$, respectively). Shaded areas are just a guide for the eye, corresponding to an amplitude of 1$\sigma$ (darker orange) and 2$\sigma$ (lighter orange) around a sigmoidal fit of the data, where $\sigma$ is the standard deviation of the residuals.}
\label{fig:correlation-topo-kin}
\end{figure}

%{\bf Broken trees correspond to slow colonies.}
To quantify the previous observation that  broken trees correspond to slower colonies, we examine the relation between the total number of inactive cells $\Nina$ and the mean division time $\overline{\tau}$ (both directly measured from the data) in Fig.~\ref{fig:correlation-topo-kin}a: despite the scatter, we find a strong and significant correlation (Spearman: $\rho=0.825$, $P$-value $ <10^{-6}$).  Such correlation does not depend on the parameters we choose to characterize topology and kinetics: we can use the number of missing cells, $\Nmiss$, or the inverse proliferation rate $1/\gamma$ (which derives from a fit to equation \eqref{tonga}) and the results do not change: in Fig.~\ref{fig:correlation-topo-kin}b, c, and d we show that all these correlations are strong and significant, meaning that the connection between lineage topology and colony kinetics is robust. We emphasize that these two traits are determined by two phenotypically {\it well separated} sub-populations: topology is uniquely determined by {\it inactive} cells, while kinetics is exclusively determined by {\it active} cells.

% Same result by other groups: Van Vliet
In Ref.~\cite{whitfield2013onset} it was found that the division time of parents of cells that both divided was shorter than the division time of parents of cells which both stopped dividing, with mixed dividing-not-dividing pairs in between.  In Ref.~\cite{rennerfeldt2019emergent} it was reported that cells belonging to slow-dividing progenies had more senescent daughters on average than cells belonging to fast-dividing progenies  (slow-dividing were defined as those producing less than $8$ cells at day $4$ and fast-dividing as those producing more than $16$ cells at day $4$).  Both observations were made at the single-cell level and were based on a binary or ternary classification of the kinetics.  Our result shows that the correlation between topology of the inactive cells and kinetics of the active cells remains strong at the level of the entire colony, and that it is robust against unfolding the binary slow-fast classification into the full spectrum of the mean division times.

%{\bf What does it mean. Proteomic}
The strong correlation between the active and inactive phenotypes suggests that the factors regulating the duration of the cell cycle are also involved in the determination of cell-cycle exit. This is consistent with the proteomic profiling of BMSCs colonies of Refs.~\cite{mareddy2007clonal} and~\cite{mareddy2010stem}.  In those studies, colonies were classified as fast-growing or slow-growing, depending on the time taken to reach 20 doublings, and it was noted that fast colonies had tripotential differentiation capacity while 
slow colonies had unipotential differentiation capacity. Proteomic analysis showed that calmodulin (CAM) and tropomyosin (TM) had enhanced expression in {\it fast} colonies \cite{mareddy2010stem}; CAM blocks Fas-mediated apoptosis \cite{ahn2004calmodulin} and more generally CAM and TM act as modulators during cell proliferation.  On the other hand, the over abundance of caldesmon (CAD) and Annexin-I in {\it slow} colonies was equally relevant, because CAD acts as a regulated break to cytokinesis, while over-expression of Annexin-I reduces cell proliferation \cite{mareddy2010stem}. Therefore, proteomic profiling indicates that fast-growing colonies express factors that may reduce the number of cell deaths (although not necessarily of G$_0$ cells) and increase proliferation, whereas slow-growing colonies have a larger abundance of proteins potentially depressing proliferation.  Our results suggest that some of the factors highlighted in Ref.~\cite{mareddy2010stem} may also be involved in triggering the G$_0$ state of BMSCs.

A simple model based on the competition of two mechanisms (one promoting proliferation, with rather large variability in speed, and another, more stable, causing senescence) could explain the observed correlation.  A slow but steady buildup of factors precluding proliferation would cause the cell to become inactive unless the proliferation-enhancing mechanism acts first and triggers cell division.  If the proliferative mechanism is slow, then there would be a higher chance that the cell becomes inactive.  However, the slow-fast classification of Ref.~\cite{mareddy2007clonal}, using 20 doublings, involves a time window {\it hugely} longer than that of our experiments ($2^{20} >$ one million).  While the consistency between our results and those of~\cite{mareddy2007clonal} suggests that the long-term proliferation potential is already detectable in early generations ($k\le7$), specific experiments directly comparing the short- and long-term behaviour of individual colonies should be performed to validate such promising connection.  

\label{sec:fail-an-uncorr}

\subsection*{Synchronisation}
\label{sec:kinet-prol}

\begin{figure}
  \centering
  \includegraphics[width=0.85\columnwidth]{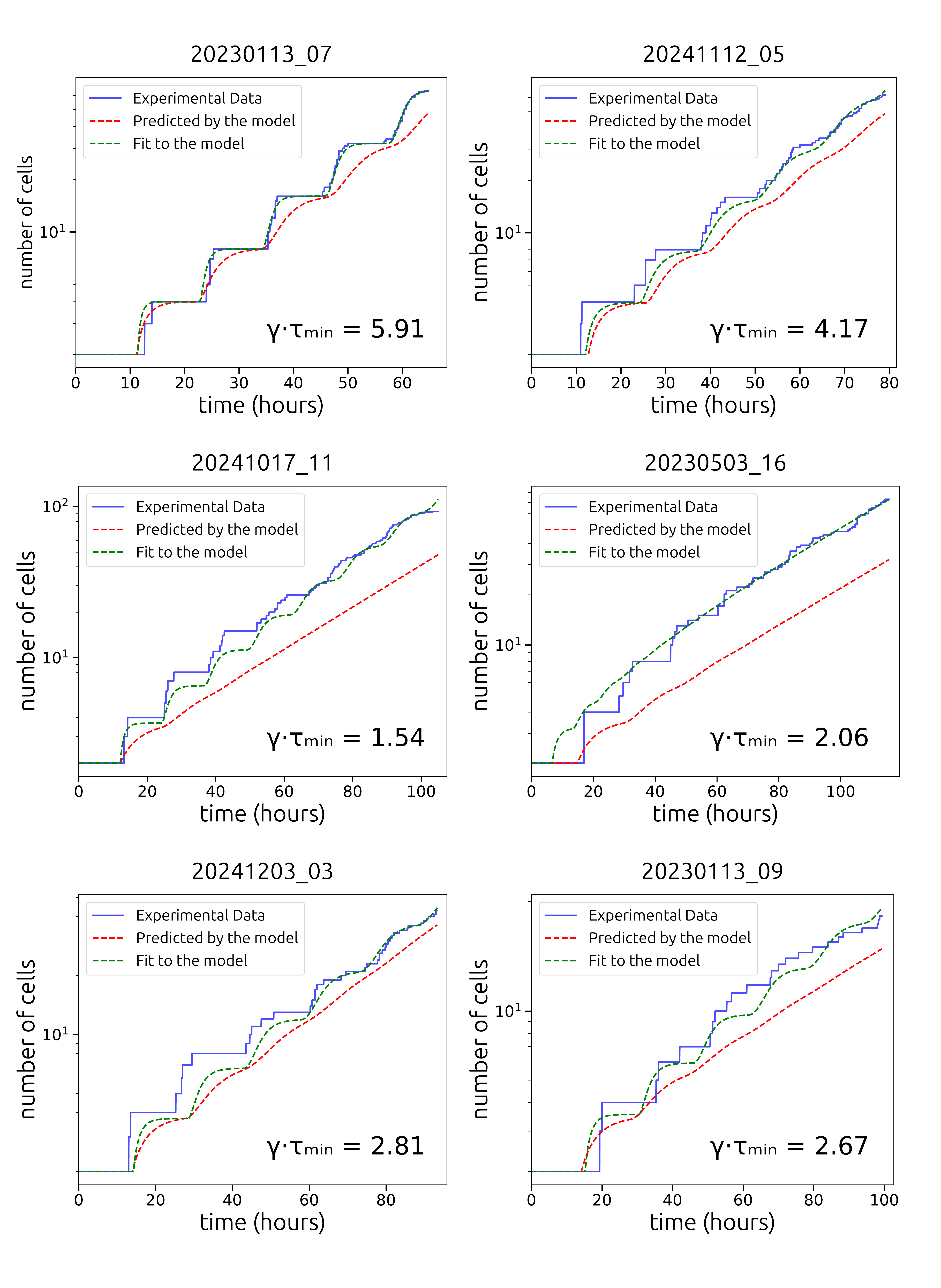}
  \caption{\textbf{Number of cells \textsl{vs.\/} absolute time.} We show $N(t)$ vs $t$ for the same colonies as in Figs.~\ref{fig:lineage-trees-topo} and~\ref{fig:colony-kinetics}.  A step-like growth of $N(t)$ is very clear for the colonies with the largest values of $\gamma\tau_\text{min}$ (with $\gamma$ and $\tau_\text{min}$ obtained from the fits to $P(\tau)$, see Fig.~\ref{fig:colony-kinetics}).  The steps indicate synchronisation of mitosis events; this synchronisation is expected even if mitosis are independent events, when the average division time is close to the minimum time, corresponding to $\gamma\tau_\text{min}\gg1$ (see text).  Synchronisation can be enhanced by the presence of correlations of division times between individuals close in the lineage tree.  Blue lines: experimental data; dashed red lines: prediction of the branching process model, with parameters taken from the lineage trees and the fitted $P(\tau)$; dashed green lines: direct fit to the branching process form of $N(t)$.}
  \label{fig:Nvst}
\end{figure}

We now examine the overall growth of the colonies, by measuring the total number of cells at a given time, $N(t)$ (all growth curves start with $N=2$ at $t=0$). This aggregate information is the result of both kinetics and topology, since how fast the colony grows depends on how fast individual cells proliferate, but also on how many of them remain active. We report $N(t)$ in Fig.~\ref{fig:Nvst}: a remarkable feature is the presence of steps, indicating a high degree of synchronisation of the mitosis, especially for early generations (in Supplementary Video S2 we show synchronisation of four cousin cells at $k=2$).  These steps are due to the dynamics of the cell cycle: 
if division happens shortly after the minimum time that must elapse before a cell can divide, then one can expect ``avalanches'' of mitosis marking the transition between different generations, which manifest as steps in the growth curve, with plateaus at $N= 2, 4, 8, 16, \dots$. 

To illustrate this mechanism, we consider the simple case of sister cells. Let us define the {\it desync}, $\Delta t$,  as the difference between the times at which the two sister cells divide. If we momentarily assume that there is {\it no} correlation between the division times (we will soon come back to this -- wrong -- assumption), it is easy to show (see Supplementary Equations) that the mean square desync is given by,
\begin{equation}
\tmean{(\Delta t)^2} = 2 \sigma_{\tau}^2  \ ,
\label{zumpa}
\end{equation}
where $\sigma^{2}_{\tau}$ is the variance of the distribution of the division times.
Therefore, in absence of correlation, synchronisation depends entirely on how narrow is the distribution of the division times. But to see sharp steps in $N(t)$ the ramp between two plateaus must be much shorter than the length of the plateaus, namely the desync must be much smaller than the average cycle length, $\sqrt{\tmean{(\Delta t)^2}} \ll \langle\tau\rangle$. From equation \eqref{zumpa}, we conclude that steps will be sharper when $\sigma_{\tau} \ll \langle\tau\rangle$.
Synchronisation will be slowly lost as generations progress, due to random fluctuations in division times, and steps will die out.

For the gap-exponential distribution that we have used to fit our data, Eq.~\eqref{tonga}, the variance is connected to the proliferation rate through the relation, $\sigma_{\tau} = 1/\gamma$, while the mean is given by, $\langle\tau\rangle = \taum+\gamma^{-1}$, where we recall that $\taum$ is the minimum division time. We conclude that (under the assumption of uncorrelated division times) the steps of $N(t)$ will be sharp when $\gamma^{-1} \ll \taum + \gamma^{-1}$, that is $ 1 \ll \gamma \taum +1$, which can only be verified if,
\begin{equation}
\gamma \taum \gg 1 \ .
\label{dinamite}
\end{equation}
This prediction is qualitatively consistent with the growth curves $N(t)$ and with the parameters of $P(\tau)$ from the gap-exponential fit: the largest values of $\gamma\tau_\text{min}$ (above 4) are found for colonies \texttt{20230113\_07} and \texttt{20241112\_05}, which have the best-defined steps with little desynchronisation, while the rest have $\gamma\tau_\text{min}\le 2.8$, with the smallest value found for \texttt{20241017\_11}.

Qualitatively explaining the nature of sharp vs smooth steps in the growth curves may seem a validation of the assumption of uncorrelated division times. However, the growth curves must be {\it quantitatively} accounted for by a satisfactory theory; we now ask whether the simple picture above is up to this task. To answer this question we need to theoretically predict the shape of $N(t)$ given a certain distribution of the division times, $P(\tau)$.

\subsection*{Branching process models}
\label{sec:branch-proc-model}

We consider an ensemble of non-interacting individuals, each of which starts a family of descendants, independent from each other, but with the same stochastic properties.  Models of this kind are called \emph{branching processes} \cite{harris1963theory} and are characterized by the  absence of correlation between individuals. The most important input of a branching process is the probability distribution of the division times, $P(\tau)$; once this is specified, one can (analytically or numerically) work out the growth curve, $N(t)$.

If one employs as an input an exponential distribution $P(\tau)$ with no gap (constant proliferation rate, independent of the cell's age), one  gets a purely exponential growth of $N(t)$, with no steps. But the actual distribution of the division times {\it does} have a gap, $\taum$. Using the gap-exponential $P(\tau)$ as an input of the branching process one obtains that $N(t)$ grows in (rounded) steps that gradually smooth out with the passing generations \cite{smith1973cells, lavalle2023fluctuations}, not unlike those we observe here. However, in the model studied in \cite{lavalle2023fluctuations} all cells were active, while inactivity is a crucial ingredient of BMSC phenomenology; hence, we would not expect the model in \cite{lavalle2023fluctuations} to reproduce our data accurately. Therefore, we have generalised the model to include inactive cells: death is modelled through a probability that at the end of its life the cell is removed, while G$_0$ cells are modelled through a probability of having an infinite lifetime (see the Supplementary Equations for details).  This new model has four parameters, which can be estimated from the experiments: $\gamma$ and $\tau_\text{min}$ from the fit to $P(\tau)$ (Fig.~\ref{fig:colony-kinetics}a), plus the death and G$_0$ probabilities, which can be extracted from the death/G$_0$ fractions in each tree. This model's predictions for $N(t)$ are shown as dashed red lines in Fig.~\ref{fig:Nvst}: the agreement is poor; in particular, the synchronisation steps are much blunter in the model than in the experiments. 
If, instead, we perform a least squares fit of the theoretical model to the experimental $N(t)$, we obtain the dashed green curves in Fig.~\ref{fig:Nvst}.  In this case the agreement is better; however, even though the fit yields a value of $\tau_\text{min}$ rather close to the experimental one, the fitted value of the proliferation rate $\gamma$ is systematically larger than its experimental counterpart; hence, the better agreement with this second procedure is only apparent, because one gets $P(\tau)$ wrong.

We conclude that the proliferation of BMSC colonies is much more synchronised than it would be expected from an uncorrelated branching process characterized by a distribution of division times equal to the experimental one; conversely, if we insist in quantitatively accounting for the observed synchronisation with an uncorrelated branching process, we end up with a distribution of the division times that is much narrower than the experimental one. What is the origin of this discrepancy?

\subsection*{Correlations and the failure of branching process models}
\label{sec:corr-fail-branch}

The branching process model makes a key assumption, which is also at the basis of equation \eqref{zumpa}, namely that the division times of different cells are uncorrelated. However, positive correlations between the division times of sister cells are known to exist and have been reported since at least the 1950s \cite{powell1955some} in several kinds of cells \cite{sandler2015lineage, mosheiff2018inheritance, seita2021}. Correlations can enhance synchronisation, therefore giving more pronounced steps in the growth curve for a given division time distribution.

Let us go back to the square desync between sister cells, but this time let us {\it not} assume that the division times are uncorrelated. Some simple algebra  (see Supplementary Equations) shows that in this case the square desync is given by,
\begin{equation}
    \tmean{(\Delta t)^2} =2 \sigma_\tau^2 \left(1  - \rho_\mathrm{P}\right) \ ,
\end{equation}
where $\rho_\mathrm{P}$ is the Pearson correlation coefficient between the sisters' division times. If $\rho_\mathrm{P}=0$, we go back to Eq. \eqref{zumpa},  telling us that synchronisation is higher the smaller the variance of the division times.  But we see that the existence of positive correlation, $\rho_\mathrm{P}>0$, has the effect of decreasing the desync, hence {\it increasing} synchronisation, even at fixed $\sigma_\tau$. It is important to appreciate these two different sources of synchronisation, namely the low variance of the division times and the large cell-cell correlation: even in the case of large variance, a strong correlation, $\rho_\mathrm{P} \sim 1$, would give near-perfect synchronisation, as cells' division times would all strongly fluctuate, but they would do that {\it together}. If cells are not sisters other terms enter into play in the equation, leading to gradual de-synchronisation as generations go by, but the effect of correlations is still that of increasing synchronisation.  Therefore, positive correlation of division times -- which the branching process model neglects -- enhances synchronisation for a given division times distribution. The discrepancy between experiments and simple branching process theory therefore indicates that equal-generation division times in BMSCs are positively correlated. 

There is a second tacit assumption of the branching process, namely that the distribution of division times, $P(\tau)$, is the same for all generations; this hypothesis may also be inadequate.  We have computed the average division time at each generation, $\overline{\tau(k)}$ (see Supplementary Fig. S2); in line with previous findings \cite{whitfield2013onset},  $\overline{\tau(k)}$ shows an increasing trend, pointing to a generation-dependent $P_k(\tau)$. The increase of the average means that our $P(\tau)$, which is consolidated over all generations, is broader than the distributions at fixed $k$, which may be a further reason why the branching process fails quantitatively.

\section*{Conclusions}

By conducting time-lapse microscopy on a set of 32 single-cell-derived colonies of human BMSCs, we discovered a sharp correlation between lineage topology (which is determined by the sub-population of inactive cells), and colony kinetics (which depends entirely on the division times of the sub-population of active cells): lineages that are broken by the presence of many inactive cells (typically G$_0$) are also characterized by an active sub-population with a slower proliferation rate and {\it vice-versa}.  This correlation connects a binary state (active-inactive) to a continuous variable (division time) that depends on the proteome expression.  Our study also reveals that simple uncorrelated population growth models do not account for the synchronisation of population growth.

As it happens with many other traits of BMSC colonies, both the number of inactive cells and the mean division time of active cells ---as well as other quantities characterising topology and kinetics--- are strongly heterogeneous.  The interesting consequence of having found a sharp correlation between topology and kinetics, is that it reduces the number of potential {\it independent} factors at the origin of the heterogeneity of both traits, hence partly reducing the complexity of the problem.  By establishing a mutual dependence of the fluctuations of the number of G$_0$ cells and the fluctuations of the division times, we effectively turn what seemed independent random variations, into a phenomenon with a single underlying causal factor.  This factor needs to explain why, within the same clonal cell population, a longer cell cycle is associated to an increase of the probability to enter the G$_0$ state.

Our findings provide fundamental biological insights about human BMSCs: we demonstrate that the propensity of individual BMSC colonies to produce senescent or apoptotic cells cannot be deduced from the characteristic of the donor, as the mean division time of the colonies is not correlated to donor sex and age and weakly correlated to the number of passages in culture. More experiments are needed to precisely assess the impact of age and sex on the BMSC growth kinetics. Rather, our findings show that it is the number of inactive cells in individual BMSC colonies that strongly correlates to the proliferation properties of the dividing cell fraction, pointing at the kinetics of growth as a potential selection criteria for in vivo application.

Notwithstanding the relevance of {\it in vitro} experiments, the gold standard to characterise the differentiation capacity of BMSC colonies is {\it in vivo} transplantation.  In Ref.~\cite{satomura1998receptor}, colonies of BMSCs were transplanted into immunocompromised mice to determine the bone-forming capacity of these colonies and assess the correlation between this capacity and their proliferation rate.  Although kinetics was not found to be strictly predictive of osteogenesis, weak correlation between proliferation rate and bone formation {\it in vivo} was anyway found.  A subsequent study \cite{sworder2015molecular} distinguished between colonies that upon {\it in vivo} transplantation formed fibrous tissue, those that formed bone, and those that formed bone/marrow organ, where only this third class was deemed multipotent.  However, the proliferation rate and its possible correlation with multipotency were not assessed {\it in vivo}.  Therefore, although it seems that high proliferation rate and hence a faster overall kinetics ---which in turn implies a smaller number of inactive cells--- may be positively correlated to the multipotency of BMSCs (likely enriched in SSCs), new {\it in vivo} studies are needed to correlate kinetics to {\it bona fide} multipotency.

\bibliography{bibliography-stem-cells-cobbs}

\section*{Acknowledgments}
It is a pleasure to acknowledge several inspiring discussions with the late Paolo Bianco and with Giorgio Parisi during the very early stages of this project. 
We thank Pietro Cirigliano, Roberto Di Leonardo, Francis Allen Farrelly, Giacomo Frangipane, Emiliano Lalli, Catia Longhi, Federica Massa, Lucia Nencioni and Shoichi Yip for technical help.

\section*{Author Contributions --- CRediT Taxonomy}

Alessandro Allegrezza --- Roles: Data curation.\\
Riccardo Beschi ---  Roles: Investigation, Data Curation.\\
Domenico Caudo --- Roles: Investigation, Data Curation.\\
Andrea Cavagna --- Roles: Conceptualization, Investigation, Data Curation, Formal Analysis, Funding Acquisition, Methodology, Project Administration, Supervision, Validation, Writing: Original Draft Preparation, Writing: Review \& Editing.\\
Alessandro Corsi --- Roles: Resources, Writing: Review \& Editing.\\
Antonio Culla --- Roles: Investigation, Data Curation.\\
Samantha Donsante --- Roles: Resources, Data Curation.\\
Giuseppe Giannicola --- Roles: Resources.\\
Irene Giardina --- Roles: Conceptualization, Data Curation, Formal Analysis, Funding Acquisition, Resources, Writing: Original Draft Preparation, Writing: Review \& Editing.\\
Giorgio Gosti --- Roles: Methodology, Validation.\\
Tom\'as S. Grigera --- Roles: Conceptualization, Formal Analysis, Methodology, Software, Writing: Original Draft Preparation, Writing:  Review \& Editing.\\
Stefania Melillo --- Roles: Conceptualization, Formal Analysis, Investigation, Data Curation, Methodology, Validation, Visualization.\\
Biagio Palmisano --- Roles: Investigation, Resources, Validation, Data curation, Writing: Original Draft Preparation, Writing: Review \& Editing.\\
Leonardo Parisi --- Roles: Formal Analysis, Investigation, Data Curation, Methodology, Software, Validation, Visualization.\\
Lorena Postiglione --- Roles: Methodology, Validation.\\
Mara Riminucci --- Roles: Resources, Writing - Original Draft Preparation, Writing: Review \& Editing.\\
Francesco Saverio Rotondi --- Roles: Data Curation, Formal Analysis, Investigation, Software, Visualization.\\

\section*{Data Availability}
The datasets generated in the current study are available from the corresponding author on reasonable request.

\section*{Funding}
This work was supported by the following grants: ERC Advanced Grant RG.BIO (Contract No. 785932) to ACa; MIUR Grant INFO.BIO (Protocol No. R18JNYYMEY) to ACa;
MIUR Grant PRIN2020 (Contract No. 2020PFCXPE-005) to IG; Grant SIOMMMS - Premi di Ricerca 2022 to BP.

\section*{Competing interests}

The authors declare no competing interests.

\vfill\eject

\appendix

\renewcommand{\theequation}{S.\arabic{equation}}
\setcounter{equation}{0}

\section*{\Large Supporting Information}

\vskip 1 cm 

\section*{Supplementary Methods}

\subsection*{Cell source}

Human bone marrow stromal cells (BMSCs) were isolated from bone samples harvested from healthy subjects undergoing elbow orthopaedic surgery through procedures approved by the local ethical board (Rif.~5313 Prot.~387/19) and upon informed consent of the patients.  Bone fragments were collected in Minimum Essential Medium with $\alpha$ modification ($\alpha$-MEM, Merck, Saint Louis, USA) supplemented with 1\% Penicillin/Streptomycin (P/S, Merck) by the surgeon and transferred to the laboratory where the isolation of the cells was performed.  Bone pieces were minced and washed with phosphate-buffered saline (PBS) in order to allow the release of bone marrow cells.  Cell suspensions were collected in culture medium consisting of $\alpha$-MEM supplemented with 20\% Fetal Bovine Serum (FBS, Thermo Fisher Scientific, Waltham, USA), 1\% P/S and 1\% L-glutamine (L-gln, Merck) and counted on a hemocytometer.  Cells were cultured at a density of $1$--$20 \cdot 10^3$ nucleated cells/cm$^2$ at $37^{\circ}$C, in a 5\% CO$_{2}$ atmosphere.  The remaining cells were cryopreserved at the cell density of $1$--$3 \cdot 10^7$ total cells (including red blood cells) per mL of Bambanker Serum-free cell freezing medium (Nippon genetics Europe) for subsequent uses.

\subsection*{Cell planting and seeding density}

After 24 hours of culture, non-adherent bone marrow cells were removed and adherent BMSCs were grown as P0 until confluency of the colonies (10-14 days). Confluent BMSC colonies were detached using 0.05\% trypsin-EDTA (Merck), counted and plated as P1. The same procedure was applied to establish P2 cultures. Time lapse experiments were performed with either P0 to P3 cultures (see Table~S1). P0 experiments were performed with freshly isolated BMSCs.  P1 experiments could be established either starting from freshly isolated cells, or from cryopreserved bone marrow cell suspensions.  P2 cultures were established starting from cryopreserved bone marrow cell suspensions.  To prepare the cultures for time lapse experiments, cells were plated at a density of 50$\,$cells/dish in $35\,$mm $\mu$-Dish Petri, model 81156 (Ibidi, Gr\"afelfing, Germany).  The culture medium was filtered through a $0.22\,\mu$m syringe filter to prevent the presence of serum residues in the culture.  Two identical dishes were generated, one starting the time-lapse experiments in the microscope stage top chamber incubator, and the other incubated in the standard incubator (Heracell, Heraeus) as positive control of cell growth.  After 6 hours, cells in the two dishes were checked to verify their attachment, and one of the two dishes was inserted in the microscope to start the time-lapse imaging, while the second dish becomes the control (see below).  At this stage, the actual number of cells found in the observational area of the $\mu$-Dish ranged between 10 to 20.

\subsection*{Time-lapse equipment}

The microscope is a phase contrast Nikon Inverted Eclipse Ti2-U, controlled by Nikon NIS-Elements Advanced Research software version 5.30.  The microscope is equipped with a Nikon 4X objective MRH20045, a Nikon 10X objective MRH20105, and a Nikon 20X objective MRH48250.  The images we use to perform the data analysis are taken with the 20X objective, but we also use the 4X objective for some auxiliary tasks.  The camera is a Nikon DS-Qi2 monochrome 14$\,$bit with a resolution of 16.25$\,$megapixels. Within the microscope there is an Okolab (Pozzuoli, Italy) stagetop chamber incubator, H301-NIKON-TI-S-ER, equipped with a two-well base H101-2 X 35-M, designed to host two dishes. The chamber incubator is mounted on the motorized translational stage Nikon Ti2-S-SS-E, thanks to which we could monitor several colonies in parallel within the same dish.  The gas mixture is controlled by an Okolab Bold-Line module H301-T-UNIT-BL-PLUS (Okolab-BL).  Redundant measurements of temperature and CO$_2$ are performed with an Okolab-LEO module.

The time lapse sampling period of experiments \texttt{20230113} and \texttt{20230503} was $20\,$min, while the sampling period of experiments \texttt{20240627}, \texttt{20240924}, \texttt{20241017}, \texttt{20241112}, \texttt{20241203}, \texttt{20250313}, \texttt{20250328}, \texttt{20250612}
and \texttt{20250626} was $15\,$mins.  This change in the sampling period was due to the need to simplify and therefore speed up the tracking procedure.

\subsection*{Culture control during time-lapse}
The conditions of the cell culture during time lapse were standard, namely humidity $90\%$, CO$_2$ 5\%, temperature $37^{\circ}$C.  {\it Gas control:} The gas mixture of air, water and CO$_2$ was produced and monitored by the Okolab-BL module; water to provide humidity was filtered and autoclaved.  To have an independent control of the CO$_2$ concentration within the incubator, we measure and store it every 5 minutes through the Okolab-LEO module: gas is pumped out of the incubator into the LEO for 3 minutes, so that an independent reliable determination of the CO$_2$ concentration can be made.  In the months before the experiment started, we calibrated this double-monitoring system in such a way to have consistent readings within the same incubator.  {\it Temperature control:} The incubator hosts two dishes. The first dish is the one that we use for the time-lapse, seeded with BMSCs, in which there are no temperature probes, in order not to perturb or contaminate the cells;  the second dish, at $6\,$cm from the cell-seeded dish, is used exclusively for temperature control (the Temperature-Control Petri, T-CP).  In the T-CP there is an amount of sterilized SIGMA water identical to the amount of medium in the cell-seeded dish ($2\,$ml). Within the T-CP there are two temperature probes; the first one is used by the Okolab-BL module to fix the temperature at 37 degrees through a digital feedback; the second probe is connected to the Okolab-LEO module that independetly records the temperature every $5\,$min.  In the months before the experiment started, we calibrated this double-monitoring system in the following way: we located the T-probe connected to the Okolab-BL module within the T-CP and the T-probe connected to the LEO within the dish that would contain the cells in a real experiment, and checked that the two readings were consistent; during this testing period the discrepancy between the two temperature probes was always lower than $0.2^{\circ}$C.

\subsection*{Selection of the originating cells}

Six hours after the primary dish is seeded with cells, the dish is inserted into the microscope incubator.  The first thing we need to do is to identify a couple of cells in the dish in order to tune the focus of the 4X objective; this procedure (which needs to be performed in live imaging) never takes more than 10 minutes, and often significantly less.  When the focus is fixed, we scan the entire dish; the scan is performed by taking $8\times 8=64$ photos covering the entire  dish.  The image resulting from this scan is then used to select the originating cells to monitor during the experiment (see below).  It is very important that this selection phase (which may be long, typically up to 30 minutes) is not performed in live imaging with the microscope lamp on, because that would dramatically increase the danger of phototoxicity (see below).  By working on the scan photograph, we can leave the cell-seeded dish dark.  Nevertheless, it is advisable to keep the selection procedure as quick as possible to avoid that potentially isolated cells candidates start dividing before the time-lapse is initiated. 

Within this study we only consider BMSC colonies that are derived from a single cell (single-cell-derived colonies in the classification of \cite{rennerfeldt2019emergent}).  Moreover, we only monitor colonies that are not infiltrated by other colonies during the time-lapse.  To meet these two requirements we seed the Petri dish at very low densities (see above), in order to have a higher probability to find isolated originating cells.  The fact that the initial cell must be isolated has a twofold importance: first, by selecting an isolated cell we give to the colony generated by that cell the necessary space to proliferate without interfering with other colonies during the time duration of the experiment; hence, here by `isolated' we mean that the distance from another originating cell must be of the same order as (or larger than) the average size of the colonies at generation $k=7$.  Secondly, by selecting only isolated cells, we are sure to avoid cases in which we have two (or more) cells close to each other at the beginning of the experiment, because in that case we cannot be sure whether the nearby cells are sisters or not, which is essential to define single-cell-derived colonies.  Given the very low seeding density, the size of our dish, and the size of the colonies at $k=7$, the isolation criterion severely limits the number of colonies we manage to successfully select and follow during each experiments; this is the chief reason why our experimental campaign lasted for so long.  We never selected more than 20 initiator cells, only a handful of which generate colonies that are finally included in the study (see below).

Once the cells have been selected from the scan image, we switch to the 20X objective and fine-tune the focus for each one of the selected cells, in order to optimise the sharpness of the images; moreover, in this phase we tune the position of each selected originating cell in order for it to be precisely at the centre of the field of view.  This procedure is quick (max $10\,$min) and needs to be done in live imaging, with the lamp on.  Finally, after the focus of the 20X objective is set, we start the time lapse.

\subsection*{Identifying the mitosis}

The precise determination of the division times is crucial for the analyses conducted in this study, hence it is essential to use a well-defined marker of mitosis  to uniquely identify a precise moment during the mitosis phase of the cell cycle.  Mitosis consists of several phases of different durations; the most distinctive and shortest is the mitotic cell rounding \cite{taubenberger2020mechanics, thery2008get}, during which the cell assumes an almost spherical shape, appearing as a bright, circular object in the image (see Fig.~1).  This trait is easily recognisable and was chosen as marker of mitosis.  The same criterion has been adopted in other studies \cite{rennerfeldt2019emergent}.  Clicking on the image corresponding to the mitotic rounding sets the stopping time of the mother cell and the starting time of the two daughter cells; these time-stamps are used to determine the division time $\tau$.  At the frame corresponding to the mitotic rounding the two daughter cells are still not distinguishable, hence the operator scrolls on through the sequence of images until the nuclei of the two daughters are clearly separated.  At that point the operator clicks on the two daughters, thus assigning a new label to each of them; these labels are connected to the mother's label by the corresponding mitosis. The process is repeated recursively for all cells, until the seventh generation is reached.  Cells of the seventh generation are no longer followed, so they only have the starting time, hence we do not have the division times of the cells of generation $k=7$.

One may ask why we use the cell rounding as marker of mitosis, instead of simply using the same frame as we use to label to two daughters.  The answer is that deciding what is the first image at which the two daughters' nuclei are clearly separated from each other has a large degree of arbitrariness and is operator-dependent; therefore, it is not advisable to use this frame as a marker of mitosis (for safety, we anyway checked that these two times are perfectly correlated to each other, see Fig.S2).  In contrast, the arbitrariness in the selection of the first image in which the two daughters are well-separated has no consequences on any observable, as it is only used to assign a label to the two daughters.

\subsection*{Monitoring the experiment}

The time-lapse experiment is monitored every 6 hours for its whole duration, according to the following check-list:
\begin{enumerate*}[label=(\roman*)]
  \item the  temperatures measured by the two probes within the T-CP must be consistent with each other and both equal to $37 \pm 0.5^{\circ}$C; 
  \item the graph in the past 6 hours of the temperatures measured by the two probes within the T-CP must show no anomaly; 
  \item the  CO$_2$ measured by the BL module at the gas source and by the LEO module within the incubator must be consistent with each other and equal to $5 \pm 0.2\%$;
  \item the graph in the past 6 hours of the CO$_2$ measured by the same two modules must show no anomaly;
  \item all photos must be sharp and all cells must be well in focus;
  \item colonies are checked for the need to expand the field of view (see below);
  \item colonies are checked for infiltration (see below);
  \item a rough count of all cells within each colony is performed.
\end{enumerate*}
Every 24 hours a scan of the entire dish at 4X is performed, to have a global view of the development of all colonies in the dish.  Since colonies' kinetics strongly fluctuate, the total experiment duration is variable: single experiments lasted between 10 to 14 days.  The time at which a colony reaches the seventh generation can be as short as 5 days for fast colonies but significantly larger for slower colonies.

\subsection*{Field of view expansion}

The field of view (FOV) around each colony is composed by $n\times n$ photographs at 20X, acquired by translating the chamber incubator with the motorized stage and stiched together by the proprietary Nikon software NIS-Elements.  At the beginning of the time lapse, when there is only the originating cell, we set $n=3$.  As the colonies grow it becomes necessary to expand the FOV, hence $n=3, 5, 7, \dots$. We always use an odd value of $n$ so that the FOVs before and after the expansion have the same center; this simplifies subsequent image analysis.  Our current version of the software does not allow to selectively expand the FOV of only certain colonies, therefore a FOV expansion from $n$ to $n+2$ implies a significant increase of the total number of photographs taken at each iteration.  This introudces some constraints related to phototoxicity that we discuss below.

\subsection*{Criteria to stop monitoring colonies}

During the experiment we can stop monitoring a colony for one of the following reasons: 
\begin{enumerate*}[label=(\roman*)]
  \item Despite the isolation criterion, a colony may come in contact with another colony during the experiment (infiltration); in that case we stop monitoring and we discard it for data analysis.  The infiltration criterion is the following: colonies A and B are mutually infiltrated if the distance between the two closest cells of A and B is equal to or smaller than the mean inter-cell distance within A and B.
  \item A selected originating cell may either die or detach from the dish before starting mitosis; in this case we stop monitoring it.
  \item When a selected originating cell does not make the first division within 4 days, we stop monitoring it.
  \item If one or more cells belonging to a colony drop out of the FOV before we manage to expand it, we stop monitoring that colony.
\end{enumerate*}
As a general rule, we keep monitoring all colonies that do not incur in the criteria above in order to avoid any selection bias.  In some cases, however, it becomes necessary to forcibly reduce the number of monitored colonies.  This is because as the colonies grow we need to expand the FOV, but as we shall see later on, the total number of photographs for each loop in the time-lapse is limited to 750 to avoid phototoxicity effects.  Hence, when FOV expansion is absolutely urgent but would exceed the total of 750 photographs and no colonies match the discard criteria above, we drop one colony randomly. 

Finally, of all colonies monitored during the experiment, we only include in our analysis those that produce {\it at least} two cells at generation $k=7$, lest the seventh generation criterion we have adopted to characterise the whole study becomes void.

\subsection*{Medium replacement}

Medium is changed every 3--4 days (typically every Tuesday and Friday, with the experiment starting on Friday), for the whole duration of the experiment.  Whenever the medium is changed in the cell-seeded dish, the sterile water is also changed in the T-CP, so to have always the same amount of fluid in the two dishes, which is essential to have a reliable temperature control.

\subsection*{Illumination parameters and phototoxicity}

To obtain brilliant and sharp images (which immensely simplifies the chain of analysis downstream) one would ideally illuminate the cells as much as possible during imaging.  However, this is not advisable due to the grave danger of phototoxicity.  In order to select the illumination parameters (light intensity, filters, exposure time) it is therefore essential to keep into account the potential phototoxic effects of a prolonged exposure of cells to light. 

Phototoxicity is a very dangerous effect, which acts in two ways: directly, by interfering negatively with the various cellular processes, among other things depressing proliferation and increasing the apoptosis rate \cite{surovtseva2021survival, laissue2017assessing, ong2013activation, yuan2017effects, wang2017red, li2023implication};  and indirectly, by altering the properties of the culture medium \cite{magni2024can}.  Phototoxicity can severely bias the data and must therefore be kept strictly under control.  The issue of phototoxicity is very complex: experimental tests are few and conducted on rather diverse cell types, by observing different cellular traits; hence it is difficult to  establish a consensus on the danger threshold of light level from the data in the literature.

The chief parameter to assess the impact of phototoxicity is {\it fluence}, defined as the energy per unit area hitting the culture surface, measured in Joule per square centimetre (J/cm$^2$).  In BMSCs, the reported critical value of fluence above which clear effects on cellular processes are observed ranges from $21\,$J/cm$^2$ \cite{surovtseva2021survival} to $12\,$J/cm$^2$ \cite{yuan2017effects}, but the danger threshold has been estimated as low as $4\,$J/cm$^2$ in ASC cells \cite{wang2017red}.  Regarding medium degradation, the most recent experiments indicate a toxicity threshold of $21\,$J/cm$^2$ \cite{magni2024can}.  From these results and given the lack of information about fluence and illumination parameters in most, if not all, time-lapse studies of BMSCs, we chose to stay well below the lowest of the above-mentioned bounds, namely $4\,$J/cm$^2$.  To do that, we proceed as follows.

In all our experiments the light intensity of the LED lamp of the microscope (Cree, Inc. XM-L2 series,\\ \texttt{XLMBWT-00-0000-0000T60E2}) is reduced by $5$ stops (i.e.\ a factor $2^5=32$) by means of a neutral filter.  By using a power meter (Thorlabs PM200, sensor Thorlabs S170C), we measured the power of the filtered light; even though the power does not depend on the objective, each objective requires a different phase filter to work, and the intensity of the light received by the sample {\it does} depend on the phase filter.  Hence, all power measurements must be performed with the appropriate phase filter.  To compute the power, the meter needs as an input the wavelength of the LED light.  Our LED's spectrum has two primary peaks (at $440\,$nm and $540\,$nm); to be conservative in our estimates of the fluence we input the power meter with the wavelength of the most energetic peak ($440\,$nm).  Power is the energy per unit time (milliwatt, mW), hence by dividing the measured power by the area of the disc of light produced by the lamp ($0.64\,$cm$^2$) we obtain the power per unit area that the sample receives, called {\it irradiance} (notice that for this to be true the light disc area must be smaller than the sensor area, which is the case in our setup; otherwise the power must be divided by the sensor area).  Irradiance is measured in mW/cm$^2$.  Once the irradiance is determined, the fluence ---total energy per unit area---  is obtained by multiplying the irradiance by the exposure time.  The exposure time depends on the specific phase of the experiments; those below are the values of the illumination parameters and relative irradiance and fluence during each one of the phases when the microscope lamp is on:
\begin{enumerate*}[label=(\roman*)]
  \item {\it Initial cell search ---} Once the cell-seeded dish is inserted into the microscope incubator we first need to identify a couple of sample cells, in order to tune the focus on them.  In this phase, the light intensity is set at $40\%$, the $5$-stop filter is on, and we use a 4X objective, giving an irradiance of $0.0086$ mW/cm$^2$; the camera sensor gain is set at $1.0$X.  The search lasts a maximum of 10 minutes (often significantly less), giving a fluence smaller than $0.006\,$J/cm$^2$. This operation is done once in the experiment.
  \item {\it Scan of the Petri dish ---} We scan the cell-seeded dish at the beginning of the experiment in order to identify candidate cells; moreover, we scan every 24 hours the dish to have a global view of the experiment's evolution.  During this task, the light intensity is set at $4\%$, the $5$-stop filter is on, and we use a 4X objective, giving an irradiance of $0.70$ $\mu$W/cm$^2$; the camera sensor gain is set at $1.0$X.  The scan is performed by taking $8\times 8=64$ photos, and the exposure time of each photo is $2\,$s, giving a fluence of $9.0 \times 10^{-5}\,$J/cm$^2$. This operation is done every 24 hours.
  \item {\it Focus fine-tuning ---} Once the originating cells are selected we fine-tune the focus for each cell.  In this phase the light intensity is set at $75\%$, the $5$-stop filter is on, and we use a 20X objective, giving an irradiance of $1.0$ mW/cm$^2$.  The procedure lasts a maximum of 10 minutes, giving a fluence smaller than $0.6\,$J/cm$^2$; the camera sensor gain is set at $4.1$X.  This operation is done once in the experiment.
  \item {\it Time lapse ---} During the time-lapse the light intensity is set at $75\%$, the $5$-stop filter is on, and we use a 20X objective, giving an irradiance of $1.0\,$mW/cm$^2$.  The exposure time of each photo is $80\,$ms, corresponding therefore to a fluence-per-photo of $8 \times 10^{-5}\,$J/cm$^2$; the camera sensor gain is set at $4.1$X.  At each iteration the system takes $n\times n$ photos around each colony, where $n=3, 5, 7, 9, \dots$.  If we are following $C$ colonies, the total number of photos at each iteration is $n^2 C$; we always keep this total number below 750 (in the most flourishing experiments, we may have 6 colonies, each one covered by a $11\times 11$ photo tiling, giving a total of 726 photos at each iteration).  In this way we have a maximum fluence of $8\times 10^{-5} \times 750 = 0.06\,$J/cm$^2$ for each iteration.  The whole loop repeats after 15 minutes.
  \end{enumerate*}
We therefore have a time-lapse fluence that is $66.6$ times lower than the most conservative bound found in the literature ($0.06\,$J/cm$^2$ vs.\ $4.0\,$J/cm$^2$).  Hence, we are confident that our observations of BMSCs are not affected by phototoxicity. 

During preparation, and for the entire duration of the time-lapse, all lights in the lab are off, including all instruments lights and displays; windows and doors are shielded.

\subsection*{Control dish}

When the cell-seeded  dish for time-lapse microscopy is prepared, a second  dish with identical procedure and seeding is also prepared and kept in a different (large) incubator (Heracell, Heraeus) as a control; this is the Control Dish (CD). The CD is not exposed to any light (besides the room and hood lights during medium changes), nor perturbation, and it undergoes medium replacements at identical intrevals as the time-lapse dish.  At the end of the experiment we perform a comparative analysis of the  dish used for the time-lapse vs.\ the CD, including a 4X scan of both.  In this way we can be certain that there are no statistically significant differences between the two dishes, and therefore conclude that the culture conditions were identical, so that the time lapse procedure has not interfered with the proliferation of the cells.  This check is of great importance, especially in connection with the risk of phototoxicity: since this risk is negligible for the CD (which is kept in the dark), comparable proliferation rate and vitality of the two dishes is a further test that no phototoxic effects are in action. All experiments used in this study passed this test.

\subsection*{Determination of the P-value}

To assess the significance of the correlation between two sets of $M$ variables, $\{x_i\}$ and $\{y_i\}$, with $i=1,\dots, M$, we proceed as follows.  First, we calculate the Spearman correlation coefficient $\rho$ on the actual experimental variables.  Then we randomly scramble the positions of the values in the first set, $\{x_i\}$, hence destroying correlations between the two variables, and we recalculate $\rho$.  We repeat the random scrambling $10^6$ times and calculate the fraction of cases that give a correlation coefficient $\rho$ larger than or equal to the experimental one; this fraction is the $P$-value of Spearman correlation.  Notice that with this procedure the most significant correlation that we can get has $P$-value less than $10^{-6}$.  The $P$-value defined in this way is therefore equal to the probability that the correlation found in the data is accidentally present also in an uncorrelated random dataset of the same size and of the same nature as the experimental one.

\subsection*{Cell tracking method}

To work out the lineage of a colony it is necessary to determine the relationships between mother and daughter cells, and to work out precisely all division times. To ensure unequivocal identification of cells and mitosis, the tracking of individual cells was performed manually via a custom semi-automatic software.  No unsupervised algorithm has been used.  To do this a C/C++ program was developed, based on the OpenCV computer vision library \cite{bradski2000opencv}.  Each of the 32 colonies was semi-manually tracked, requiring an average of 13 to 15 hours of work per colony.  The tracking program has a graphical user interface consisting of two windows: the first displays the acquired images and the second shows the lineage tree that is progressively formed during the tracking process.  Since all colonies are single-cell derived, the first step of the procedure is to tag the originating cell in the first frame of the acquisition by mouse-clicking on it; in that way the starting root link of the colony's lineage tree is generated.  The originating cell is special in the lineage, because we have no information about the mitosis generating it, so no division time is associated to the originating cell.  The next step is to connect the generating cell to its two daughter cells: the originating cell is selected in the lineage tree, which associates a label to that cell; the operator then scrolls through the sequence of images following the cell until reaching the images showing the initiation of the mitosis.

\section*{Supplementary Notes}

\subsection*{Robustness of the G$_0$ criterion}

During tracking, two special cases can be met:
\begin{enumerate*}[label=(\arabic*)]
\item A cell commits apoptosis and dies; this process is very clear in the imaging, hence there is no ambiguity in its identification.  The cell is tagged as `dead' in the lineage tree and graphically indicated with a red dot.
\item A cell stops dividing, going into the G$_0$ phase.  We have defined a G$_0$ cell as a cell that has not divided for a period of 84 hours since its birth; when this condition is met, we stop tracking that cell and we tag it as G$_0$ in the lineage tree.
\end{enumerate*}
In only three cases we were not able for technical reasons to follow a G$_0$ cell up to the full $84\,$h criterion: in colony \texttt{20241112\_02} one cell was tagged as G$_0$ after $68\,$h; in colony \texttt{20250313\_14} two cells were tagged as G$_0$ after $57\,$h.  The morphology of these cells, though, was very typical of the G$_0$ state.  Moreover, all three cells belonged to the last full generation, $k=6$, hence even in the unlikely case that their G$_0$ status changed, the topology of their trees would not be significantly altered.

The topology of a given tree entirely depends on the number and distribution of inactive cells; but while dead cells are unambiguously identified, G$_0$ cells (which are 95\% of inactive cells) are less easily so.  In our study we have employed a reasonable but still arbitrary operational definition of what a G$_0$ cell is.  We must be sure that by changing the threshold $t_c$ we do not produce trees with significantly different topologies.  This objection is not pedantic: if the threshold time is chosen too close to the bulk of the division times of the population, we would mistakenly classify as G$_0$ cells that are not, and consequently we would produce lineage topologies heavily dependent on the arbitrary value $t_c$.  In order to know that we are not running this risk, we cannot limit ourselves to noticing that $84$ hours is much larger than the mean division time of $20$ hours; precisely because of the great heterogeneity of BMSC colonies.  This is why in the main text we checked for each single colony that the threshold time $t_c$ that we are using to classify cells as G$_0$ is very distant not simply from the mean, but also from the bulk of the distribution of the division times of that colony.

Despite this separation, one could argue that the the G$_0$ criterion still influences some of our results.  Indeed, a possible (wrong) explanation of the correlation between kinetics of the active cells and the frequency of G$_0$ cells would be the following: if a colony is slow, the divisions times of its cells are on average large, so that once in a while a cell may divide after an exceedingly long time, larger than the threshold $t_c$ that we use to tag cells as G$_0$. Hence, the G$_0$ tag would not correspond to a {\it qualitatively} different cell state, but just to a {\it quantitative} epiphenomenon of having chosen an arbitrary threshold $t_c$.  Were this the case, the correlation between the number of inactive cells (which is primarily determined by the G$_0$ cells) and the division time (which is determined by the active cells) would be trivial ---in fact, even tautological--- as both quantities would derive from the {\it same} actual probability distribution of division times, which we would be artificially separating into two parts: the active cells, $\tau < t_c$, and the inactive cells, $\tau > t_c$.  This argument, however, is not consistent with the data, as we now show at a quantitative level. 

Given a certain colony, we can use the gap-exponential probability distribution $P(\tau)$ fitted to the actual experimental histogram of the division times of that colony (Fig.~3a), and calculate the probability to find a cell that has a division time larger than $t_c$ (i.e.\ a cell that would therefore be erroneously tagged as G$_0$ in our experiment, according to the argument above). This probability is given by, 
\begin{equation}
  P_{\text{G}_0} =  e^{-\gamma(t_c-\tau_\text{min})} \ .
\end{equation}
Using the actual experimental value $t_c=84\,$h and the fitted values of $\gamma$ and $\taum$ in each one of our colonies, we find that this probability is always extremely small, ranging between $10^{-17}$ to $10^{-2}$.  This means that if in each colony we extracted the cell division times using its own probability and tagged as G$_0$ all cells with $\tau>t_c$, we would obtain a number of G$_0$ cells approximately equal to $P_{\text{G}_0} \times 128$, which yields {\it zero} in 29 out of 32 colonies, in stark contrast with their actual total number of G$_0$ cells, which is $310$; in fact, even in the three colonies in which $P_\mathrm{G_0} \times 128$ is non-zero, we would obtain $7$, $2$ and $1$ G$_0$ cells, instead of the actual $22$, $18$, and $22$ G$_0$ cells that those colonies actually have.  We conclude that cells classified as G$_0$ in our experiments are {\it not} merely the very slow cells; the inactive cells sub-population is thus a {\it bona fide} qualitatively different group from that of the active cells.  Therefore, the tautological explanation of the correlation between topology and kinetics does not stand.

\subsection*{A caveat on the sign of the correlation between topology and kinetics}

It is important to note that the {\it  positive} correlation between the number of inactive cells and the mean division time (Fig.5 of the main text) can only hold for colonies large enough to have some cells at generation $k=7$ (as those included in this study); if, on the contrary, we selected very small colonies where {\it all} cells become G$_0$ early on in the development, so that no cells arrive at the seventh generation, we would enter a qualitatively different regime where $\Nina$ would be {\it proportional} to the size of the colony and where $N_\mathrm{missing}$ would always be equal to $128$. If one is interested in that regime, other metrics should be used to describe the topology of the tree. 

\section*{Supplementary equations}

\subsection*{Synchronisation vs correlation}

To illustrate the mathematical mechanism behind synchronisation, we analyse explicitly the case of sister cells. Consider a mother cell that divides at an absolute time $t_{m}$, which is also the  birth time of its two daughters. If we call $\tau_{s1}$ and $\tau_{s2}$ the division times of the two sisters, sister 1 will divide at a time $t_{s1}= t_{m} + \tau_{s1}$ and sister 2 at $t_{s2} = t_{m} + \tau_{s2}$; perfect synchronisation of the daughters occurs when $t_{s1}=t_{s2}$. The square {\it desync} is the square difference of the absolute division times of the two sister cells, $(\Delta t)^2 = \left(t_{s1} - t_{s2}\right)^2$. We can average the square desync over all divisions within the same generation, a quantity that we will indicate as $\langle (\Delta t)^2 \rangle$. By using the definition of $t_{s1}$ and  $t_{s2}$ given above, we obtain,
\begin{equation}
 \tmean{(\Delta t)^2} = 
\tmean{ \left(t_{s1}-t_{s2}\right)^2}=
 \tmean{ \left( t_{m} + \tau_{s1} -  t_{m} - \tau_{s2} \right)^2}=
 \tmean{ \tau_{s1}^2} +  \tmean{ \tau_{s2}^2} -2 \tmean{ \tau_{s1} \tau_{s2}} \ .
 \label{carell}
\end{equation}
Once we average over all events, all single cell means will be the same, hence $\tmean{ \tau_{s1}^2}= \tmean{ \tau_{s2}^2}= \tmean{ \tau^2}$. In absence of correlation we have, $ \tmean{ \tau_{s1} \tau_{s2}} =  \tmean{ \tau_{s1} } \tmean{\tau_{s2}} = \tmean{ \tau}^2$. In this way the square desync becomes, 
\begin{equation}
 \tmean{(\Delta t)^2} = 
 2\left(\tmean{ \tau^2} -  \tmean{ \tau}^2 \right)\ ,
 \label{stanW}
 \end{equation}
which is just twice the variance of the distribution of the division times, $\sigma^{2}_{\tau}$; hence the desync in absence of correlation is given by,
\begin{equation}
\tmean{(\Delta t)^2} = 2 \sigma_{\tau}^2  \ .
\label{zumpa}
\end{equation}
Instead, if we do not assume that the sisters' division times are uncorrelated, we can rewrite the equation \eqref{carell} in the following way, 
\begin{equation}
  \tmean{(\Delta t)^2} =  \tmean{\tau_{s1}^2} + \tmean{\tau_{s2}^2} - 2\tmean{\tau_{s1}}\tmean{\tau_{s2}} 
  + 2\tmean{\tau_{s1}}\tmean{\tau_{s2}}  - 2  \tmean{\tau_{s1}\tau_{s2}} = 
  2\left( \tmean{\tau^{2}} - \tmean{\tau}^2 \right) - 
  2\left(  \tmean{\tau_{s1}\tau_{s2}}  - \tmean{\tau_{s1}}\tmean{\tau_{s2}} \right) \ ,
  \end{equation}
where we recall that $\tmean{\tau^{2}} - \tmean{\tau}^2=\sigma^2_\tau$ is the variance of the division times.
If we now introduce the Pearson correlation coefficient, $\rho_\mathrm{P}$, between the division times, 
\begin{equation}
\rho_\mathrm{P} = \frac{\tmean{\tau_{s1}\tau_{s2}}  - \tmean{\tau_{s1}}\tmean{\tau_{s2}}}{\sigma_\tau^2}  \ ,
\end{equation} 
we finally obtain that the square desync with correlated sisters' division times is given by,
\begin{equation}
    \tmean{(\Delta t)^2} =2 \sigma_\tau^2 \left(1  - \rho_\mathrm{P}\right) \ .
\end{equation}

\subsection*{The branching process model}

Here we give the solution of the branching process used in the main text to model the growth of the number of cells in a colony.  A branching process \cite{harris1963theory} is a stochastic process for the evolution of a population where one assumes \citep[Ch.~III]{van_kampen_stochastic_2007} that
\begin{enumerate*}[label=(\roman*)]
\item each individual (here, a cell) starts its own family of descendants;
\item all the families have the same stochastic properties;
\item the families do not interact with each other (no inter-marriage).
\end{enumerate*}
At any given time one has $N(t)$ cells, whose life can end with some probability per unit time $\gamma(\tau)$, which in general depends on each individual's age (and is thus different for each), and in this sense the process is non-Markovian.   At the end of its life, the individual is replaced by $r$ other identical individuals with probability $p_{r}$; here we allow only $r=0,2$ (i.e.\ apoptosis or cell division).  Alternatively, a Markovian description can be obtained if the state of the population is described not simply by the number of objects present but by a list of objects with an age and a lifetime \cite[Ch.~VI]{harris1963theory}.

One defines the conditional probability $P(N,t |m,0)$ of having $N$ cells in the colony at time $t$ given that there were $m$ cells at the beginning ($t=0$).  A consequence of the fact that different families do not interact is that we can write $P(N,t |m,0)$ as the convolution of $m$ factors $P(N,t|1,0)$,
\begin{equation}
    P(N,t |m,0) = \sum_{s_1=0}^N \ldots \sum_{s_{m-1}=0}^{N-s_{1}-\ldots-s_{m-2}}P(N-s_{1}-\ldots-s_{m-1},t |1,0)P(s_1,t |1,0)\ldots P(s_{m-1},t |1,0)
\end{equation}
Now we introduce the generating function \cite{harris1963theory, van_kampen_stochastic_2007},
\begin{equation}\label{eq: generating function}
    F(z,t|m,0) = \sum_{n=0}^{\infty}z^n P(n,t|m,0) ,
\end{equation}
with $t\geq 0$ and $|z|\leq 1$. It can be shown that the generating function of a convolution is the product of the generating functions \cite{van_kampen_stochastic_2007}, so that
\begin{equation}\label{eq: generating function of a convolution}
    F(z,t|m,0) = [F(z,t|1,0)]^m.
\end{equation}
We can then focus on $P_n(t) = P(n,t|1,0)$ and its generating function, which we write simply as  $g(z,t) = F(z,t|1,0)$.

The probability that a cell of age $\tau$ ends its life (through either division or apoptosis) in the next  infinitesimal time interval $d\tau$ is $ \gamma(\tau)d\tau$.  Calling $w(\tau)$ the probability for the cell to reach age $\tau$ without death or division, it holds that
\begin{equation}
    dw(\tau) = -\gamma(\tau) w(\tau) d\tau
\end{equation}
(note that this is different from the probability to divide or die exactly at age $\tau$, defined below).  Solving the differential equation for $w(\tau)$,
\begin{equation}\label{eq:wtau}
    w(t) = e^{-\int_0^t dt' \gamma(t')},
\end{equation}
and the probability to reach age $\tau$ and undergoing an event between $\tau$ and $\tau+d\tau$ is
\begin{equation}
    p(\tau)\,d\tau = w(\tau) \gamma(\tau)\,d\tau  = -\frac{dw(\tau)}{d\tau} \,d\tau.
    \label{eq:ptau-wtau}
\end{equation}
With these definitions, an integral equation for the generating function can be found \cite{harris1963theory}:
\begin{equation}\label{eq: generating function differential equation}
    g(z,t) = zw(t) - \int_0^t h[g(z,t-t')]\, dw(t'), 
\end{equation}
where $t\geq 0$, $|z|\leq 1$, $h(s)$ is the generating function of the probability for the number of descendants,
\begin{equation}\label{eq:pr-gen-function}
    h(s) = \sum_{r=0}^{\infty} p_{r} s^r.
\end{equation}
Following the procedure used in \cite{lavalle2023fluctuations} we finally obtain an equation for the average number of cells as function of absolute time:
\begin{equation}\label{eq:eq-for-N}
    \langle N(t) \rangle - 1 = \int_0^t [m \langle N(\tau) \rangle - 1]\gamma(t-\tau)w(t-\tau)\,d\tau ,
\end{equation}
where $m = h'(s)|_{s=1}$.  The solution $\protect\langle N(t) \protect\rangle$ can be shown to be unique and bounded in every finite interval $[0,t]$.

Equation~\eqref{eq:eq-for-N} is quite general; to attempt to solve it, explicit expressions for $\gamma(\tau)$ and $p_{r}$ must be given.  For the rate $\gamma(\tau)$ we choose
\begin{equation}\label{eq:gamma-t}
    \gamma(\tau) = \Theta(\tau-\tau_\text{min})\gamma,
\end{equation}
i.e.\ zero probability for $\tau\le\tau_\text{min}$, and a constant rate for larger ages.  Through~\eqref{eq:wtau} and~\eqref{eq:ptau-wtau}, this leads to the gap-exponential distribution of cell cycle duration used in the main text,
\begin{equation}\label{eq:gap-exponential}
    p(t) = \gamma \Theta(t-\tau_\text{min}) e^{-\gamma(t-\tau_\text{min})}.
\end{equation}
For the probability of descendants we choose
\begin{equation}
  \label{eq:pr}
  p_{r} = p_{0} \delta_{r,0} + (1-p_{0}) \delta_{r,2},
\end{equation}
i.e.\ the cell can either die or divide (hence $p_0$ is the probability of apoptosis).  In this model, then, apoptosis only occurs at the end of the cell cycle.  Given that apoptosis is more likely to occur after the end of the G$_{1}$ phase \cite{pucci2000} (which is typically the longest), and that apoptosis occurs with low probability in our experiments, this limitation should not be problematic.

We still need to add the probability of entering the G$_{0}$ phase.  We do this by modifying $p(t)$: the probability of undergoing division or death in the interval $[t_1, t_2]$ is $\int_{t_{1}}^{t_{2}}p(t)\,dt$.  Expression~\eqref{eq:gap-exponential} is normalized such that $\int_0^\infty p(t)\,dt=1$, i.e.\ all cells die or divide.  The most convenient way to introduce G$_{0}$ is not through a new transition but simply allowing some probability for a cell to never finish the cycle: we thus repalace~\eqref{eq:gap-exponential} by
\begin{equation}\label{eq:gap-exponential2}
    p(t) = a \gamma \Theta(t-\tau_\text{min}) e^{-\gamma(t-\tau_\text{min})}  \ ,
\end{equation}
with $a<1$, so that $1-a$ is the probability to never divide or die, i.e.\ to become G$_{0}$.

Now the equation for $\langle N(t)\rangle$ can be solved using Laplace transforms, adapting the procedure of \cite{lavalle2023fluctuations} for the present case with two additional parameters.  Introducing the Laplace transforms
\begin{align}
    n(s) &= \int_0^{\infty}e^{-st}\protect\langle N(t) \protect\rangle \, dt , \\
    y(s) &= \int_0^{\infty}e^{-st}w(t)\gamma(t) dt = \int_0^{\infty}e^{-st}p(t) \, dt
    =  a\cdot\frac{\gamma}{s+\gamma}e^{-s \tau_\text{min}},
\end{align}
where $s \in \mathbb{C}$, equation~\eqref{eq:eq-for-N} is solved in the Laplace domain as
\begin{equation}\label{eq: equation for n(s)}
    n(s) = \frac{1}{s}\frac{1-y(s)}{1-m\cdot y(s)},
\end{equation}
and, going back to the time domain,
\begin{equation}
    \protect\langle N(t) \protect\rangle = \frac{1}{2\pi i}\int_{\alpha - i\infty}^{\alpha + i\infty}\left[\frac{1}{s} + \frac{\gamma(m-1)a\cdot e^{-s\tau_\text{min}}}{s\phi(s)}\right]e^{st}\, ds,
\end{equation}
where $\phi(s) = s+\gamma(1 - ma\cdot e^{-s\tau_\text{min}})$ and $\alpha$ is a real number such that the integration contour lies to the right of all the poles of the integrand.  The integration is done using the residue theorem; the integrand has poles at $\hat{s} = 0$ and at
\begin{equation}
    \hat{s}_k = -\gamma + \frac{1}{\tau_\text{min}}W_k(m\gamma a\tau_\text{min} e^{\gamma \tau_\text{min}}),
\end{equation}
where $W_k$ is the k-$th$ branch of the Lambert, or product log, function.  The residue theorem gives
\begin{equation}
    \protect\langle N(t) \protect\rangle =  1 + \Theta(t-\tau_\text{min})\cdot\left[-\frac{(m-1)a}{(ma-1)} + \sum_{k=-\infty}^{+\infty}\gamma(m-1)a \frac{e^{\hat{s}_k(t-\tau_\text{min})}}{\hat{s}_k[1 + \tau_\text{min}(\hat{s_k} + \gamma)]}\right].
\end{equation}
Finally, from~\eqref{eq:pr} and~\eqref{eq:pr-gen-function} we have $h(s) = p_0 + (1-p_0)s^2$, so that
\begin{equation}
     m = h'(s)|_{s=1} =  2(1-p_0),
\end{equation}
and therefore
\begin{equation}\label{eq: N(t) alternative}
    \langle N(t)\rangle = n_0 + n_0\cdot \Theta(t-\tau_\text{min})\left[-\frac{(1-2p_0)a}{2a(1 - p_0) - 1} + \sum_{k=-\infty}^{+\infty}\gamma(1-2p_0)a \frac{e^{\hat{s}_k(t-\tau_\text{min})}}{\hat{s}_k[1 + \tau_\text{min}(\hat{s_k} + \gamma)]}\right] \ .
\end{equation}
This is the function we use to predict and fit the experimental growth curves in the main text.\\
The red and green dashed curves shown in Fig. 6 are obtained by plugging the correct set of parameters ($\gamma$, $\taum$, $p_0$, $a$) for each colony in the previous expression and computing $\langle N(t)\rangle$ accordingly. The four parameters used for the red dashed line  were obtained directly from the experimental data: $\gamma$ and $\taum$ are obtained by fitting the experimental probability distribution of the division times $P(\tau)$, while $p_0$ and $a$ are obtained by simply applying, for each colony,
\begin{equation}
    p_0 = \frac{\text{\# dead cells}}{\text{\# dead cells + \# active cells}}
\end{equation}
\begin{equation}
    a = 1 - \frac{\text{\# $G_0$ cells}}{\text{\# $G_0$ cells + \# active cells + \# dead cells}}
\end{equation}
On the other hand, the four parameters used for the green dashed line are obtained through a least square fit of equation (\ref{eq: N(t) alternative}) on the experimental $N(t)$ in the temporal range $[t_0, t_0 + 5\bar{\tau}]$, where $t_0$ is the absolute time at which the first cells of the colony undergoes mitosis and $\bar{\tau}$ is the average division time of the specific colony we are considering.

\cleardoublepage

\section*{Supplementary table}

\begin{figure}[h!]
  \centering
  \includegraphics[width=\textwidth]{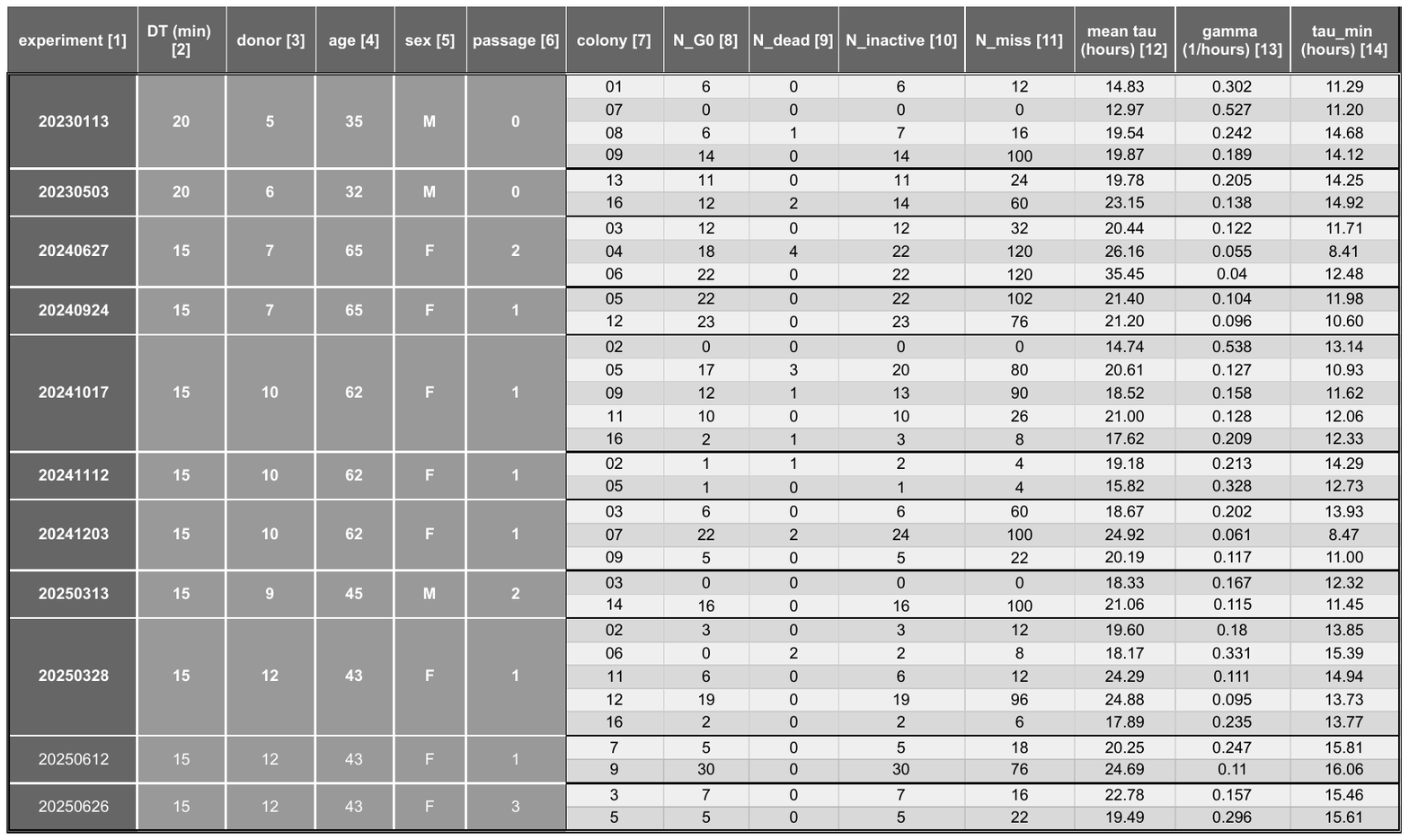}
  \label{TableS1}
  \caption*{ {\bf Table S1: Detailed parameters of all colonies.}
  [1] Date of the experiment;
  [2]  time interval between frames in the time lapse;
  [3] donor's label;
  [4] donor's age;
  [5] donor's sex;
  [6] passage;
  [7] label of the colony within that experiment;
  [8] number of G$_0$ cells,  $N_{G_0}$;
  [9] number of dead cells, $N_\mathrm{dead}$;
  [10] number of inactive cells, $N_\mathrm{inactive} = N_{G_0} + N_\mathrm{dead}$;
  [11] number of missing cells at generation $k=7$, $N_\mathrm{missing}$;
  [12] mean division time of the colony, $\bar\tau$;
  [13] proliferation rate,  $\gamma$, obtained from the gap-exponential fit;
  [14] minimum division time, $\tau_\mathrm{min}$, obtained from the gap-exponential fit.}
\end{figure}

\cleardoublepage

\section*{Supplementary videos}

\paragraph*{\bf Video S1.}
\label{S1_Video}
{\bf Development of a colony starting from a single cell.} 
Time-lapse video showing the development of colony \texttt{20250328\_16}, starting from a single cell and progressing until confluence is reached. 
This colony was selected as it exhibits moderate growth and is representative of the average behaviour observed in the colonies analysed in this study. 
The video spans slightly more than 10 days and 14 hours of real time, with each second in the video corresponding to 6 hours. 
The timer displayed in the bottom right corner begins at 00d:00h:00m, corresponding to the frame in which the initial cell undergoes mitosis to give birth to two daughter cells. 
As the colony expands and cells move across the field of view, the video zooms out to ensure all cells remain visible. 
These changes in magnification can be tracked using the scale bar in the bottom left corner.
A colour shift in the time from white to yellow, occurring between days 5 and 6, marks the final frame included in the quantitative analysis. At this point, all cells had either become inactive or reached the seventh generation.

\paragraph*{\bf Video S2.}
\label{S2_Video}
{\bf Synchronisation of cells' mitosis.} 
Time-lapse video showing the near-perfect synchronisation of the mitosis of the four cousin cells of generation  $k=2$. Frames are separated by 15 minutes intervals.

\cleardoublepage

\section*{Supplementary Figures}
\vskip 1 cm 
\subsection*{Figure S1}
\begin{figure}[h!]
\centering
\begin{subfigure}{0.48\textwidth}
  \centering
  \includegraphics[width=\textwidth]{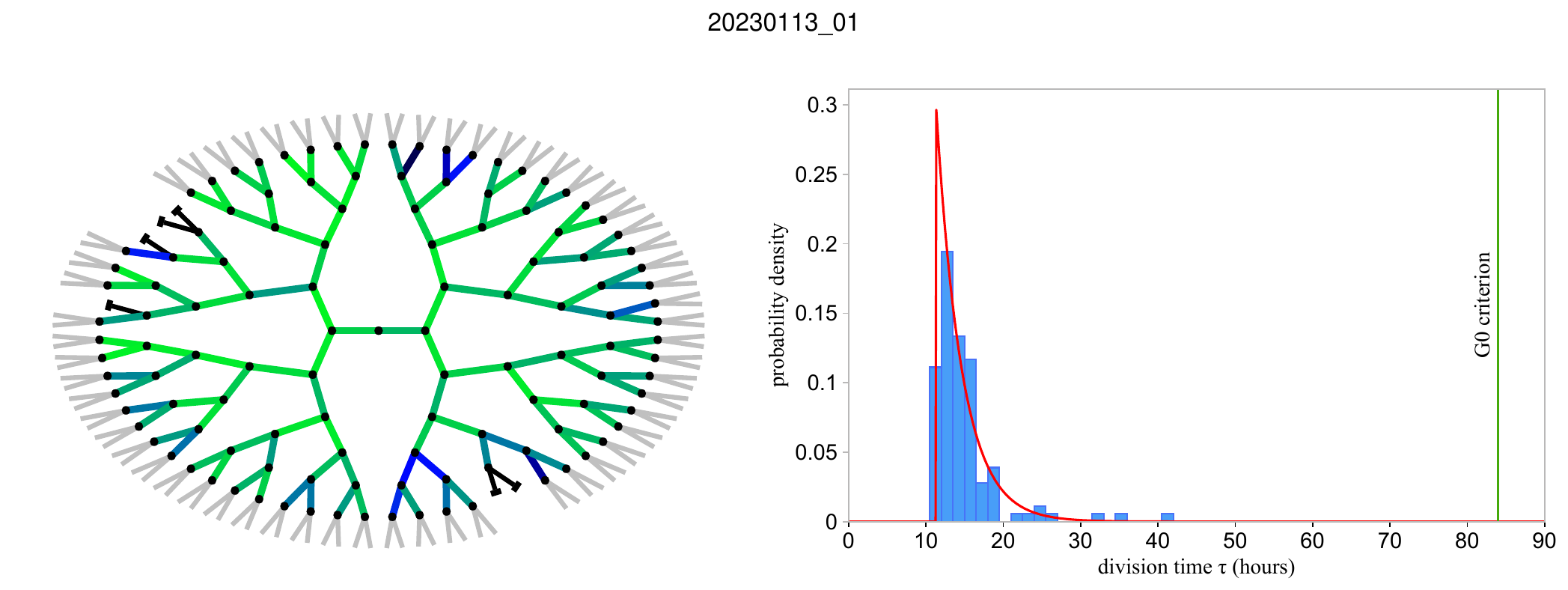}
\end{subfigure}
\hfill
\begin{subfigure}{0.48\textwidth}
  \centering
  \includegraphics[width=\textwidth]{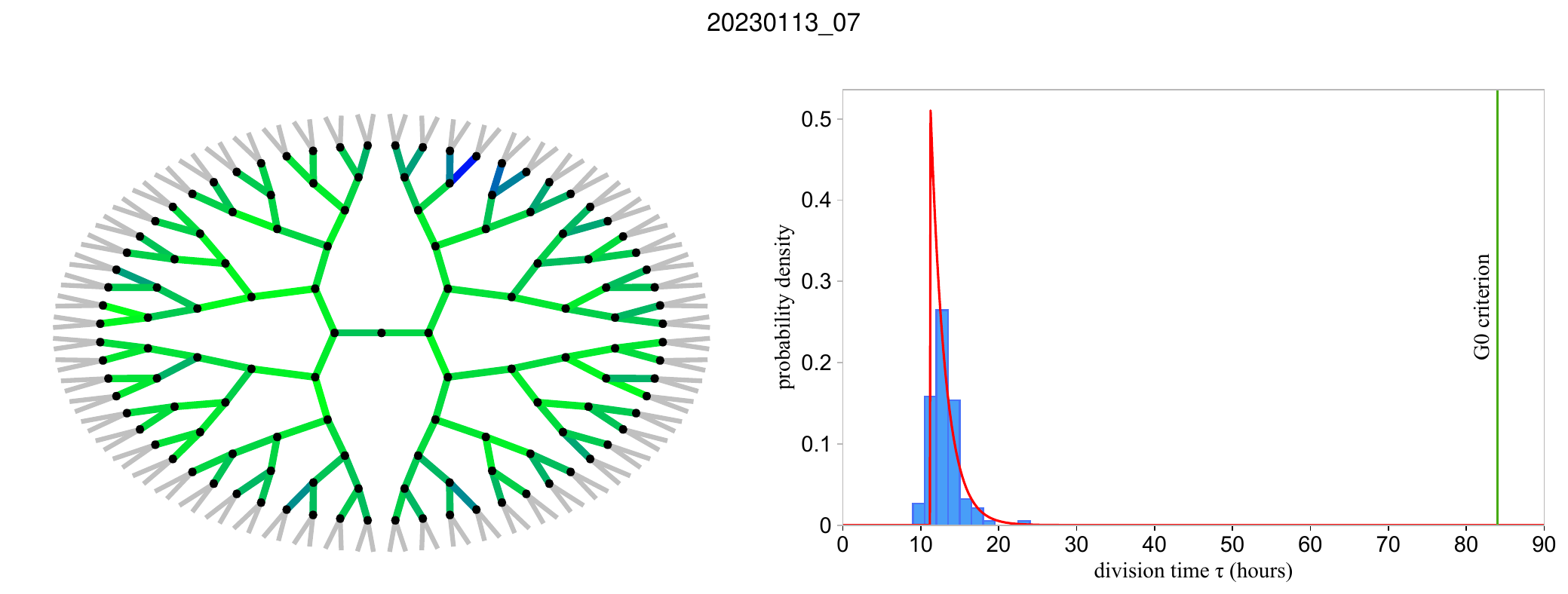}
\end{subfigure}
\vspace{2em} 

\begin{subfigure}{0.48\textwidth}
  \centering
  \includegraphics[width=\textwidth]{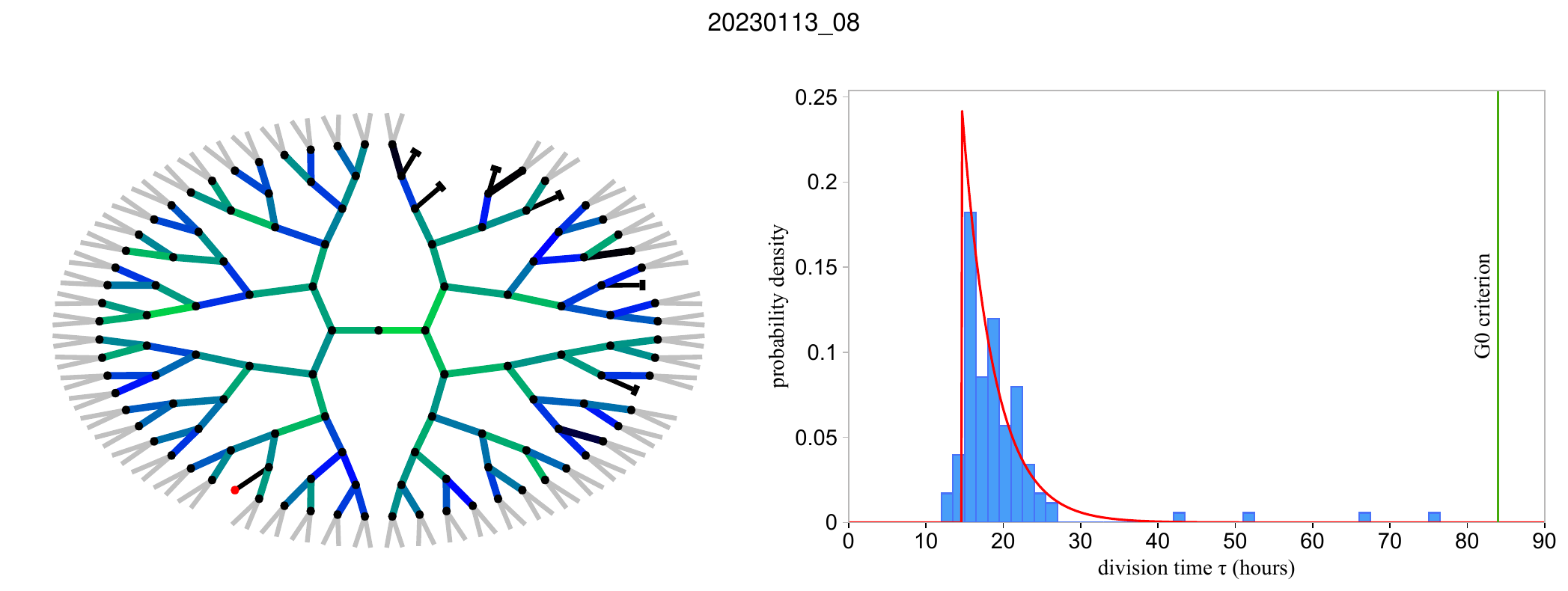}
\end{subfigure}
\hspace{0.pt}
\begin{subfigure}{0.48\textwidth}
  \centering
  \includegraphics[width=\textwidth]{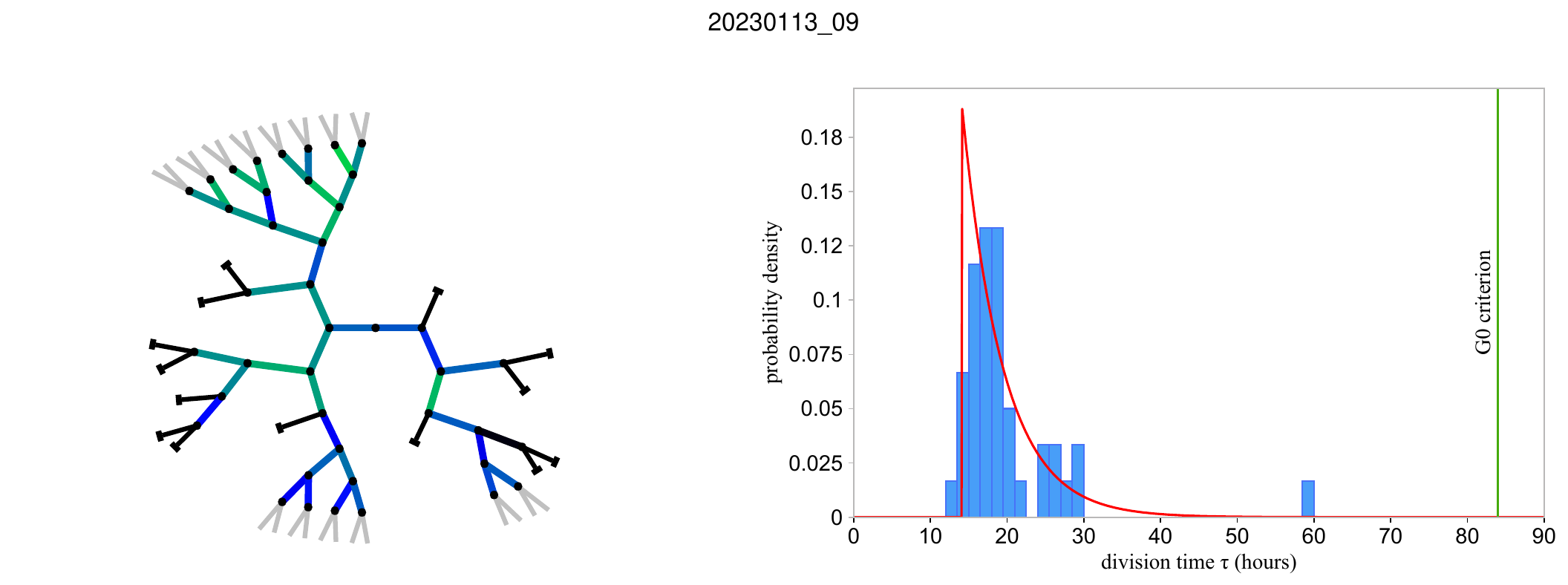}
\end{subfigure}

\vspace{2em}

\begin{subfigure}{0.48\textwidth}
  \centering
  \includegraphics[width=\textwidth]{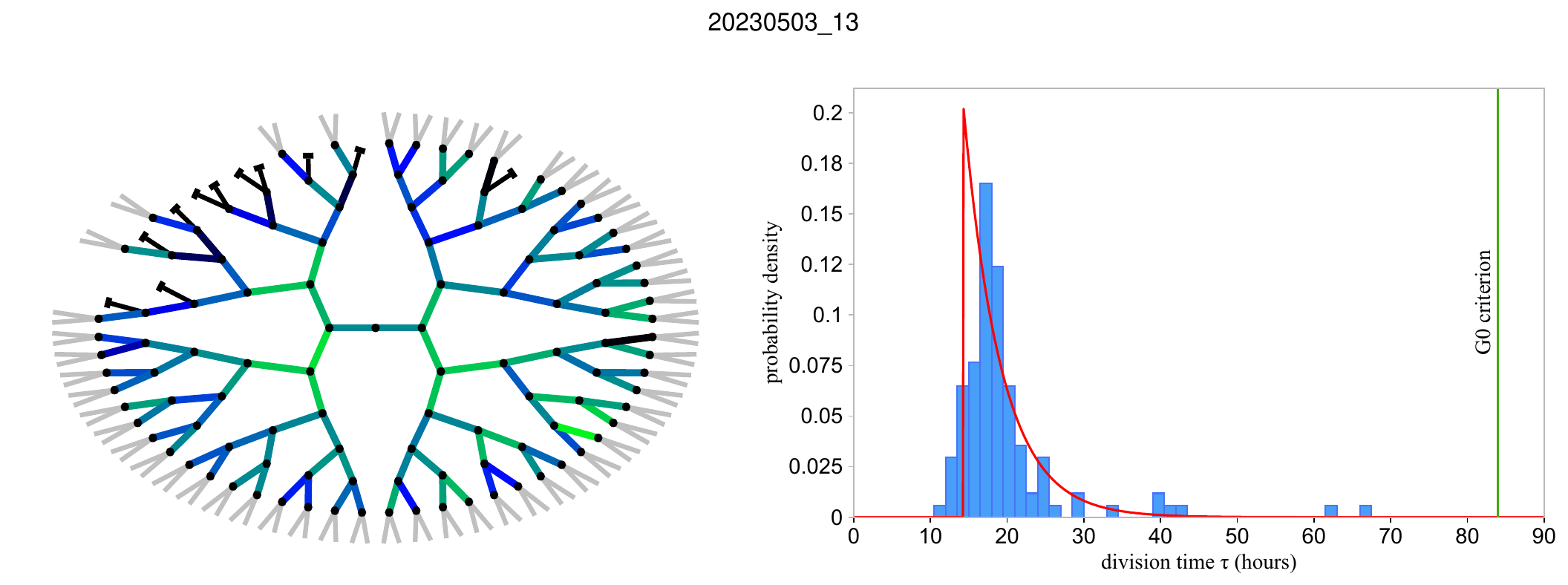}
\end{subfigure}
\hspace{0.pt}
\begin{subfigure}{0.48\textwidth}
  \centering
  \includegraphics[width=\textwidth]{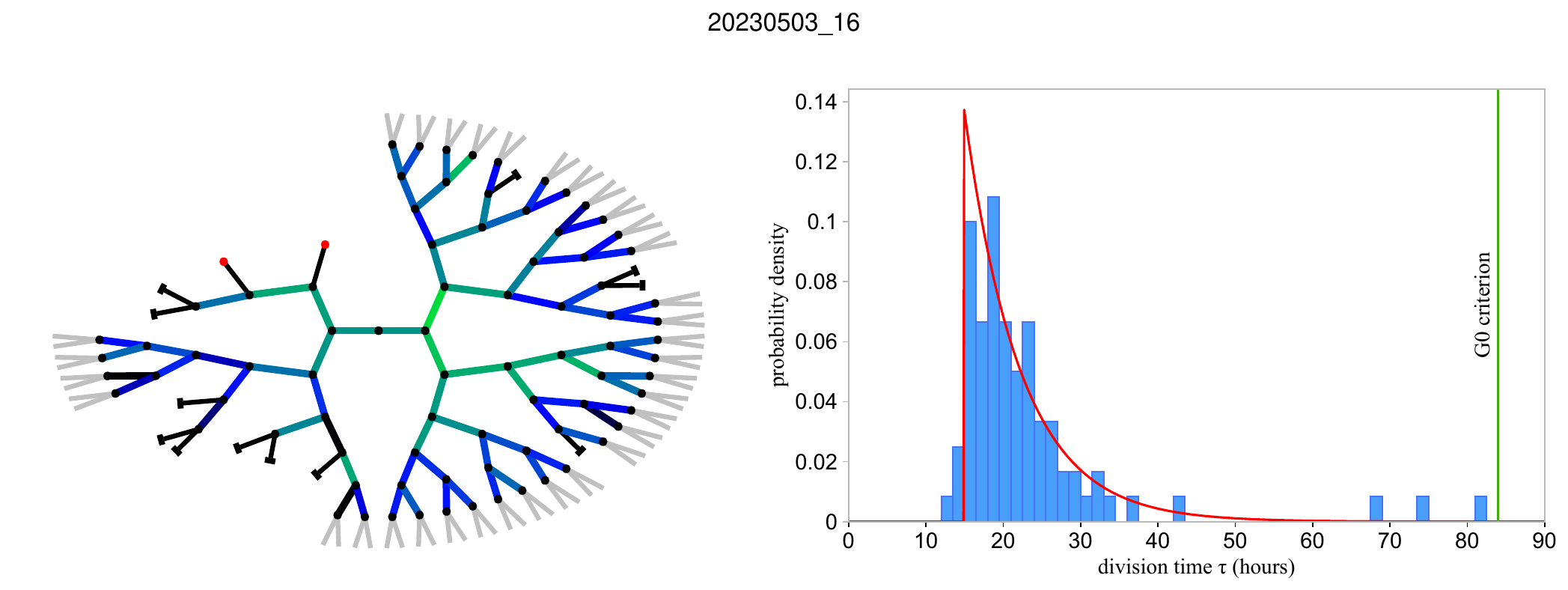}
\end{subfigure}

\vspace{2em}

\begin{subfigure}{0.48\textwidth}
  \centering
  \includegraphics[width=\textwidth]{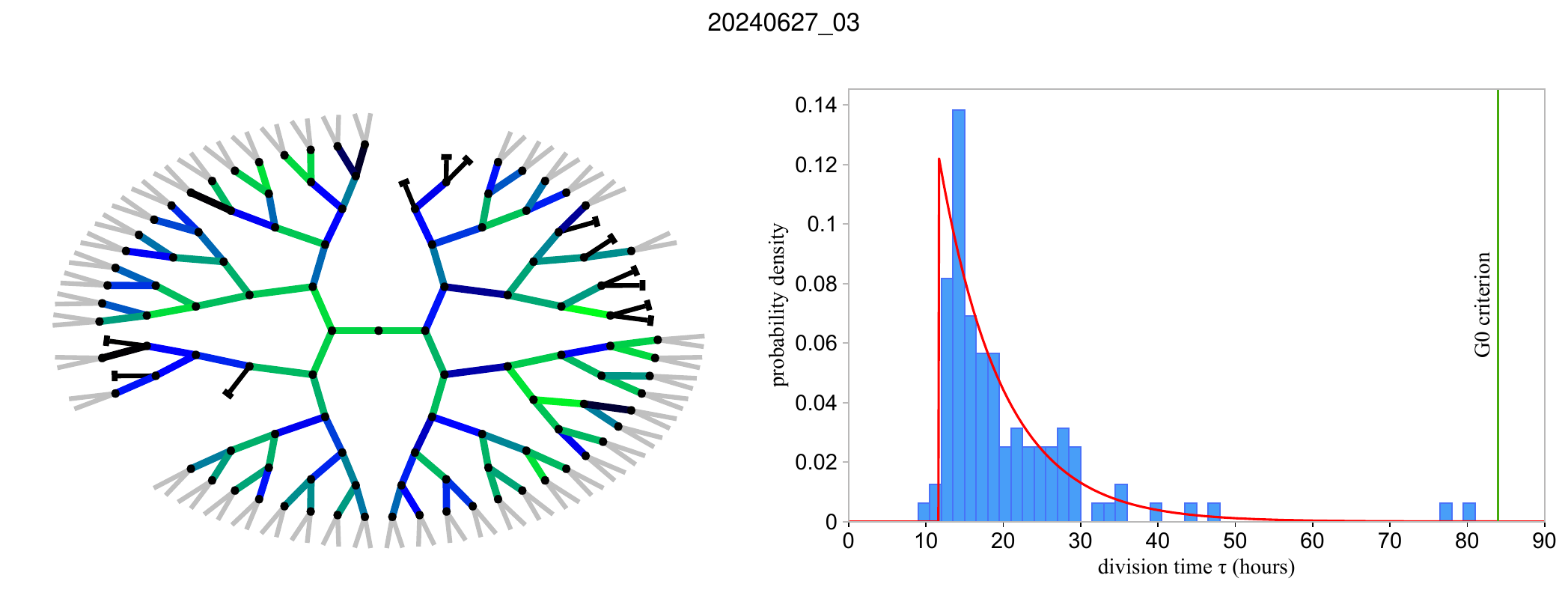}
\end{subfigure}
\hspace{0.pt}
\begin{subfigure}{0.48\textwidth}
  \centering
  \includegraphics[width=\textwidth]{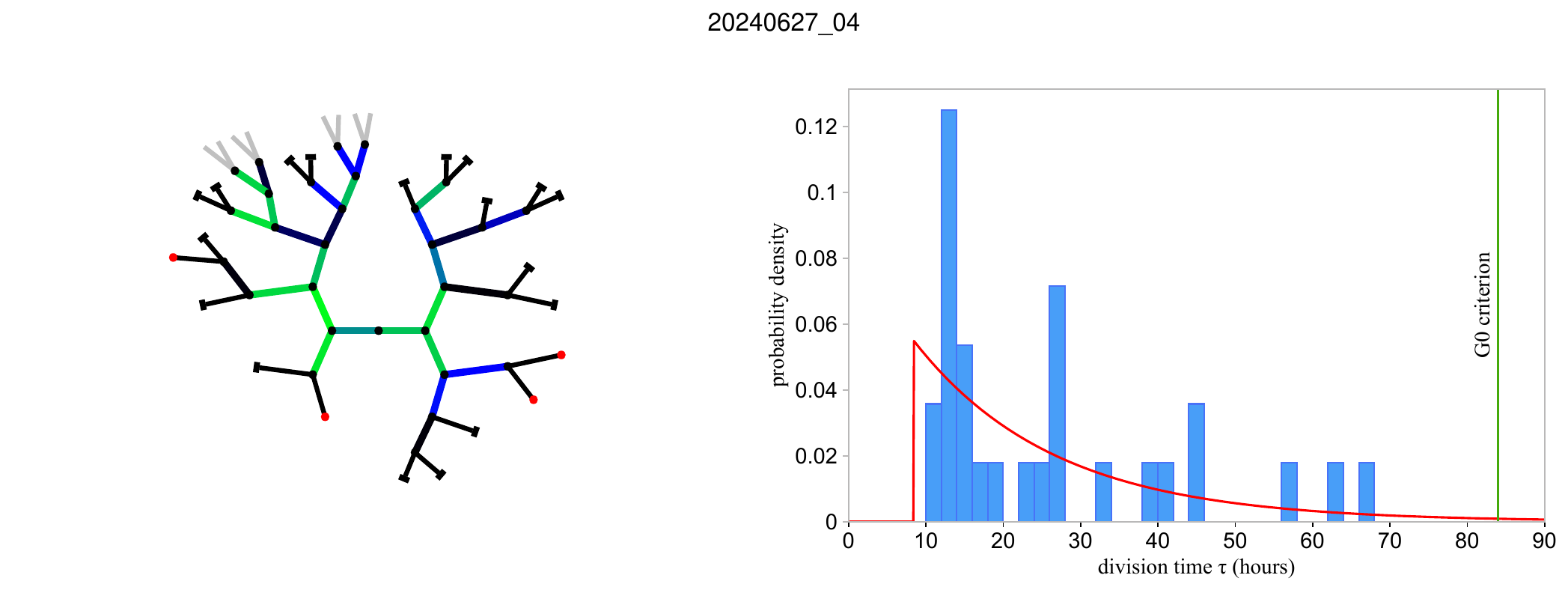}
\end{subfigure}

\captionsetup{width=\dimexpr\textwidth\relax, font=small, justification=justified, skip=2cm}
\caption*{{\bf Fig S1. Lineage trees and division times histograms of all colonies.} Abstract representation of the lineages of all the BMSC colonies in our dataset, including those reported in Fig.2. For each colony we also report the corresponding normalised histogram of the cells' division times, analogous to the ones in Fig.3.}
\label{S1_Fig}
\end{figure}

%%%%%
\newpage
\subsection*{{\bf Figure S1 (continued)}}
\begin{figure}[h!] 
\centering

\begin{subfigure}{0.48\textwidth}
  \centering
  \includegraphics[width=\textwidth]{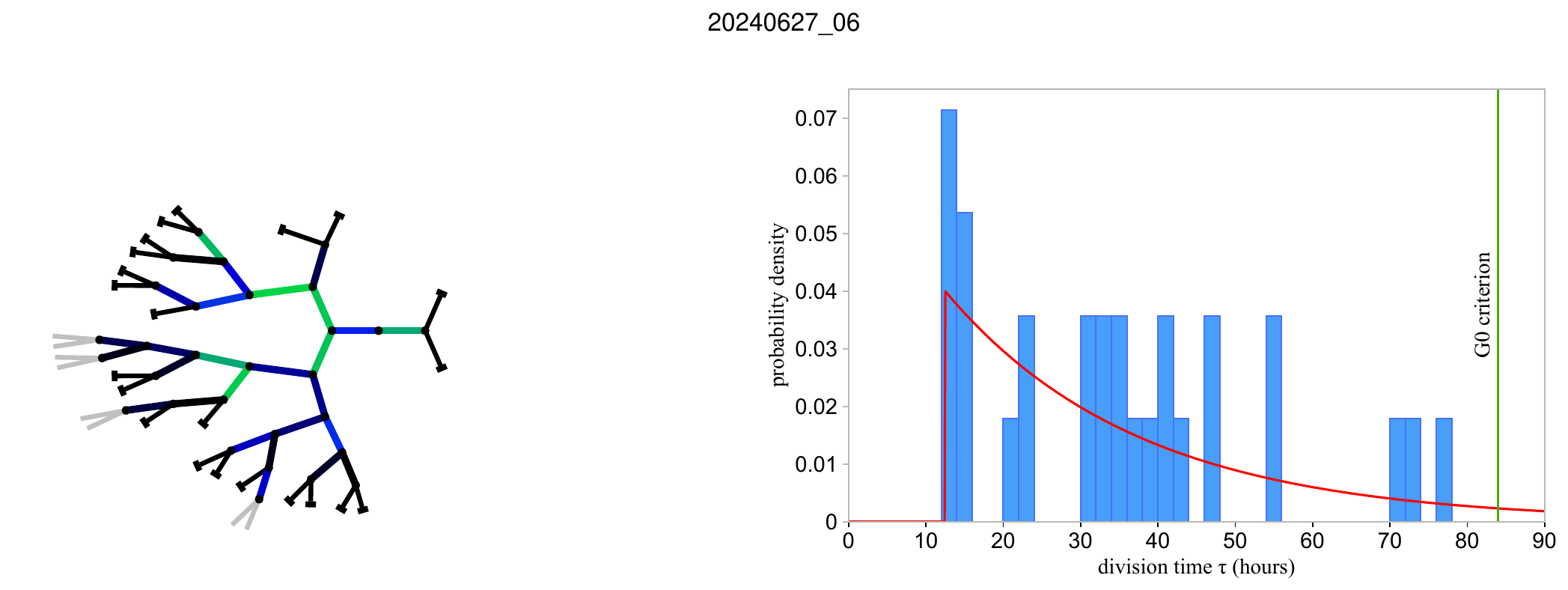}
\end{subfigure}
\hspace{0.pt} 
\begin{subfigure}{0.48\textwidth}
  \centering
  \includegraphics[width=\textwidth]{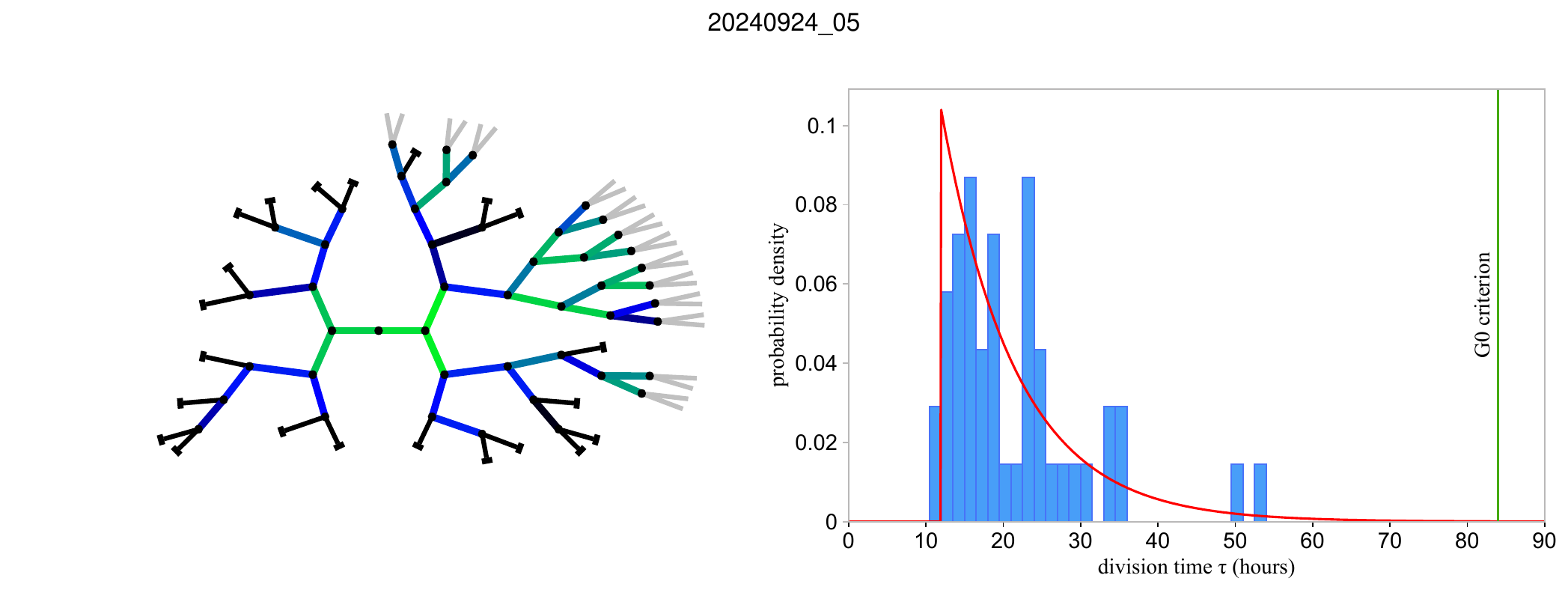}
\end{subfigure}

\vspace{0.5em}

\begin{subfigure}{0.48\textwidth}
  \centering
  \includegraphics[width=\textwidth]{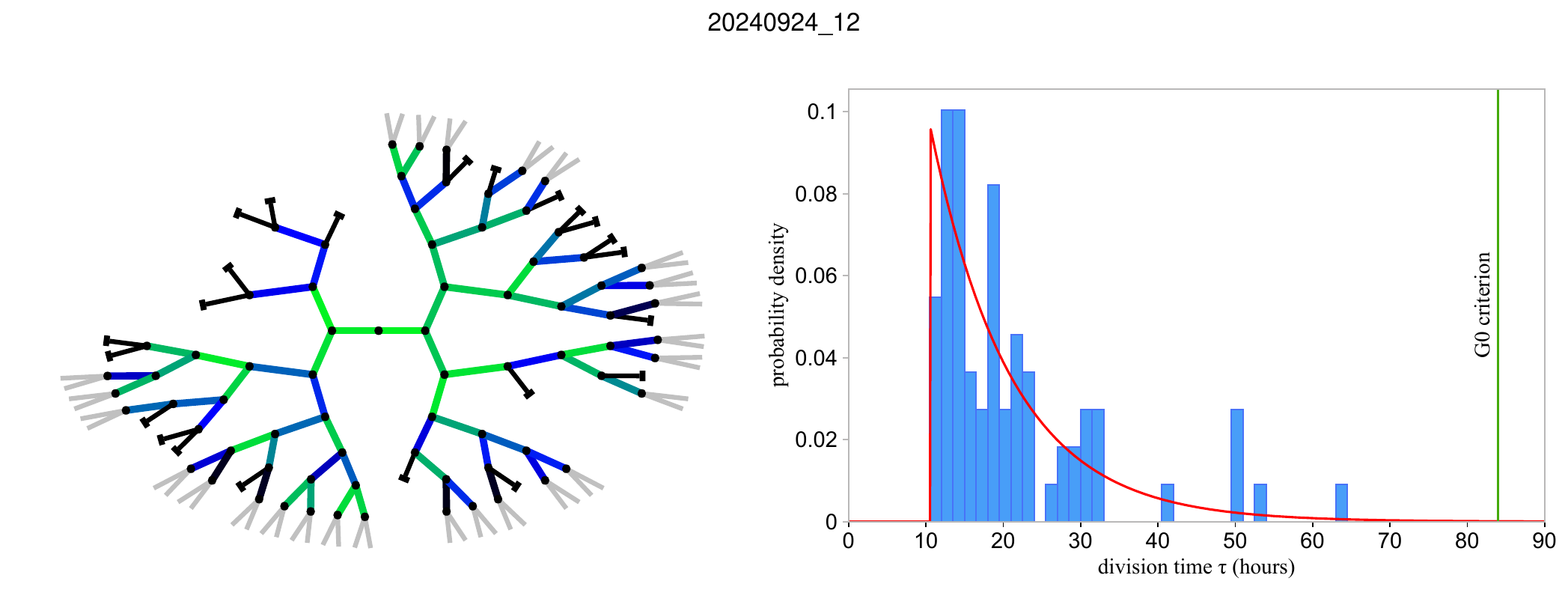}
\end{subfigure}
\hspace{0.pt}
\begin{subfigure}{0.48\textwidth}
  \centering
  \includegraphics[width=\textwidth]{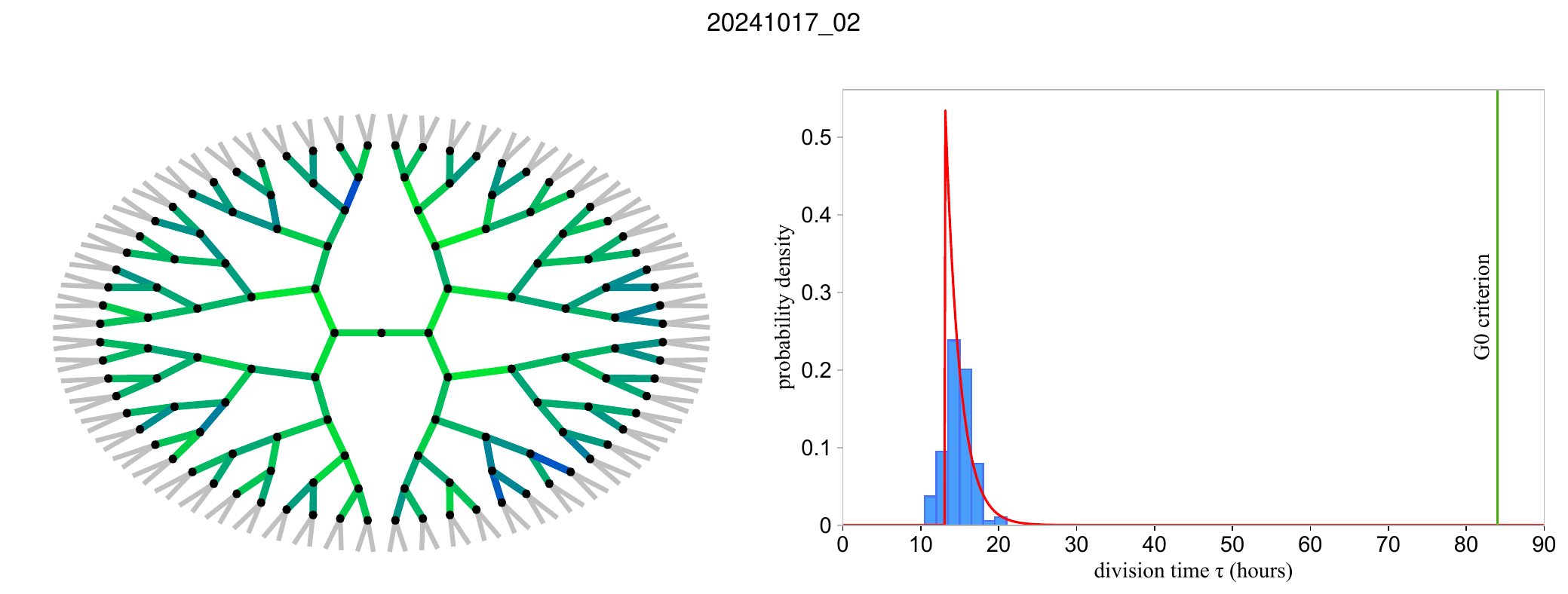}
\end{subfigure}

\vspace{0.5em}

\begin{subfigure}{0.48\textwidth}
  \centering
  \includegraphics[width=\textwidth]{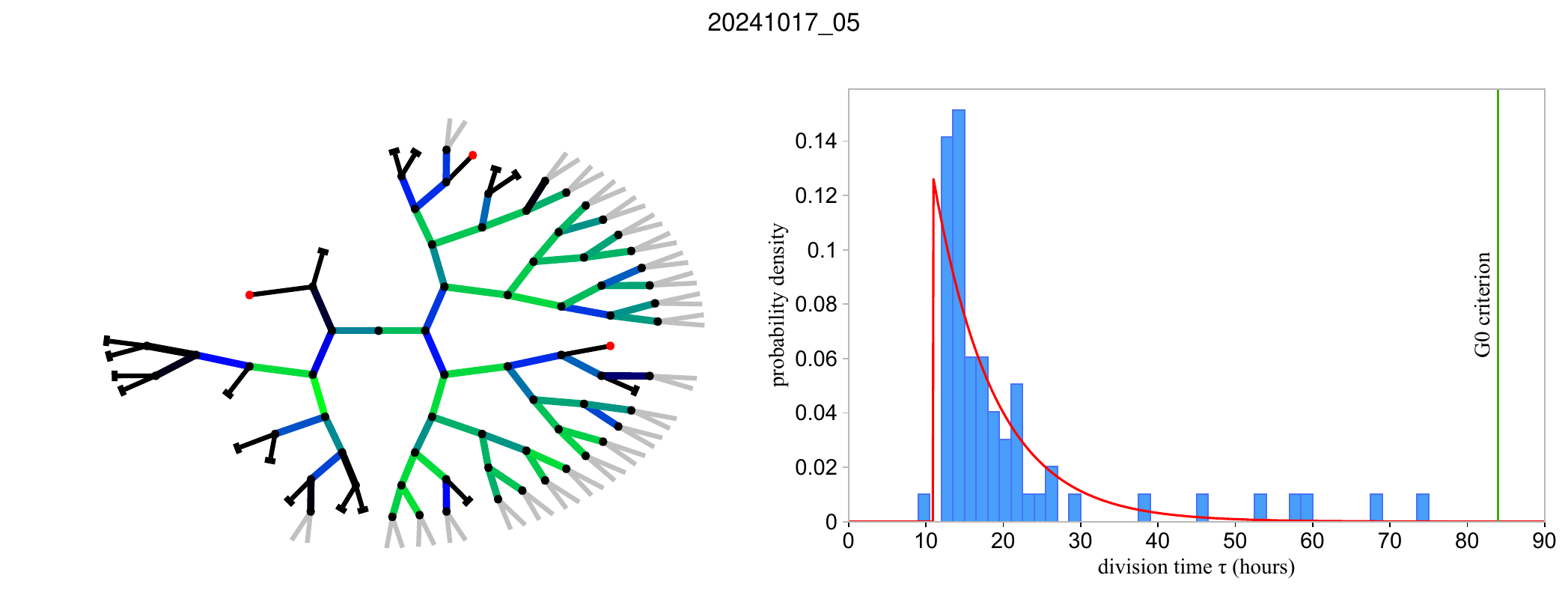}
\end{subfigure}
\hspace{0.pt}
\begin{subfigure}{0.48\textwidth}
  \centering
  \includegraphics[width=\textwidth]{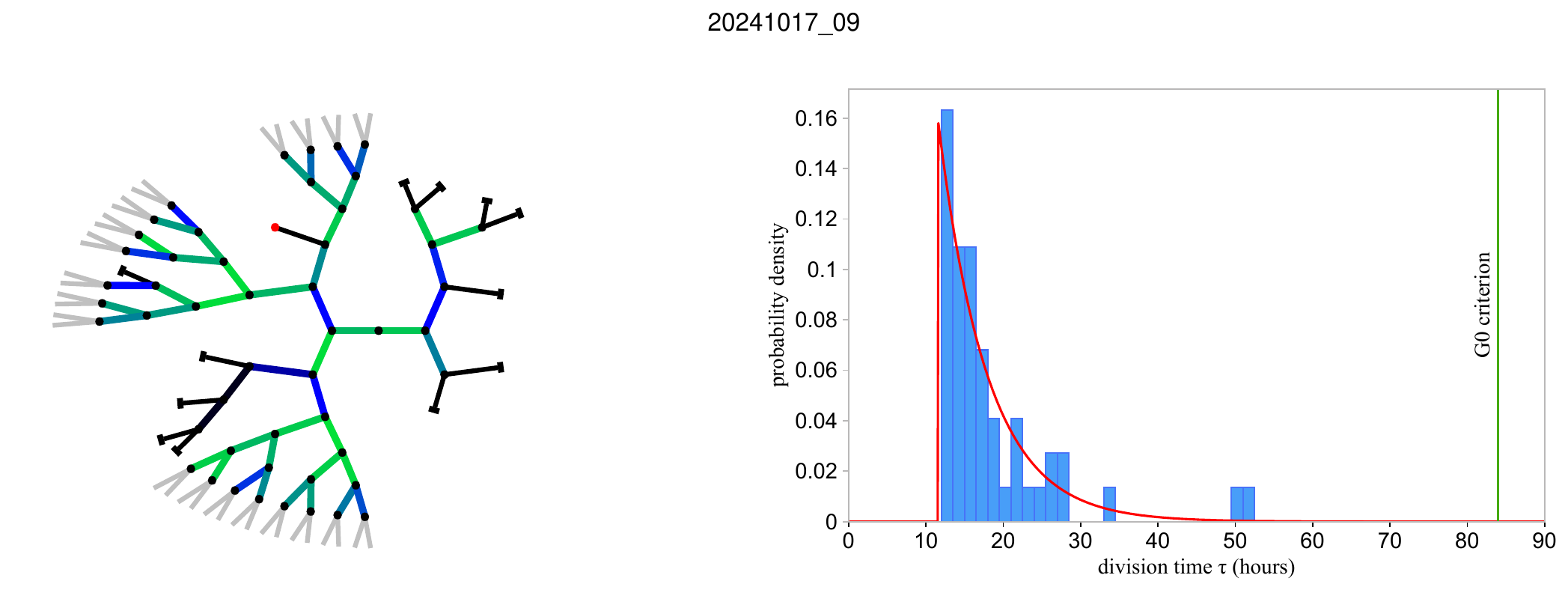}
\end{subfigure}

\vspace{0.5em}

\begin{subfigure}{0.48\textwidth}
  \centering
  \includegraphics[width=\textwidth]{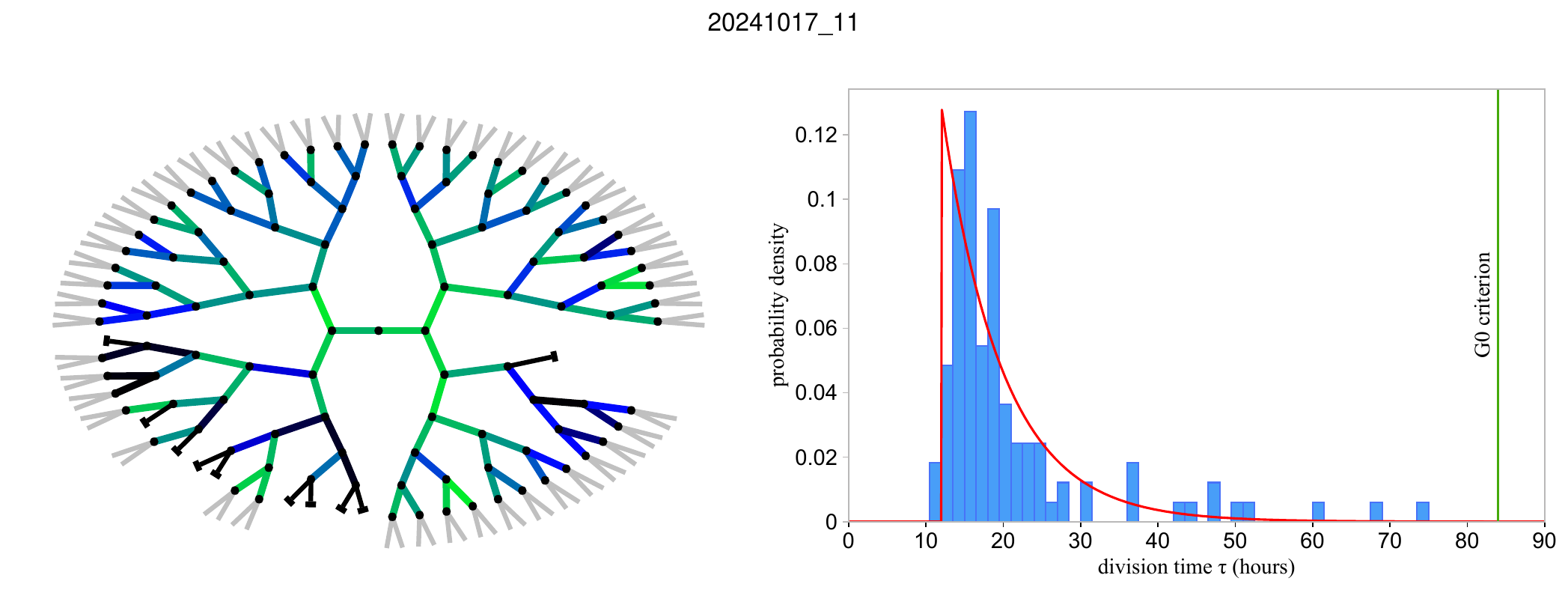}
\end{subfigure}
\hspace{0.pt}
\begin{subfigure}{0.48\textwidth}
  \centering
  \includegraphics[width=\textwidth]{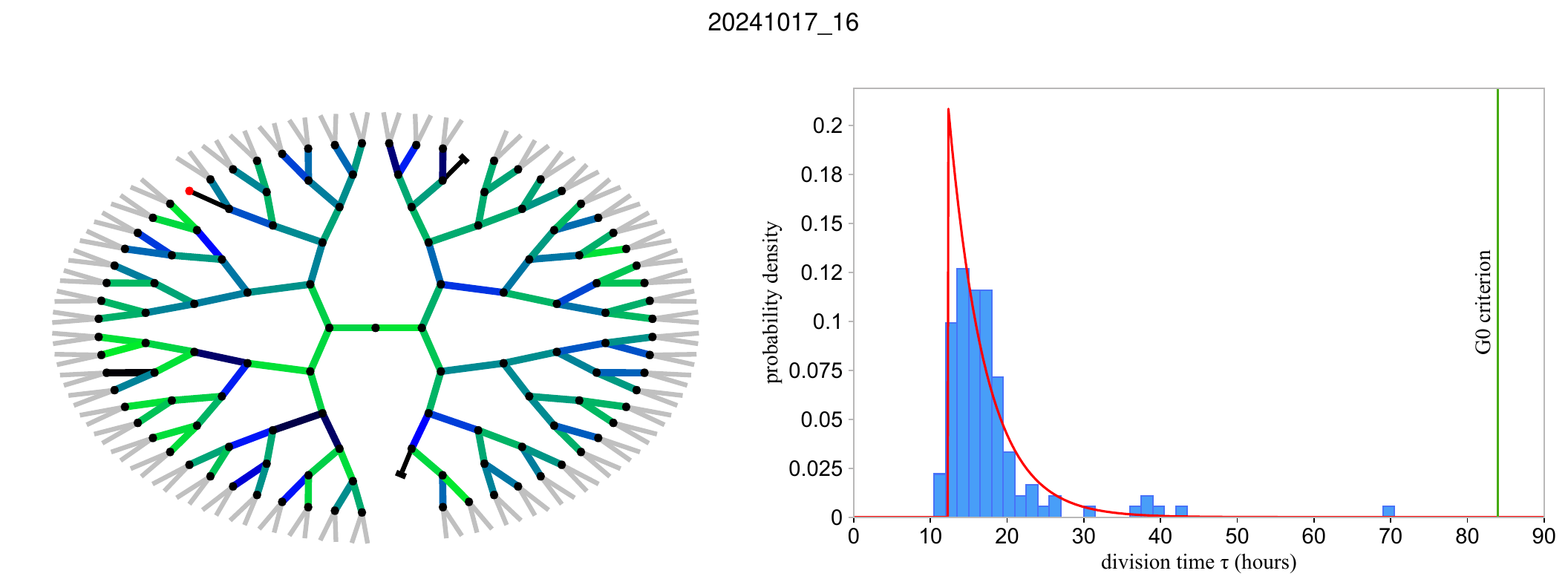}
\end{subfigure}

\vspace{0.5em}

\begin{subfigure}{0.48\textwidth}
  \centering
  \includegraphics[width=\textwidth]{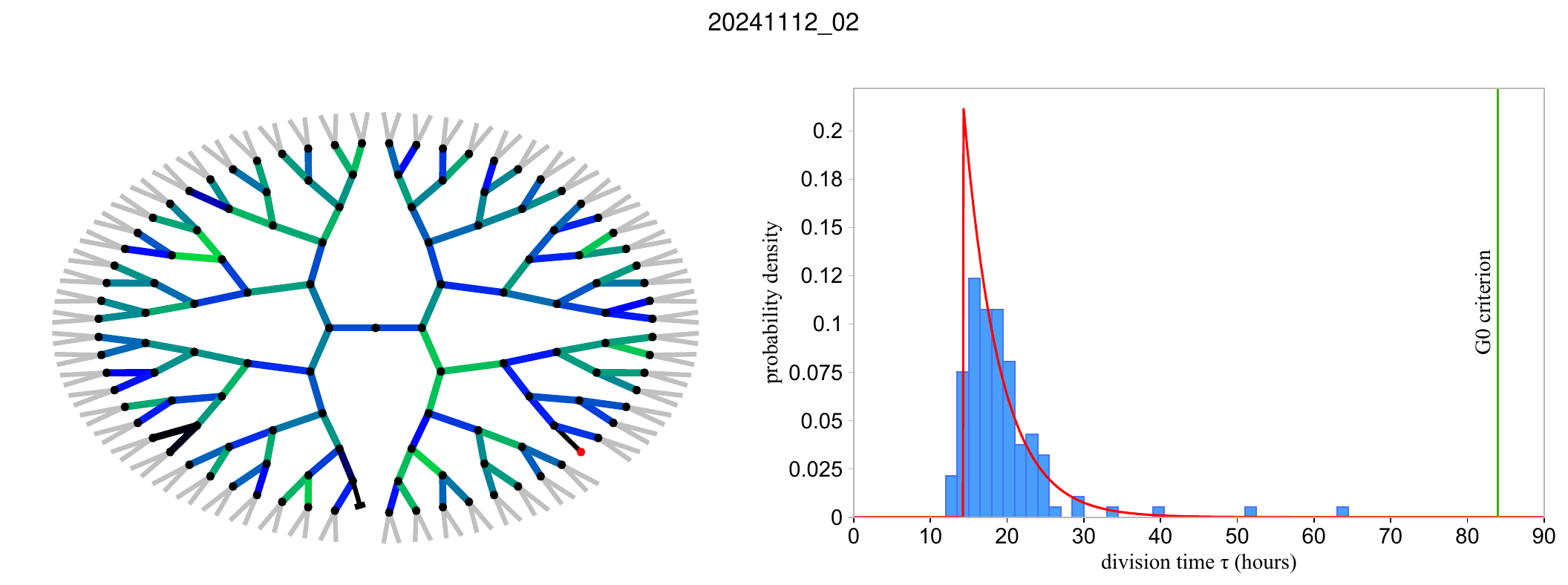}
\end{subfigure}
\hspace{0.pt}
\begin{subfigure}{0.48\textwidth}
  \centering
  \includegraphics[width=\textwidth]{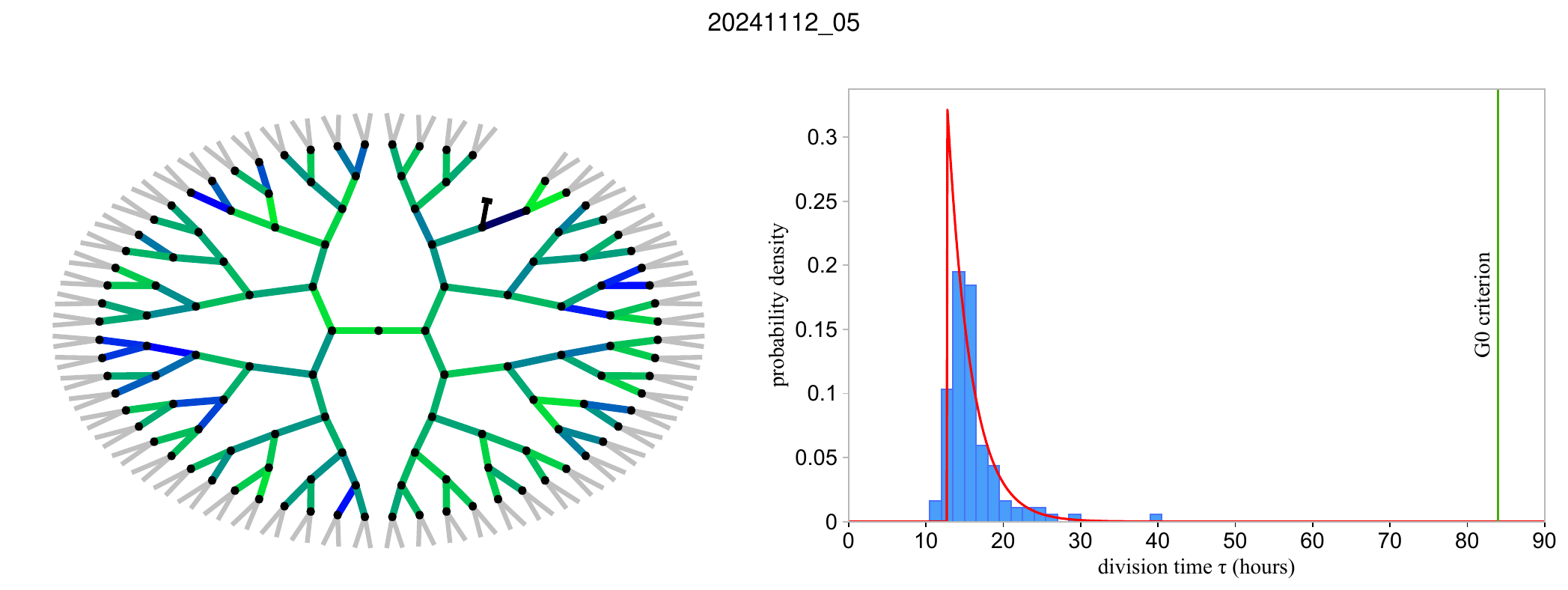}
\end{subfigure}

\vspace{0.5em}

\begin{subfigure}{0.48\textwidth}
  \centering
  \includegraphics[width=\textwidth]{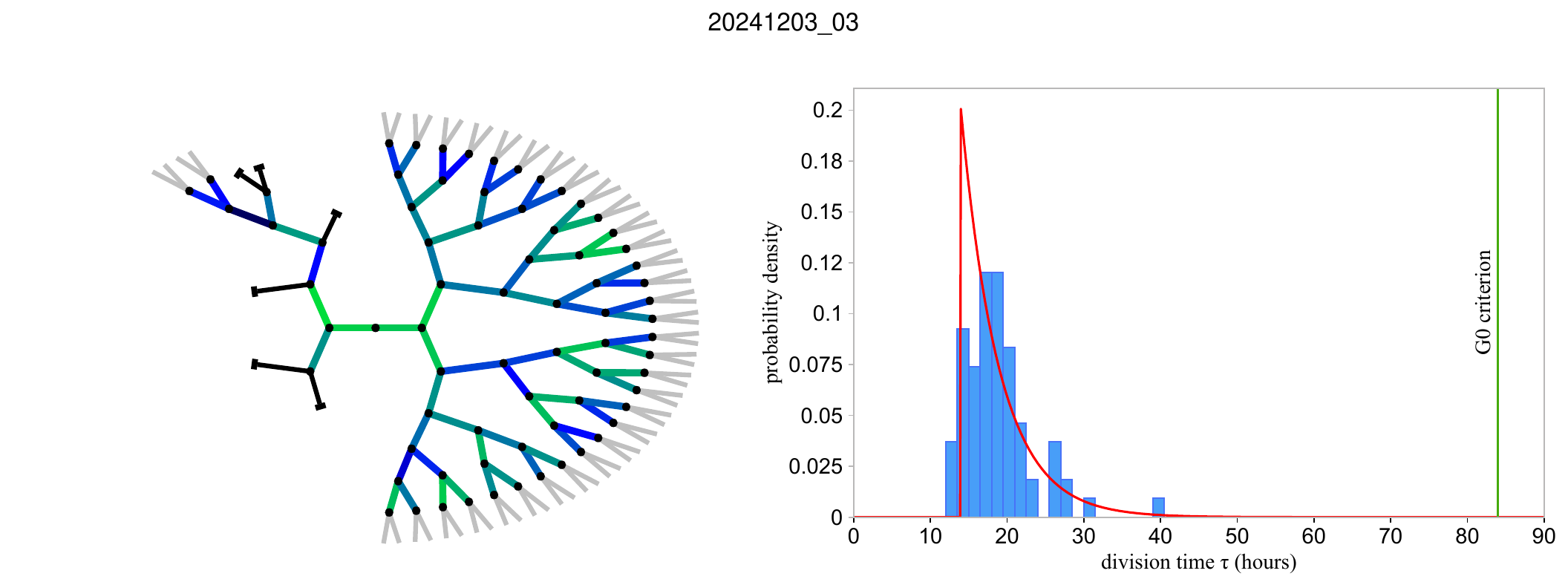}
\end{subfigure}
\hspace{0.pt} 
\begin{subfigure}{0.48\textwidth}
  \centering
  \includegraphics[width=\textwidth]{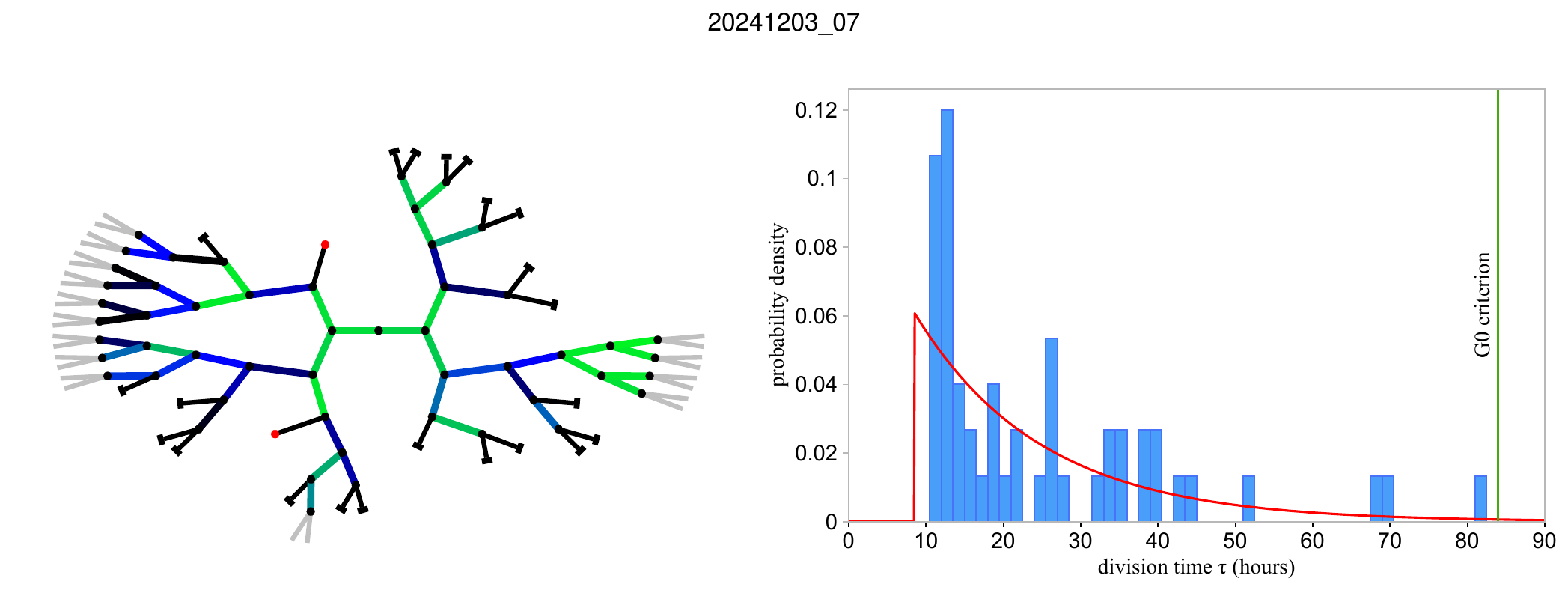}
\end{subfigure}

\captionsetup{width=\dimexpr\textwidth\relax, font=small, justification=justified, skip=0cm}
\caption*{}
\end{figure}

%%%%%%%
\newpage
\subsection*{{\bf Figure S1 (continued)}}

\begin{figure}[h!] 
\centering

\begin{subfigure}{0.48\textwidth}
  \centering
  \includegraphics[width=\textwidth]{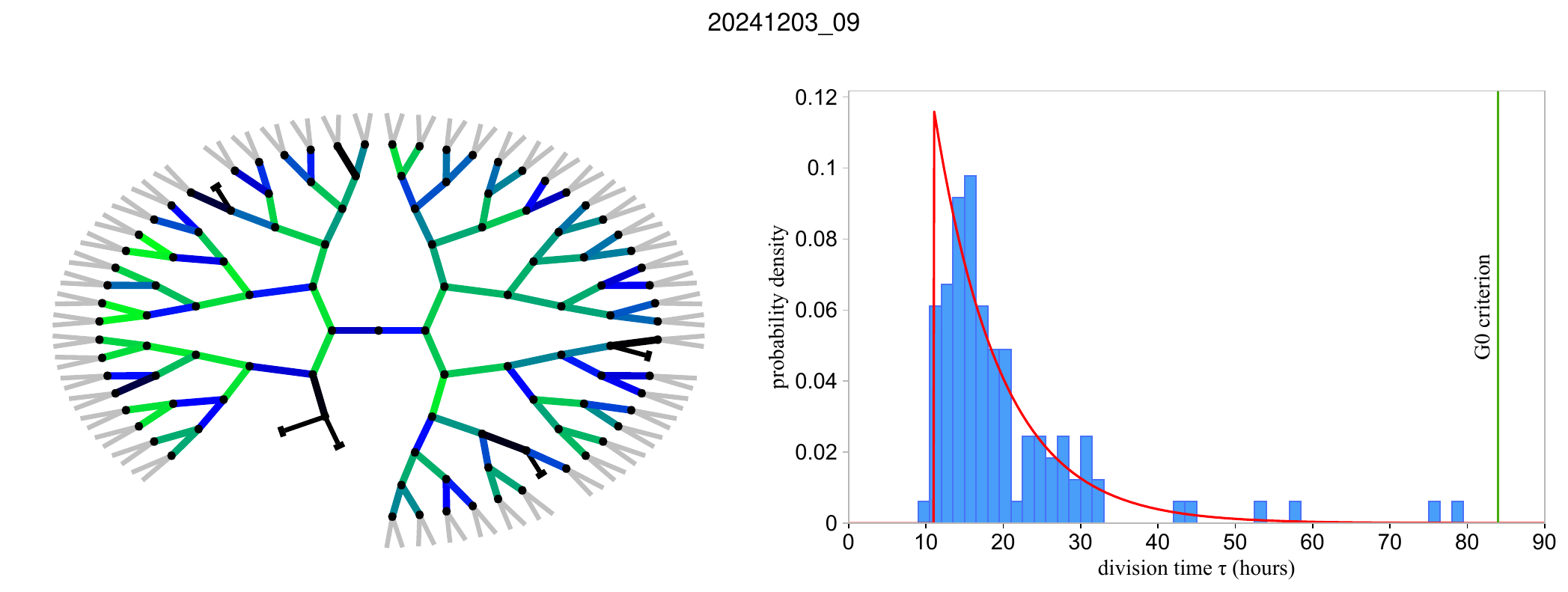}
\end{subfigure}
\hspace{0.pt}
\begin{subfigure}{0.48\textwidth}
  \centering
  \includegraphics[width=\textwidth]{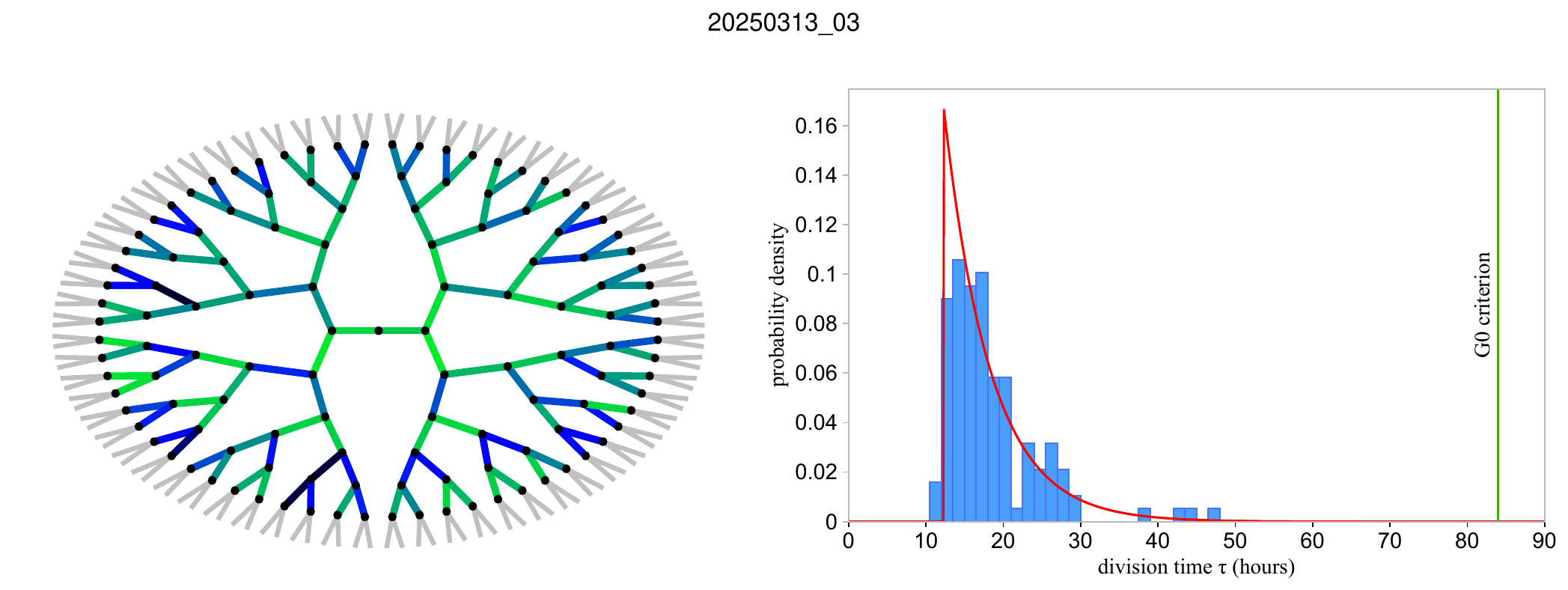}
\end{subfigure}

\vspace{0.5em}

\begin{subfigure}{0.48\textwidth}
  \centering
  \includegraphics[width=\textwidth]{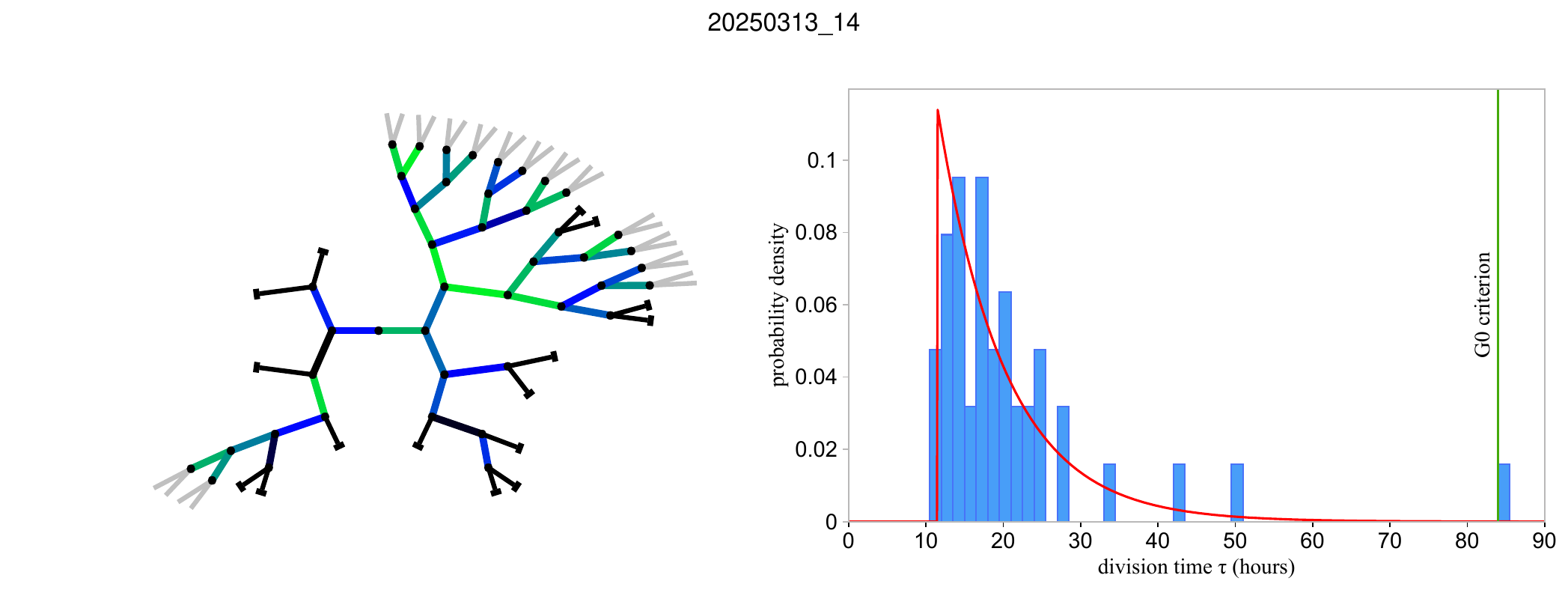}
\end{subfigure}
\hspace{0.pt}
\begin{subfigure}{0.48\textwidth}
  \centering
  \includegraphics[width=\textwidth]{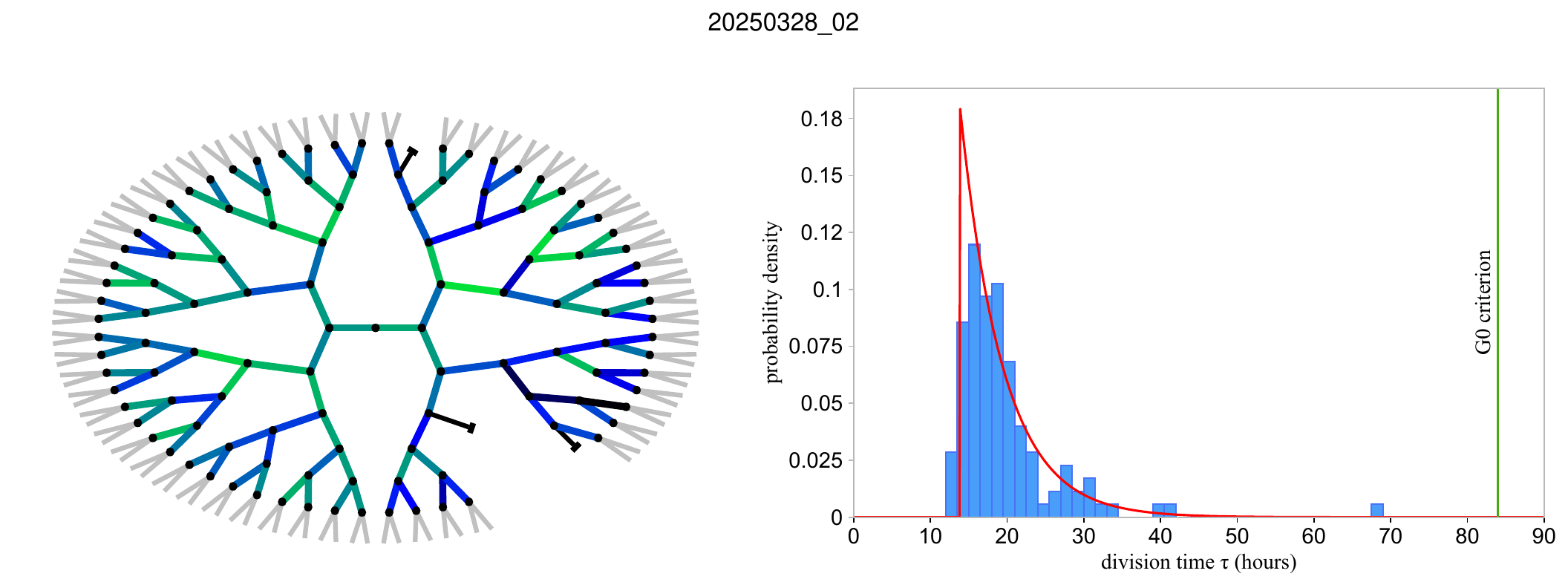}
\end{subfigure}

\vspace{0.5em}

\begin{subfigure}{0.48\textwidth}
  \centering
  \includegraphics[width=\textwidth]{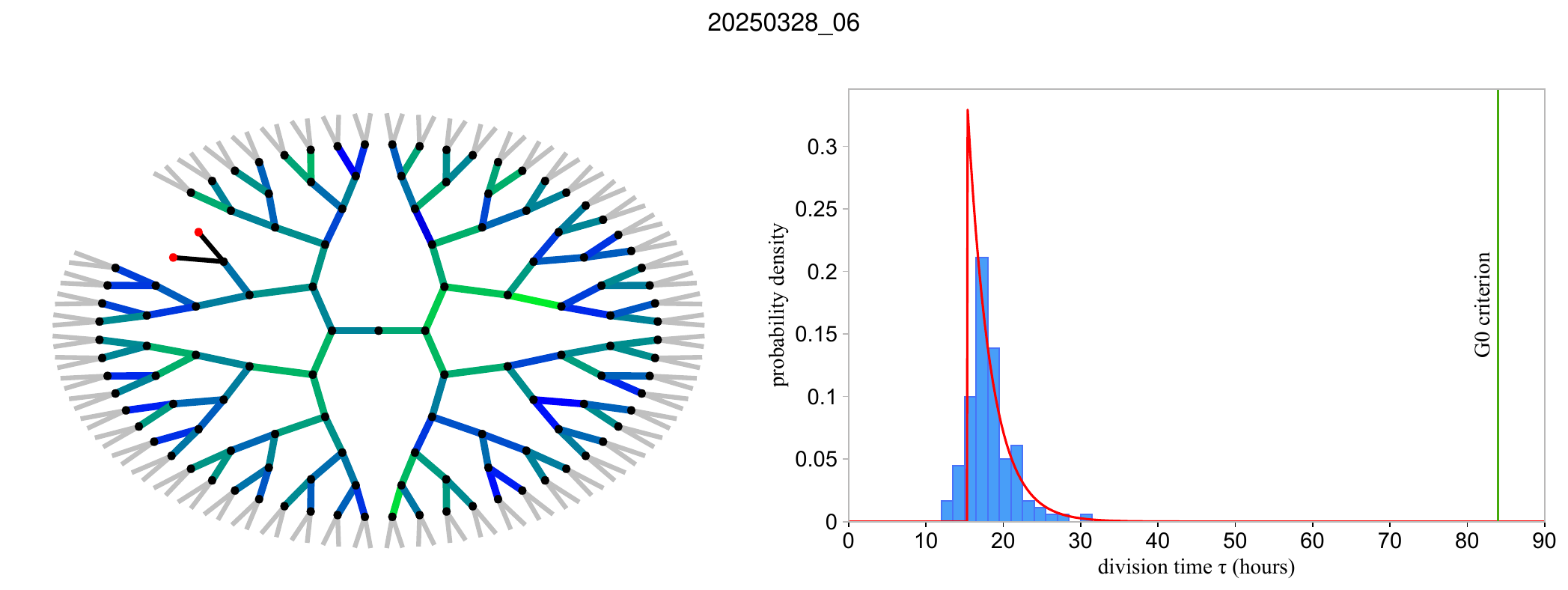}
\end{subfigure}
\hspace{0.pt}
\begin{subfigure}{0.48\textwidth}
  \centering
  \includegraphics[width=\textwidth]{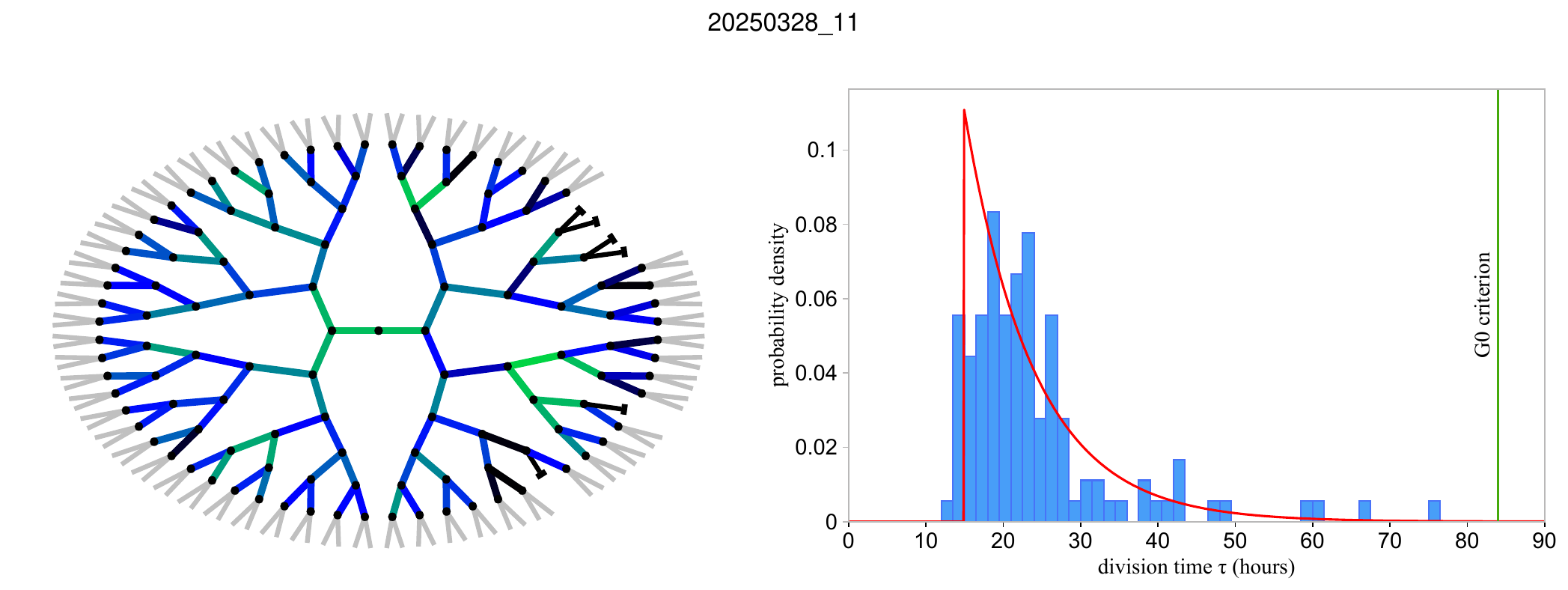}
\end{subfigure}

\vspace{0.5em}

\begin{subfigure}{0.48\textwidth}
  \centering
  \includegraphics[width=\textwidth]{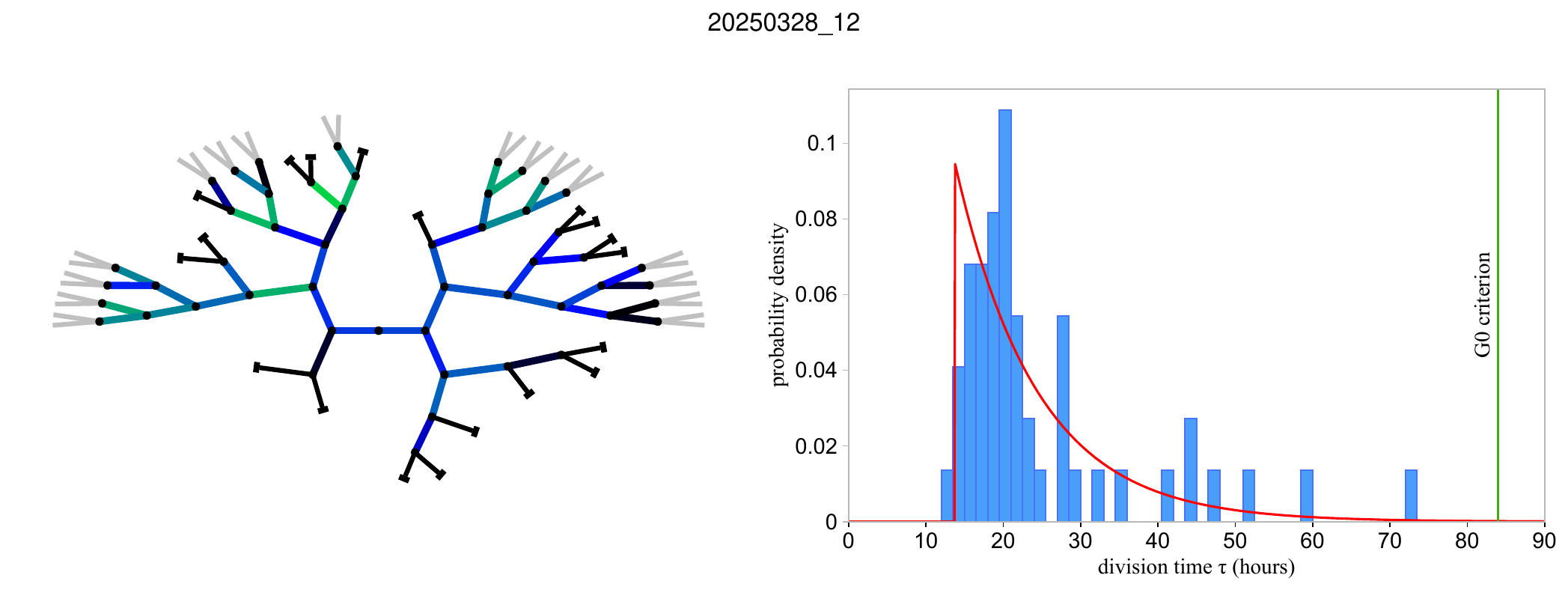}
\end{subfigure}
\hspace{0.pt}
\begin{subfigure}{0.48\textwidth}
  \centering
  \includegraphics[width=\textwidth]{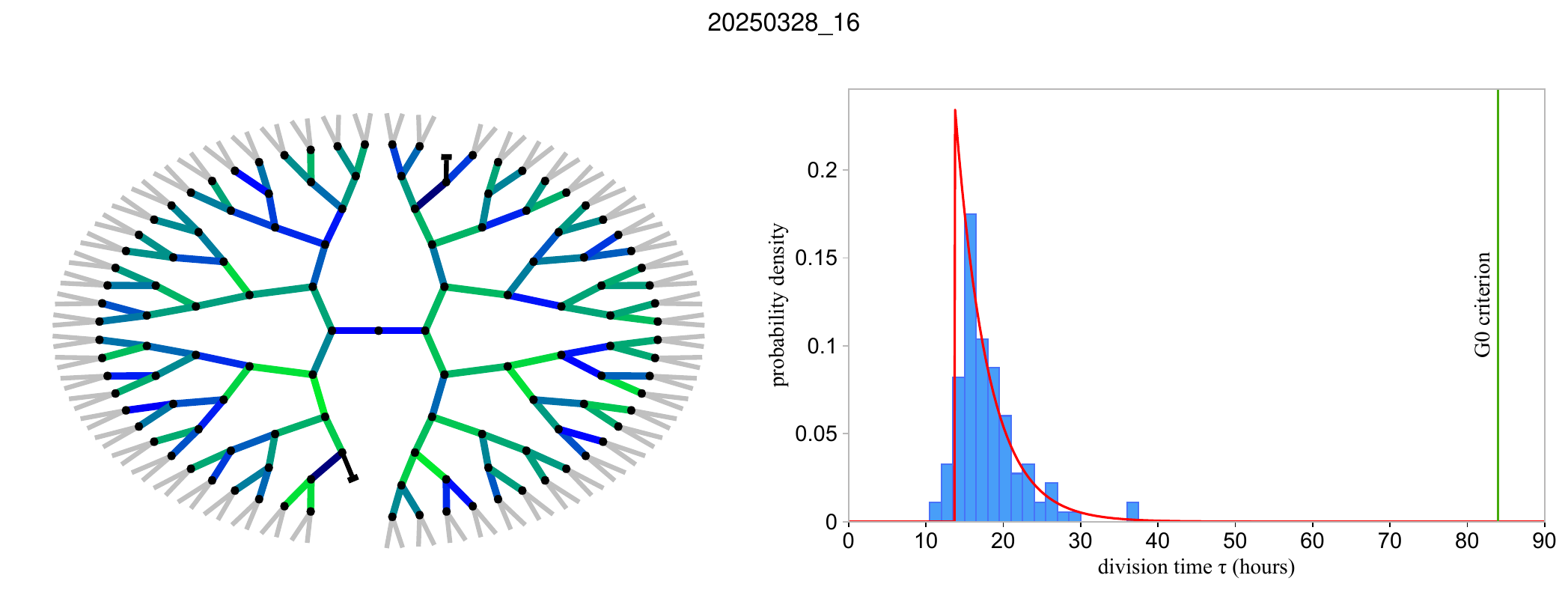}
\end{subfigure}

\vspace{0.5em}

\begin{subfigure}{0.48\textwidth}
  \centering
  \includegraphics[width=\textwidth]{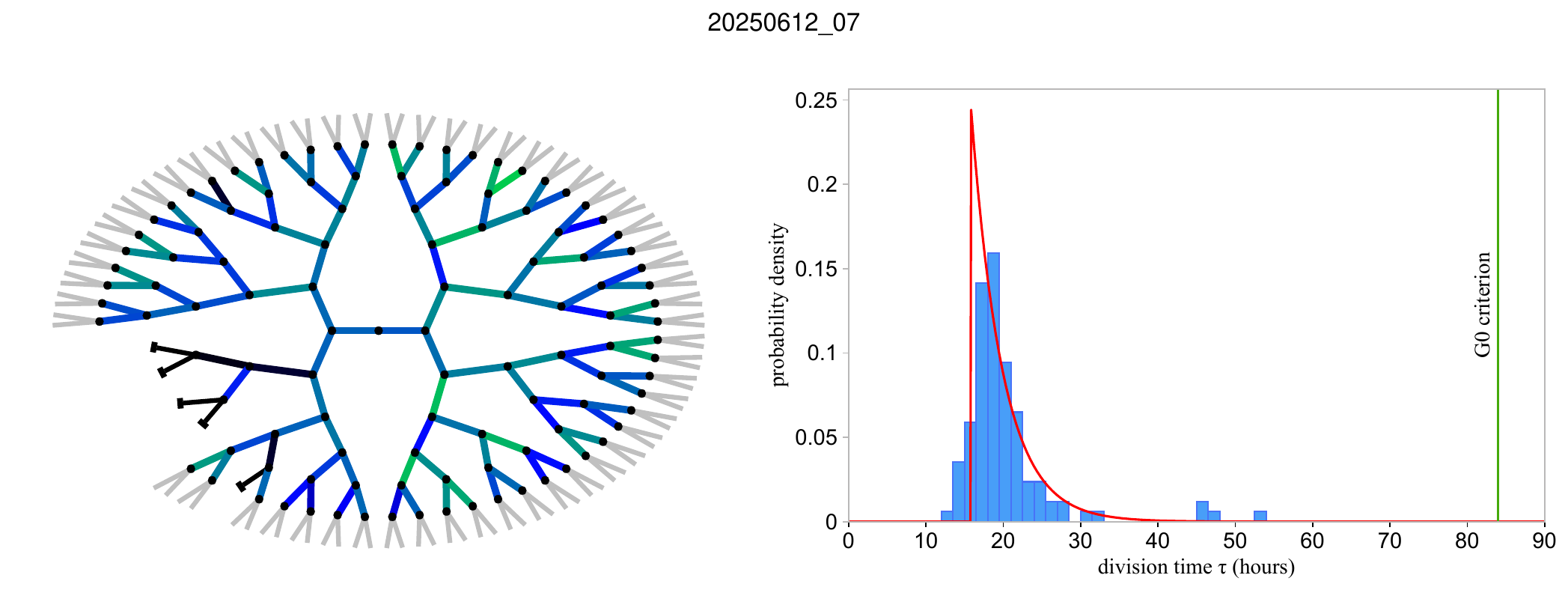}
\end{subfigure}
\hspace{0.pt}
\begin{subfigure}{0.48\textwidth}
  \centering
  \includegraphics[width=\textwidth]{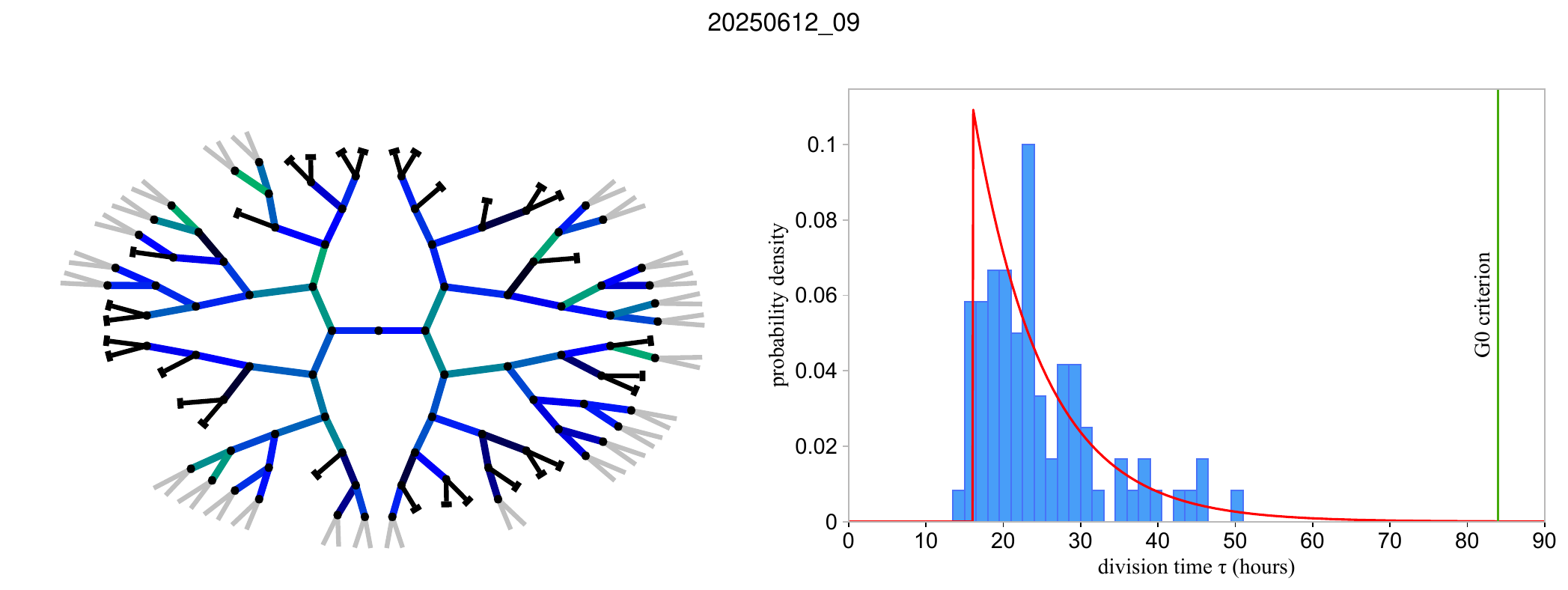}
\end{subfigure}

\vspace{0.5em}

\begin{subfigure}{0.48\textwidth}
  \centering
  \includegraphics[width=\textwidth]{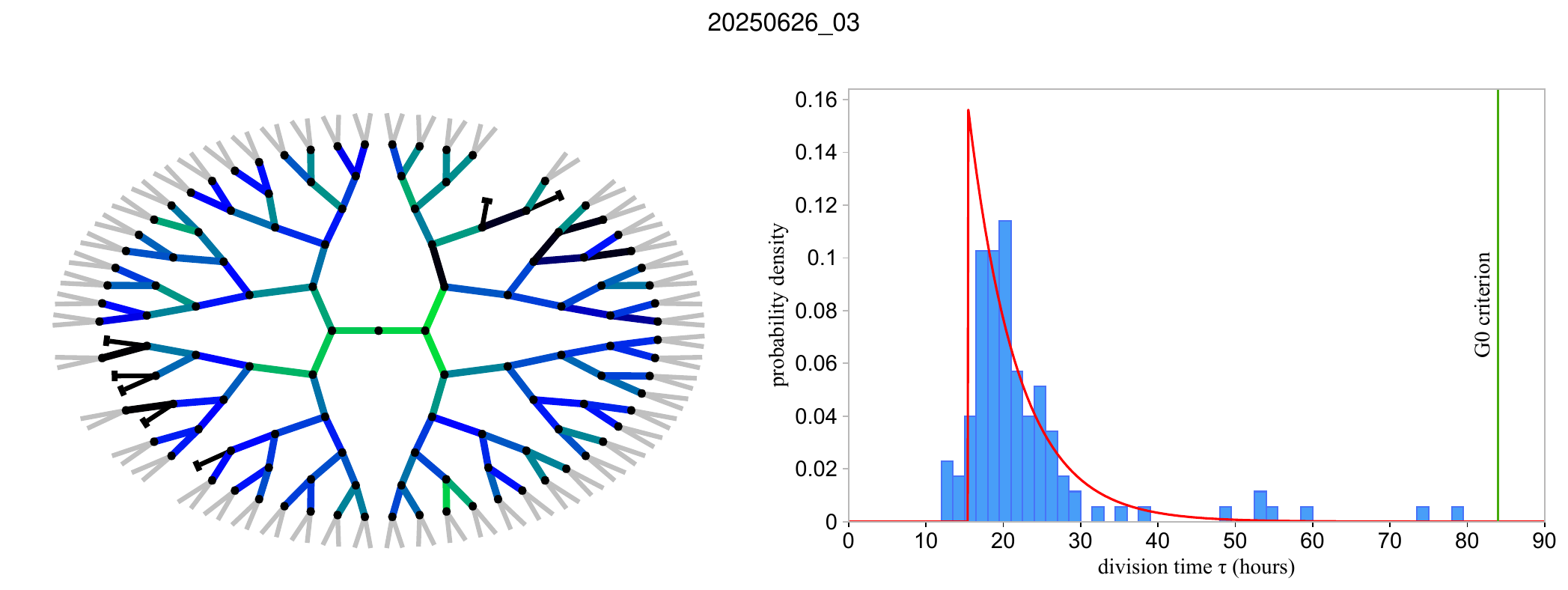}
\end{subfigure}
\hspace{0.pt}
\begin{subfigure}{0.48\textwidth}
  \centering
  \includegraphics[width=\textwidth]{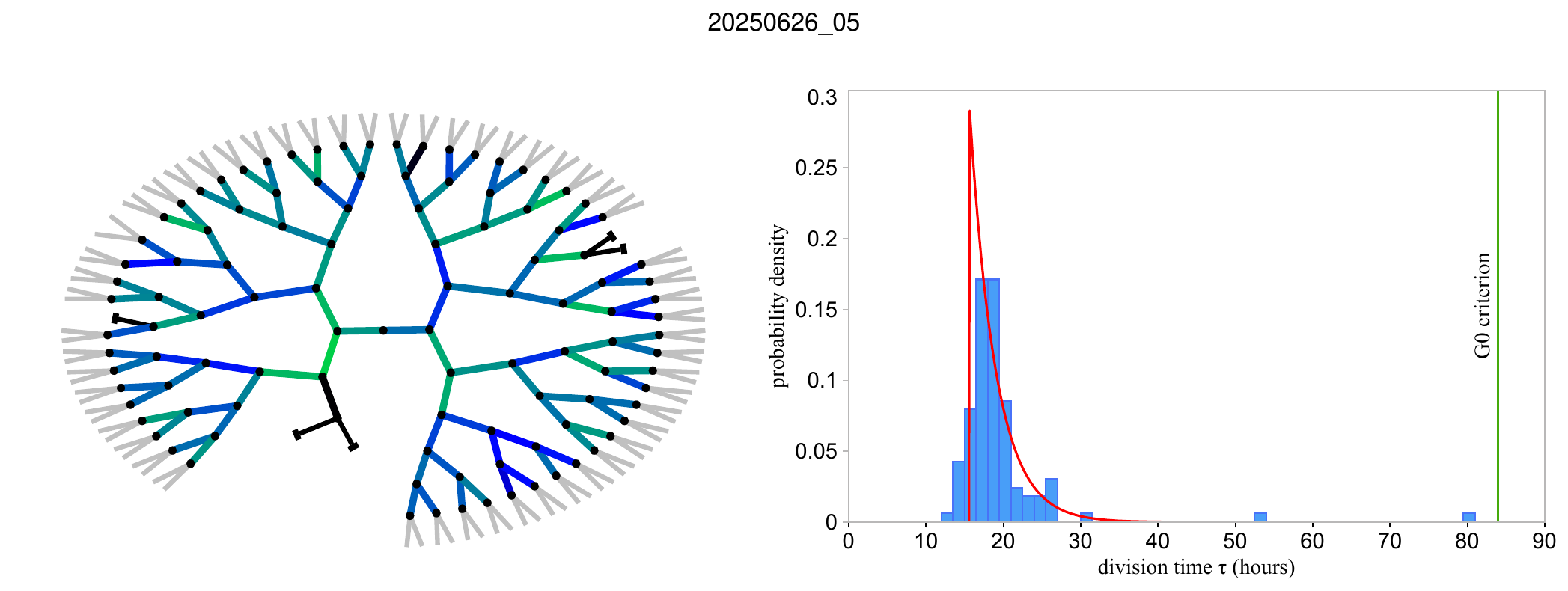}
\end{subfigure}

\captionsetup{width=\dimexpr\textwidth\relax, font=small, justification=justified, skip=0cm}
\caption*{}
\end{figure}

\newpage
\subsection*{Figure S2}
\begin{figure}[h!]
  \centering
  \includegraphics[width=0.9 \textwidth]{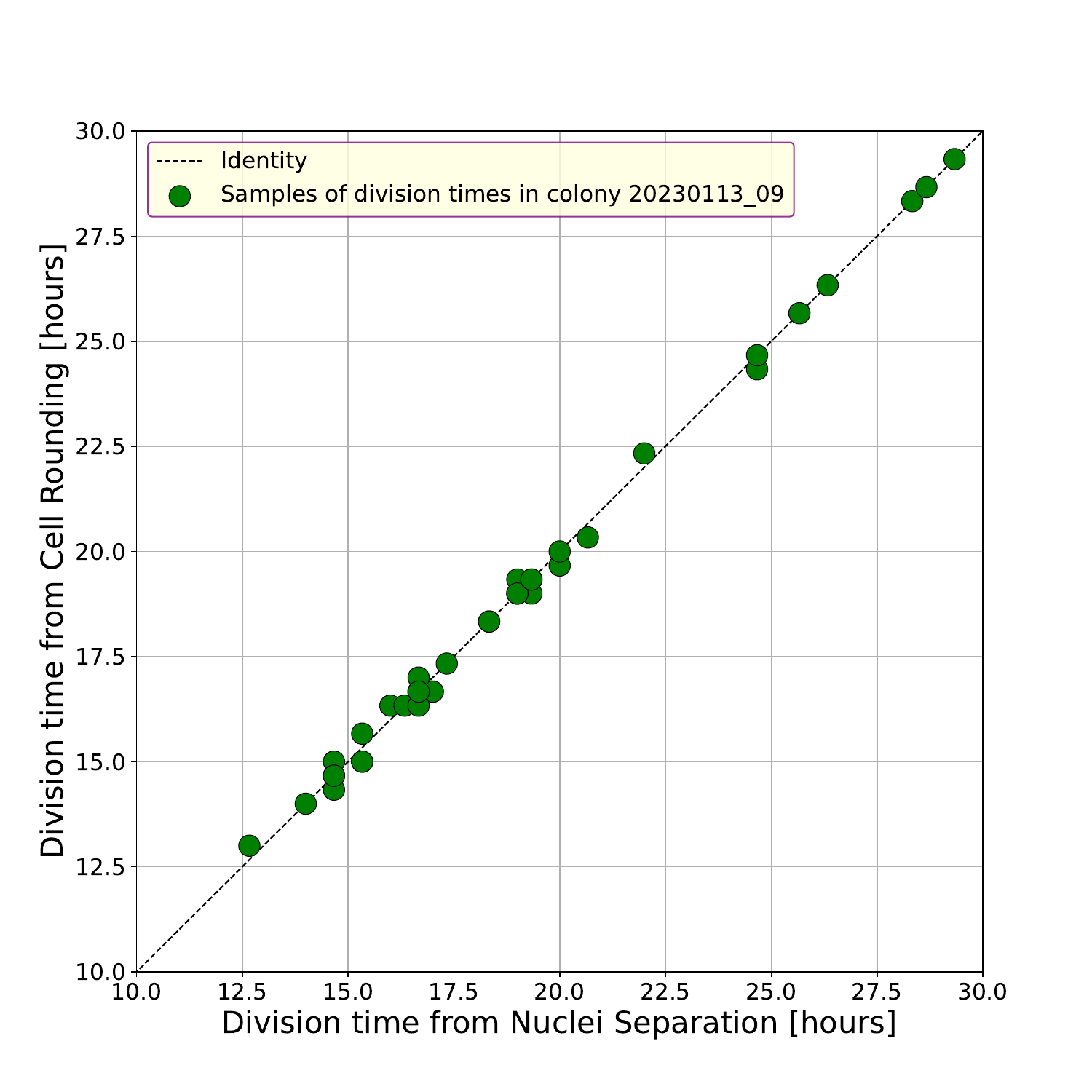}
  \caption*{ {\bf{Figure S2}: Different ways to determine the division times.} 
  We determine the division time, i.e. the duration of the cell cycle, through the mitotic cell rounding, a very short phase during which the cell assumes an almost spherical shape, appearing as a bright, circular object. This trait is easily recognisable in the image, which is why we use it an unambiguous time stamp of the cell cycle. An alternative time stamp of mitosis is the first image at which the two daughters' nuclei are clearly separated from each other; this criterion, however, has a larger degree of arbitrariness and it is operator-dependent, because the exact time at which the two nuclei become clearly separated depends on the morphology of those particular cells.  In this figure we show that these two times are anyway perfectly correlated to each other.  In colony \texttt{20230113\_09} we use both determinations of the division time in a subsample of mitotic events and show that the two definitions are completely equivalent.}
 \label{fig:retta}
\end{figure}

\newpage

\subsection*{Figure S3}
\begin{figure}[h!]
  \centering
  \includegraphics[width=0.9 \textwidth]{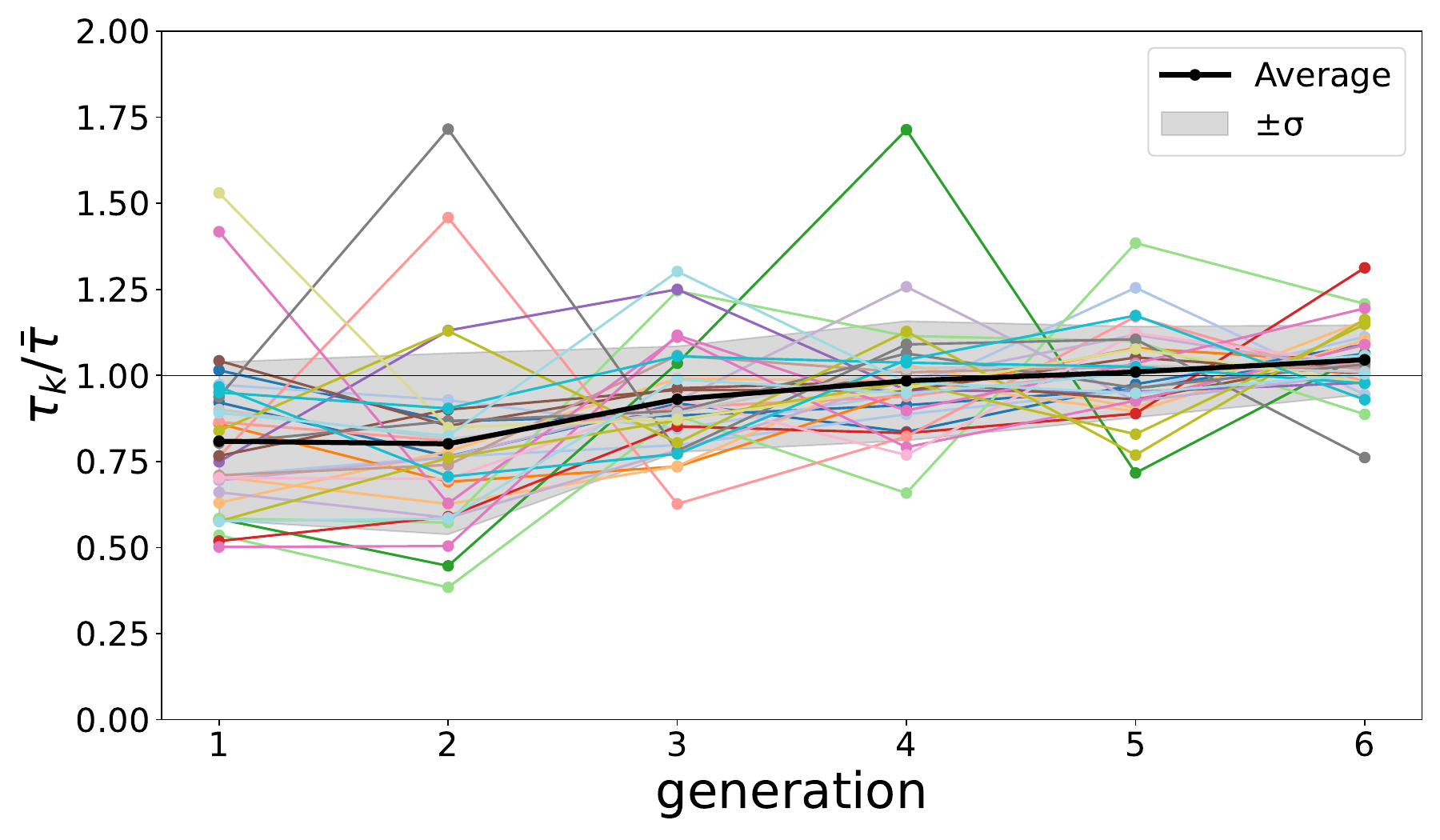}
  \caption*{ \textbf{ Figure S3: Average division time vs.\ generation.}  For each colony, the average division time at fixed generation, $\tau_{k}$, divided by the colony's global average division time, $\overline{\tau}$, is plotted against generation number, $k$.  The black curve is the average of $\tau_{k}/\overline{\tau}$ over all colonies.  Despite fluctuations, a global trend toward increasing division times with generation is observed.  Some colonies show a dip in division time at generation $k=2$, but this is not detectable in the global average (the drop between $k=1$ and $k=2$ observed in some of the colonies may be ascribed to trypsinisation \cite{staudte1984}).
  Notice that later generations have more cells than earlier ones, hence they weight more in the determination of $\overline{\tau}$; this is the reason why the black curve has a smaller offset with respect to 1 in later generations than in earlier ones.}
  \label{fig:tauav-vs-k}
\end{figure}

\newpage

\end{document}